\documentclass[11pt,a4paper]{article}

\usepackage{jcappub}
\usepackage{graphicx,wasysym}
\usepackage{rotating}
\usepackage{color}
\usepackage{bm}
\usepackage{hyperref}
\usepackage{url}

\newcommand{\be}{\begin{equation}}
\newcommand{\ee}{\end{equation}}

\newcommand{\GeV}{{\rm GeV}}

\newcommand{\TeV}{{\rm TeV}}

\newcommand{\kpc}{{\rm kpc}}

\newcommand{\cm}{{\rm cm}}

\begin{document}

\hspace*{100mm}{\large \tt FERMILAB-PUB-13-320-A}
\vskip 0.2in

\title{Constraints on dark matter annihilations from diffuse gamma-ray emission in the Galaxy}

\author[a,b]{Maryam Tavakoli}
\author[c]{,Ilias Cholis}
\author[a]{,Carmelo Evoli}
\author[d,e]{,Piero Ullio}

\affiliation[a]{{II.} Institut f\"ur Theoretische Physik, Universit\"at Hamburg, Luruper Chaussee 149, 22761 Hamburg, Germany}
\affiliation[b]{School of Astronomy, Institute for Research in Fundamental Sciences (IPM), P.O. Box 19395-5531, Tehran, Iran}
\affiliation[c]{Fermi National Accelerator Laboratory, Center for Particle Astrophysics, Batavia, IL 60510, USA}
\affiliation[d]{SISSA, Via Bonomea, 265, 34136 Trieste, Italy}
\affiliation[e]{INFN, Sezione di Trieste, Via Bonomea 265, 34136 Trieste, Italy}

\emailAdd{maryam.tavakoli@desy.de}
\emailAdd{cholis@fnal.gov}
\emailAdd{carmelo.evoli@desy.de}
\emailAdd{ullio@sissa.it}

\date{\today}

\abstract{
Recent advances in $\gamma$-ray cosmic ray, infrared and radio astronomy have allowed us to develop a significantly better understanding 
of the galactic medium properties in the last few years. In this work using the DRAGON code, that numerically solves the CR propagation 
equation and calculating $\gamma$-ray emissivities in a 2-dimensional grid enclosing the Galaxy, we study in a self consistent manner 
models for the galactic diffuse $\gamma$-ray emission. Our models are cross-checked to both the available CR and $\gamma$-ray data. 
We address the extend to which dark matter annihilations in the Galaxy can contribute to the diffuse $\gamma$-ray flux towards different directions 
on the sky. Moreover we discuss the impact that astrophysical uncertainties of non DM nature, have on the derived $\gamma$-ray limits. Such 
uncertainties are related to the diffusion properties on the Galaxy, the interstellar gas and the interstellar radiation field energy densities. Light 
$\sim$10 GeV dark matter annihilating dominantly to hadrons is more strongly constrained by $\gamma$-ray observations towards the inner  
parts of the Galaxy and influenced the most by assumptions of the gas distribution; while TeV scale DM annihilating dominantly to leptons has its 
tightest constraints from observations towards the galactic center avoiding the galactic disk plane, with the main astrophysical uncertainty  being the 
radiation field energy density. In addition, we present a method of deriving constraints on the dark matter distribution profile from the diffuse $\gamma$-ray 
spectra. These results critically depend on the assumed mass of the dark matter particles and the  type of its end annihilation products.
}

\maketitle

\section{Introduction}

The nature of dark matter (DM) consisting of 85$\%$ of the total matter in the Universe \cite{Hinshaw:2012aka, Ade:2013zuv} remains to date a prominent subject 
of research. The general class of weakly interacting massive particles (WIMPs), can be probed either by their direct detection 
\cite{Aprile:2013doa, Ahmed:2009zw, Agnese:2013rvf, Aalseth:2012if, Bernabei:2010mq, Angloher:2011uu, Akerib:2012ys, Behnke:2010xt, Armengaud:2012pfa}, at colliders (e.g. \cite{Aaltonen:2012jb, Fox:2012ee, Chatrchyan:2012tea, Bell:2012rg, ATLAS:2012ky, Dutta:2013sta, Lin:2013sca}), or through indirect detection probes via the 
identification of WIMP annihilation yields (for some reviews on indirect detection probes see \cite{Cirelli:2012tf, Serpico:2011wg, Bringmann:2012ez, Daw:2010ud} 
and for a review on WIMP DM candidates see, e.g. \cite{Bertone:2010zz}).  The latter has been gaining more and more relevance the last years due
 to the wealth of data from experiments as \textit{PAMELA}, \textit{AMS}-02, \textit{Fermi}-LAT and \textit{Planck} 
 to name the most recent. 
There is a vast literature on possible targets and messengers for indirect WIMP detection, sometimes addressing possible indication of a signal, but in most cases presenting upper limits. 
An issue, which is however not always transparent in this class of analyses, is how the various astrophysical uncertainties affect background estimates and their impact on derived limits. 

The present analysis, is an update on one of the prominent targets for indirect detection, i.e. the search for WIMP annihilations in the halo of our own Galaxy. 
It aims at presenting updated limits from currently available gamma- and cosmic-ray data, as well as discussing the relevance of the uncertainties involved. 
For such a goal, we implement a framework in which cosmic-ray propagation and $\gamma$-ray emissivities are treated numerically with the use of the  DRAGON code~\cite{DRAGONweb,Evoli:2008dv};  treating self-consistently both the standard astrophysical components and the terms associated to WIMP annihilations. 
The models that we are considering when including only the standard astrophysical sources, are in general agreement with both local cosmic ray data and with measurements of the $\gamma$-ray flux made available by the  \textit{Fermi}-LAT $\gamma$-ray telescope. 

In \cite{FermiLAT:2012aa, Timur:2011vv, Cholis:2011un, Tavakoli:2011wz}, broad studies have been performed to quantify the astrophysical 
uncertainties on the estimate of the diffuse galactic $\gamma$-ray background. While in \cite{Ackermann:2012rg, Cirelli:2013mqa} the impact of background astrophysical 
uncertainties in deriving limits on DM annihilation from $\gamma$-rays towards the galactic center (GC) have been discussed. Among these uncertainties  one of the
 most important,  is the distribution of gas in the Milky Way, for which the large scale properties can be constrained by a combined analysis of cosmic-ray and $\gamma$-ray spectra \cite{Cholis:2011un}.
 
In this work, we use the most recent models for the distribution of the interstellar atomic and molecular hydrogen gas \cite{Tavakoli:2012jx, Pohl:2007dz} (discussed in sections~\ref{subsec:Gas} and~\ref{subsec:DarkGas}).
Using these distributions, we estimate the interaction rate of cosmic rays during their propagation within the Galaxy and evaluate diffuse $\gamma$-ray spectra.
We then explore the possible contribution of WIMP annihilations to diffuse $\gamma$-ray fluxes from different sky regions. We also discuss the impact that astrophysical uncertainties have over $\gamma$-ray induced limits on models of DM, over different patches that cover the entire sky. To that goal and excluding the very inner part of the Galaxy, we select the regions of interest to place constraints on the general properties of annihilating DM. 
(section~\ref{subsec:GammaLimits}). Along with the analysis devoted to $\gamma$-ray constraints, we discuss also the limits from antiprotons and leptons (see sections~\ref{subsec:pbarLimits} and~\ref{subsec:Leptons} respectively). 
In deriving limits on DM annihilation, an orthogonal uncertainty to that of the background contribution is that of DM distribution in the Galaxy. 
In section~\ref{sec:DMprofile}, we discuss the constraints which can be derived on the galactic DM profile assuming generic annihilation channels.
Discussion and comparison of other indirect DM search probes are presented in section~\ref{sec:multiwavelength}, with our summary and conclusions
 given in Section~\ref{sec:conclusions}. 
  

\section{Diffuse $\gamma$-ray Background}
\label{sec:background}

At energies above 100 MeV most of the observed photons are connected to diffuse emission. 
The bulk of these photons is from the decay of neutral pions, which themselves are produced by inelastic collisions of cosmic ray protons and helium nuclei with the interstellar gas, from bremsstrahlung emission of cosmic ray electrons and positrons in the interstellar gas and from inverse Compton scattering of cosmic ray $e^{\pm}$ off the interstellar radiation field. 
In \cite{Cholis:2011un}, by using the numerical DRAGON code, we have studied the impact of different assumptions about the properties of the interstellar medium and cosmic ray propagation on the evaluated diffuse $\gamma$-ray spectrum.
In the following, we present our assumptions for modeling the diffuse $\gamma$-ray background. 


\subsection{Diffusion Properties}

The combined analysis of the spectra of cosmic rays and diffuse $\gamma$-rays suggests a slight preference for thicker diffusion zones while there is a weak dependence on the variation of the diffusion coefficient in the radial direction \cite{Cholis:2011un}. 
Therefore, we ignore the radial dependence of the diffusion coefficient and assume that the diffusion coefficient exponentially increases outward the galactic plane only in the vertical direction,
\be
D(z,R) = D_0\beta^{\eta}\left(\frac{R}{R_0}\right)^{\delta}\exp{ \left( \frac{|z|}{z_d} \right)},
\label{eq:Diffusion}
\ee
where $R$ is the particle's rigidity $R = p/q$ ($p$ the momentum and $q$ the charge). 
Following the reference propagation model in \cite{Cholis:2011un}, the diffusion spectral index $\delta$ and the diffusion scale height $z_d$ are chosen to be equal to 0.5 and 4 kpc respectively.
Convection is neglected, since models with strong convective winds in the entire Galaxy are not favored by the combined fit to the cosmic ray and the \textit{Fermi} $\gamma$-ray data \cite{Cholis:2011un}. 

The normalization of the diffusion coefficient $D_0$, the parameter $\eta$ and the Alfv\'en velocity $v_A$ are fitted to the local spectrum of the ratio of the secondary to the primary cosmic ray nuclei, boron to carbon (B/C). 
The spectral properties of cosmic ray primary sources are fitted against the local flux of protons measured by \textit{PAMELA} \cite{Adriani:2011cu} and CREAM \cite{Yoon:2011zz}; the flux of $e^-+e^+$ observed by \textit{Fermi} \cite{Ackermann:2010ij}, MAGIC \cite{BorlaTridon:2011dk} and  H.E.S.S \cite{Aharonian:2008aa,Aharonian:2009ah}; the flux of electrons measured by \textit{PAMELA} \cite{PAMELA:2011xv} as well as the spectrum of positron fraction measured by \textit{PAMELA} \cite{Adriani:2010ib, Adriani:2008zr} and \textit{AMS}-02 \cite{2013PhRvL.110n1102A} \footnote{Recently, \textit{AMS} has presented on-line the preliminary data for proton, He, B/C and lepton CRs. There are some disagreements with older CR data in some of these cases. Yet that does not affect our calculations and results by more than a few $\%$.}.
Pulsars are included to fit the leptons spectra at high energies (for more details see \cite{Cholis:2011un}).
The predicted spectrum of antiprotons by this propagation model is in good agreement with the \textit{PAMELA} data \cite{Adriani:2010rc}. 


\subsection{Interstellar Gas}
\label{subsec:Gas}

The distribution of gas in the interstellar medium has a strong influence on the evaluated $\gamma$-ray spectra. 
Fine structures of gas distribution along lines of sight can be observed from their $\pi^0$ and bremsstrahlung emission contributing to the diffuse $\gamma$-rays.
In fact to interpret small scale features in high angular resolution maps of the \textit{Fermi} $\gamma$-ray telescope, a detailed model for the distribution of gas is demanded.

One of the main constituents of the interstellar gas is atomic hydrogen $H_I$ which is traced by the 21cm line emission. 
We use the most recent model for the three dimensional distribution of $H_I$ in the Milky Way which is constructed in \cite{Tavakoli:2012jx}.  
This model has been derived by using the combined LAB survey data \cite{Kalberla:2005ts} which is currently the most sensitive 21cm line survey with the most extensive spatial and kinematic coverage.  
The spin temperature $T_s$, which determines the opacity of the 21cm line, is assumed to be constant all over the sky.
To convert the observed brightness temperature distribution to the volume density distribution a purely circular rotation curve is assumed.   
This assumption is expected to be reasonable except for the galactic center region where the Bulge exists.
The distance ambiguity inside the solar circle is solved by estimating the vertical distribution of gas around the galactic plane as a Gaussian function.

Molecular hydrogen $H_2$ is another important ingredient of the interstellar gas which is traced by 2.6mm line emission of CO. 
We use the most recent model for the distribution of molecular hydrogen gas in the Galaxy which is given in \cite{Pohl:2007dz}. 
This model is obtained by the kinematic deconvolution of the composite CO survey of \cite{Dame:2000sp} using a new gas flow model \cite{Bissantz:2002ge}.	
The CO to $H_2$ conversion factor $X_{CO}$, which relates the $H_2$ column density $N_{H_2}$ to the velocity integrated intensity of the CO line, depends on the metallicity and the ultraviolet background radiation \cite{Shetty:2010dm, Glover:2010uz}.
It has been shown that  $X_{CO}$ increases with Galactocentric radius, although its radial gradient is largely uncertain \cite{Nakagawa:2005tw,Strong:2004td, Boselli:2001wj,Israel:1997wn, Israel:2000mx, Nakanishi:2006zf,1996PASJ...48..275A,1995ApJ...452..262S}.
Testing different physically motivated assumptions on the radial variation of $X_{CO}$, we choose the conversion factor of \cite{Boselli:2001wj} as our reference. 
The impact of using different models for the distribution of $X_{CO}$ on the diffuse $\gamma$-ray spectra is discussed in detail in Appendix \ref{sec:XCOfactor}.


\subsection{Dark Gas}
\label{subsec:DarkGas} 

The 21cm and 2.6mm emission lines suffer from absorption by interstellar gasses and may not trace all the neutral hydrogen gas within the Galaxy. 
On the other hand, $\gamma$-rays which are produced by interactions of cosmic rays with the interstellar gas do not suffer any attenuation up to energies of several TeV  within the Galaxy. 
Thus, there is a potential mis-match between the \textit{Fermi} $\gamma$-ray data and the predictions of diffuse galactic $\gamma$-rays associated with hydrogen gas distribution. 
This has already been shown for the \textit{EGRET} $\gamma$-ray sky map in \cite{2005Sci...307.1292G} and has been quoted as the ``dark gas" 
(see also \cite{FermiLAT:2012aa} for the evaluation of the dark gas related to the \textit{Fermi} $\gamma$-ray data). 
Since dust is mixed with both phases of the neutral hydrogen gas, its thermal emission can provide an alternative tracer for the interstellar gas distribution.
Therefore, we fit a linear combination of the $H_I$ column density map of \cite{Tavakoli:2012jx} and the CO column density map of \cite{Pohl:2007dz} to the 100 $\mu m$ SFD dust map of \cite{1998ApJ...500..525S} by means of maximum likelihood.
To that end, the method which is described in \cite{Dobler:2009xz} is followed. 
We apply a $2^{\circ}$ FWHM smoothing in all maps.  
There are large uncertainties in the evaluation of dark gas in regions close to the galactic disk \cite{2005Sci...307.1292G,FermiLAT:2012aa}.
Thus, we mask out the region of $|b| < 1^{\circ}$ in the maximum likelihood fits and in the evaluation of the dark gas contribution to $\gamma$-ray spectra.    


\subsection{Consistency with Data}
\label{subsec:data}

In this analysis, we use the 4-year \textit{Fermi}-LAT data which are taken from August 2008 to August 2012.
The Pass 7 (v9r27p1) ``ULTRACLEAN'' event class ensures minimal cosmic ray contamination \footnote{http://fermi.gsfc.nasa.gov/ssc/data/analysis/scitools/}.
$\gamma$-ray events with energies of 200 MeV up to 200 GeV are binned in 27 logarithmically spaced energy bins. 
The exposures and fluxes of front and back-converted events are separately calculated. 
Then their contributions are summed to obtain the total flux.
To account for contributions from known point and extended sources, the 2-year catalogue of the \textit{Fermi}-LAT is used (see \cite{Collaboration:2011bm} and references therein).
The contribution of the isotropic extragalactic background is taken from the model in \cite{Abdo:2010nz}.  
The contributions of the \textit{Fermi} bubbles/\textit{Fermi} haze and Loop I are also included using the model in \cite{Su:2010qj}.

\begin{figure}
\hspace{-1.2cm}
\includegraphics[scale=0.78]{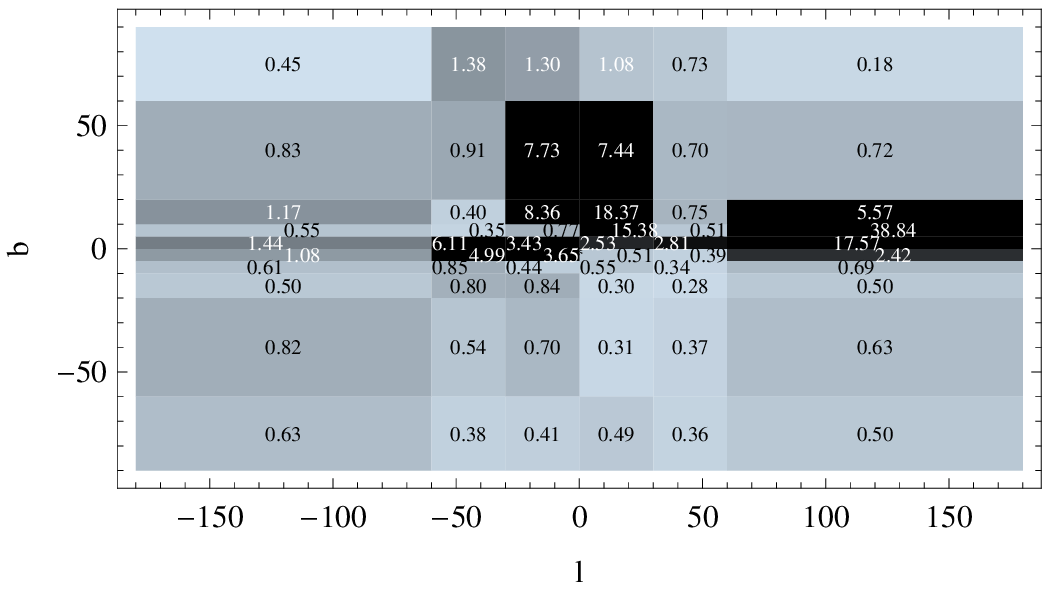}
\includegraphics[scale=0.78]{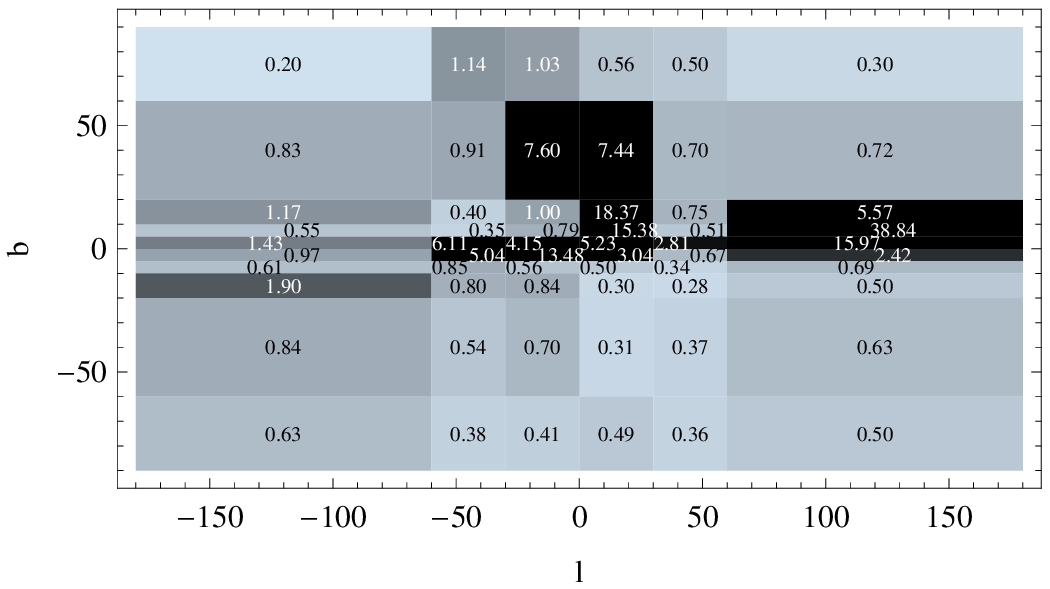}\\
\hspace*{-1.2cm}
\includegraphics[scale=0.78]{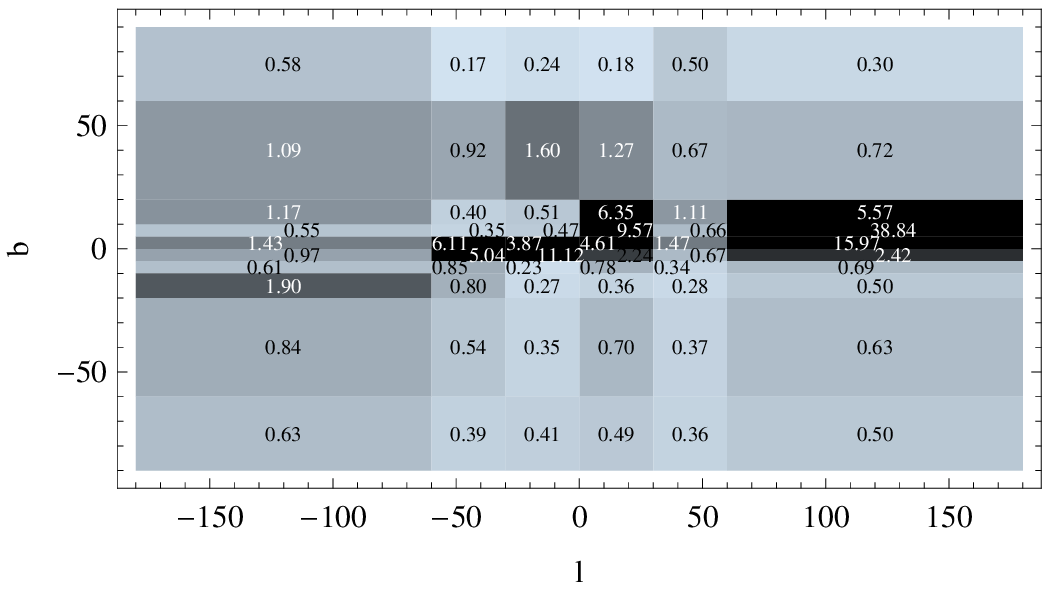}
\includegraphics[scale=0.78]{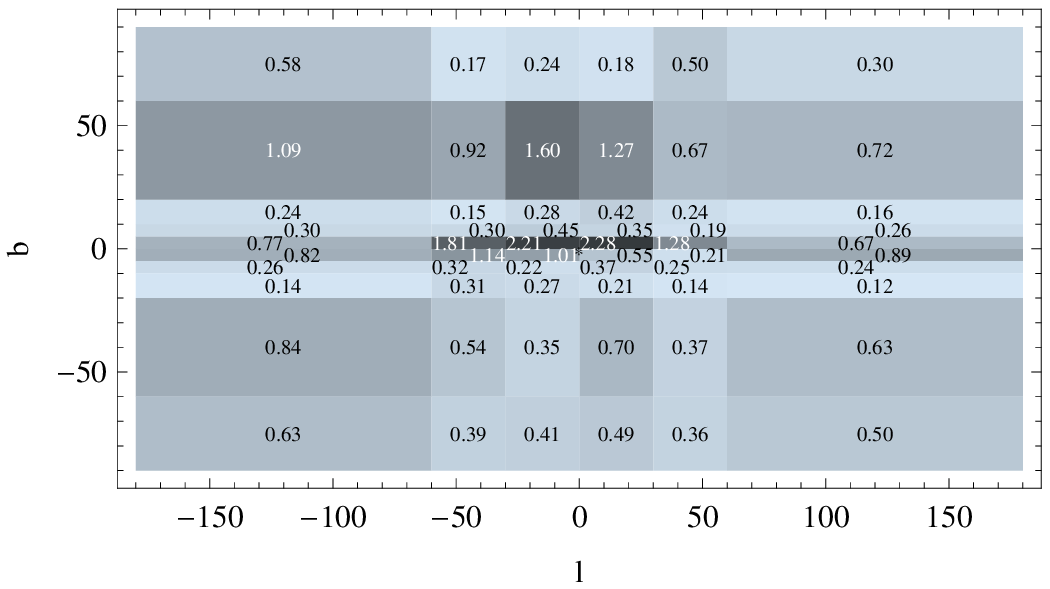}\\
\caption{Level of agreement between the predicted $\gamma$-ray background and the \textit{Fermi}-LAT data in the energy range of 1 GeV to 200 GeV.
\textit{Top left}: the $\gamma$-ray spectra are obtained by summing contributions from $\pi^0$ decay, inverse Compton scattering, bremsstrahlung emission, point/extended sources and extragalactic background.
\textit{Top right}: the contribution of the dark gas is added to the $\gamma$-ray spectra.
\textit{Bottom left}: contributions of the \textit{Fermi} bubbles/haze, Loop I and the northern arc are also included in the spectra of $\gamma$-rays .
\textit{Bottom right}:  the normalization of the total gas within $|b| <20^{\circ}$ is allowed to be free to account for under/over estimation of the gas distribution.
The goodness of prediction in each window is determined by the value of the reduced $\chi^{2}$.}
\label{fig:chi2FullSky4cases}
\end{figure}

We break the sky into 60 windows which are defined by the limits
$0^{\circ} < \pm b \leq 5^{\circ}$
$5^{\circ} < \pm b \leq 10^{\circ}$,
$10^{\circ} < \pm b \leq 20^{\circ}$,
$20^{\circ} < \pm b \leq 60^{\circ}$ and
$60^{\circ} < \pm b \leq 90^{\circ}$ in latitude and
$0^{\circ} < l \leq 30^{\circ}$, $30^{\circ} < l \leq 60^{\circ}$,
$60^{\circ} < l \leq 180^{\circ}$, $180^{\circ} < l \leq 300^{\circ}$,
$300^{\circ} < l \leq 330^{\circ}$ and $330^{\circ} < l \leq 360^{\circ}$ in longitude.
In Fig.~\ref{fig:chi2FullSky4cases}, the level of agreement between the predicted $\gamma$-ray spectra and the \textit{Fermi}-LAT data in the energy range of 1 GeV to 200 GeV for those windows are shown for various sets of assumptions.
The goodness of fit is determined by the value of reduced $\chi^2$ which is calculated by
\be 
\chi^2_{reduced}=\frac{1}{n}\Big(\Sigma_i\frac{(\Phi_i^{data} + \beta s_i + \gamma q_i - \Phi(E_i)^{theory})^2}{\sigma_i^2} + \beta^2 + \gamma^2\Big),
\ee
where $n$ is the number of degrees of freedom, $\Phi_i^{data}$ and $\Phi(E_i)^{theory}$ are, respectively, the measured and the predicted flux of $\gamma$-rays at energy $E_i$; $s_i$ and $q_i$ are the systematic errors related to, respectively, the exposure and energy resolution uncertainties; $\beta$ and $\gamma$ are the nuisance parameters \footnote{http://www.desy.de/~blobel/banff.pdf} and finally $\sigma_i^2$ is the variance related to the statistical errors of the \textit{Fermi} data and the errors of the spectral data of point/extended sources. 
The nuisance parameters are fitted to the \textit{Fermi} spectral data.
Since the exposure can be different at different directions, the value of $\beta$ varies among windows while the value of $\gamma$ is constant all over the sky and is fitted once.

In the top left panel of Fig.~\ref{fig:chi2FullSky4cases}, $\gamma$-ray spectra are obtained by summing contributions from $\pi^0$ decay, inverse Compton scattering, bremsstrahlung emission, point/extended sources and isotropic extragalactic background. 
Including the contribution of dark gas to the $\pi^0$ decay and bremsstrahlung emission components results in a better fit at low latitudes (Fig.~\ref{fig:chi2FullSky4cases}, top right).
It must be noted that the normalization of dark gas contribution is free. 
It is chosen in such a way to avoid overshooting $\gamma$-ray spectral data in any of the windows.
In the bottom left panel of  Fig.~\ref{fig:chi2FullSky4cases} we add the contributions of the \textit{Fermi} bubbles/\textit{Fermi} haze, Loop I and northern arc on top of contributions from the background and dark gas. 
The contribution of the \textit{Fermi} bubbles impacts the spectra only in windows with $|b| < 60^{\circ}$ and $|l| < 30^{\circ}$ 
\footnote{The borders of windows under study are not set to match those of the bubbles/haze whose contribution are within $|b| < 50^{\circ}$, $|l| < 20^{\circ}$.}. 
Recently, additional structures have been discovered in those regions of the sky \cite{Su:2012gu}.  
Therefore, the normalization of the spectrum of this component is allowed to be free among the windows.
Since there are uncertainties in the exact morphology of the bubbles at low latitudes, their contribution to windows with $|b| < 5^{\circ}$ and $|l| < 30^{\circ}$ is also allowed. 
The spectra of Loop I and the northern arc contribute to the northern hemisphere only
\footnote{For the northern arc we consider the same spectrum as for the Loop I for simplicity.}. 
The angular limits on these two objects are taken from \cite{Su:2010qj}.
The predicted and the observed spectra agree well in all sky regions except for some windows with $-5^{\circ}<b<20^{\circ}$. 
The predicted $\gamma$-ray spectra in those windows are below the \textit{Fermi} spectra.
This under prediction could be because of an under-estimation of dark gas contribution.
As discussed in section~\ref{subsec:DarkGas}, we masked $|b| < 1^{\circ}$ out in extracting the dark gas sky map. 
Changing the size of the mask can change the fits in low latitudes.
Another reason could be related to the absorption of the 21cm line emission.  
The opacity of the 21cm line in \cite{Tavakoli:2012jx} has been assumed to be constant all over the sky. 
However, it could vary in different regions of the sky because of different temperatures. 
Fitting the spin temperature $T_s$ to $\gamma$-ray spectra is beyond the scope of this paper. 
Nevertheless, to account for uncertainties in the gas distribution, the normalization of the total gas contribution is allowed to be free in the range of 1/2 to 2 within $|b| <20^{\circ}$.
This freedom can fix the under prediction of the model in those windows and it leads to a very good fit to the \textit{Fermi} spectral data in all regions under study (Fig.~\ref{fig:chi2FullSky4cases}, bottom right).
Apart from uncertainties in estimating the gas distribution,  there can be another reason for our under prediction which relates to the distribution of the interstellar radiation field (ISRF).
It affects the diffuse $\gamma$-ray component which is produced by inverse Compton scattering of cosmic ray electrons and positrons off the interstellar radiation field.
We use the model of \cite{Porter:2005qx} as our reference. 
The impact of different assumptions about the distribution of the ISRF on the diffuse $\gamma$-ray spectra is discussed in Appendix \ref{sec:ISRF}. 



\section{Limits on WIMP Annihilation Cross Section}
\label{sec:bounds}

To constrain WIMP annihilation cross section, we study DM models in which WIMPs dominantly annihilate into individual generic channels. These annihilation channels are $\mu^+\mu^-$, $\tau^+\tau^-$, $b\bar{b}$, $W^+W^-$ and $t\bar{t}$ with mass ranges which probe different parts of the $\gamma$-ray spectrum.
The distribution of DM in the Galaxy is assumed to follow the Einasto profile \cite{Graham:2005xx}, which is parameterized as,
\begin{equation}
\rho_{\chi}(r) = \rho_{Ein} \exp\left[-\frac{2}{\alpha}*\left((\frac{r}{r_{c}})^{\alpha}-1\right)  \right],
 \label{eq:Einasto}
\end{equation}
where $\alpha = 0.22$, $r_{c} = 15.7~\kpc$ and $\rho_{Ein}$ is set such that the local DM density is equal to $0.4~\GeV\cm^{-3}$ \cite{Catena:2009mf,Salucci:2010qr}.


\subsection{Limits from Diffuse $\gamma$-rays}
\label{subsec:GammaLimits}

WIMPs annihilation can lead to production of $\gamma$-rays in a number of ways. 
One is the prompt emission which includes the final state radiation, the virtual internal bremsstrahlung and the decay of $\pi^0$s which are, in turn, produced by hadronization or decay of WIMPs annihilation products. 
The inverse Compton scattering and bremsstrahlung of leptonic final products also yield $\gamma$-rays. 
Those contributions are significant for leptonic annihilation channels in which high energy electrons and positrons are produced. 

The diffuse $\gamma$-ray background which is modeled in section \ref{sec:background} serves as reference to extract limits on DM annihilation rate. 
To get conservative limits at high galactic latitudes, we replace the spectrum of extragalactic background \cite{Abdo:2010nz} with the minimal non-DM extragalactic $\gamma$-ray background radiation (EGBR) which is modeled in Appendix \ref{sec:EGBR} and ignore contribution from the \textit{Fermi} bubbles/\textit{Fermi} haze, Loop I and the northern arc.
Therefore, our background includes contributions from $\pi^0$ decay, bremsstrahlung emission
\footnote{The contribution of dark gas to the $\pi^0$ decay and bremsstrahlung emission components is also included.}, 
inverse Compton scattering, point and extended sources and the minimal non-DM isotropic background. 
The flux of our minimal non-DM EGBR is suppressed to the spectrum of extragalactic background modeled in \cite{Abdo:2010nz} by a factor of $\simeq$2.
This leads to more conservative limits especially at higher latitudes where the contribution of EGBR is significant.

\begin{figure}
\hspace{-1.2cm}
\includegraphics[scale=0.78]{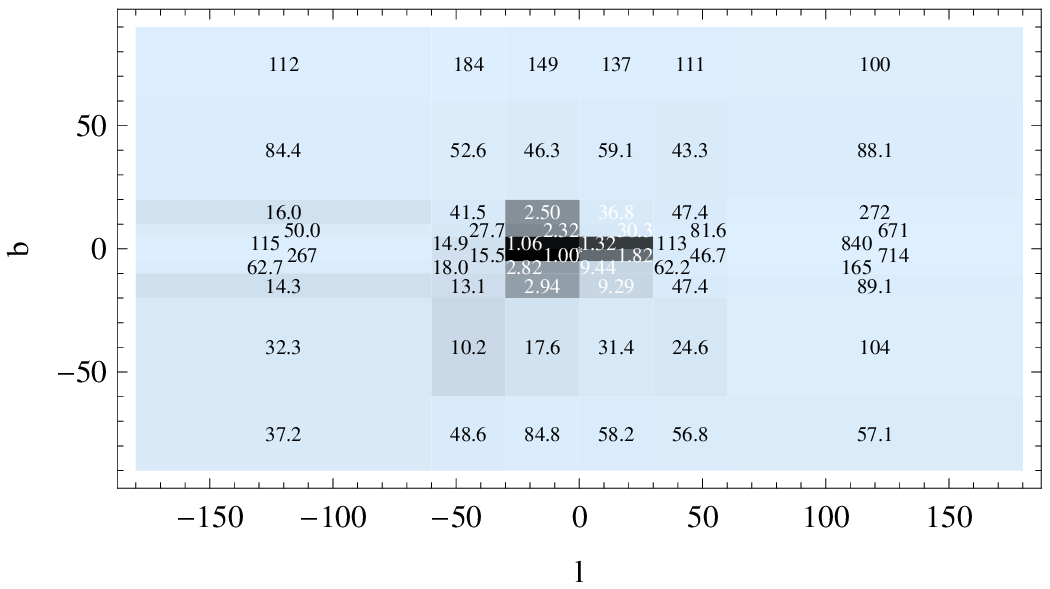}
\includegraphics[scale=0.78]{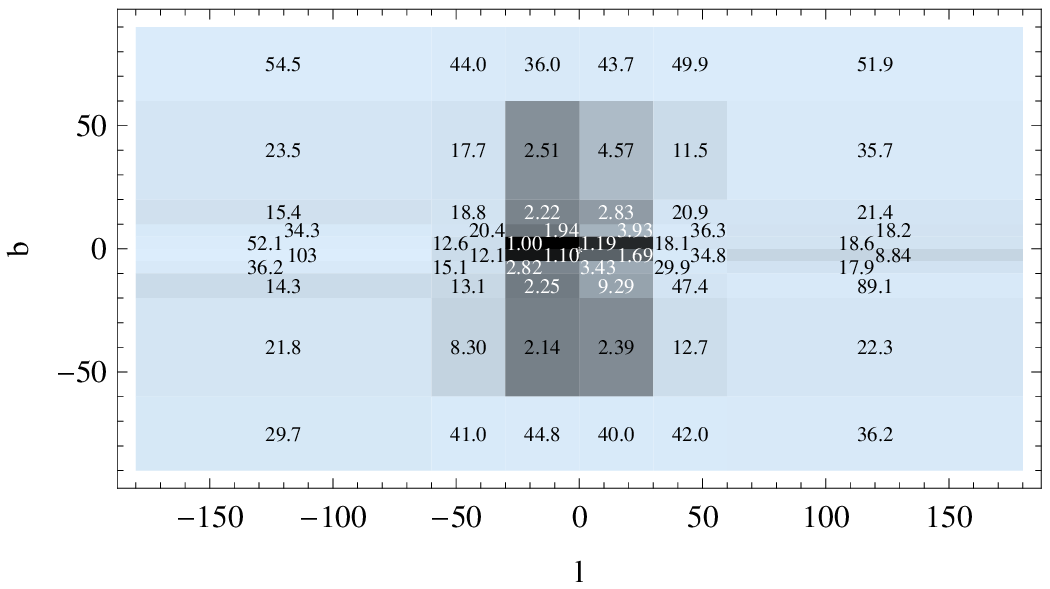} \\
\hspace*{-1.2cm}
\includegraphics[scale=0.78]{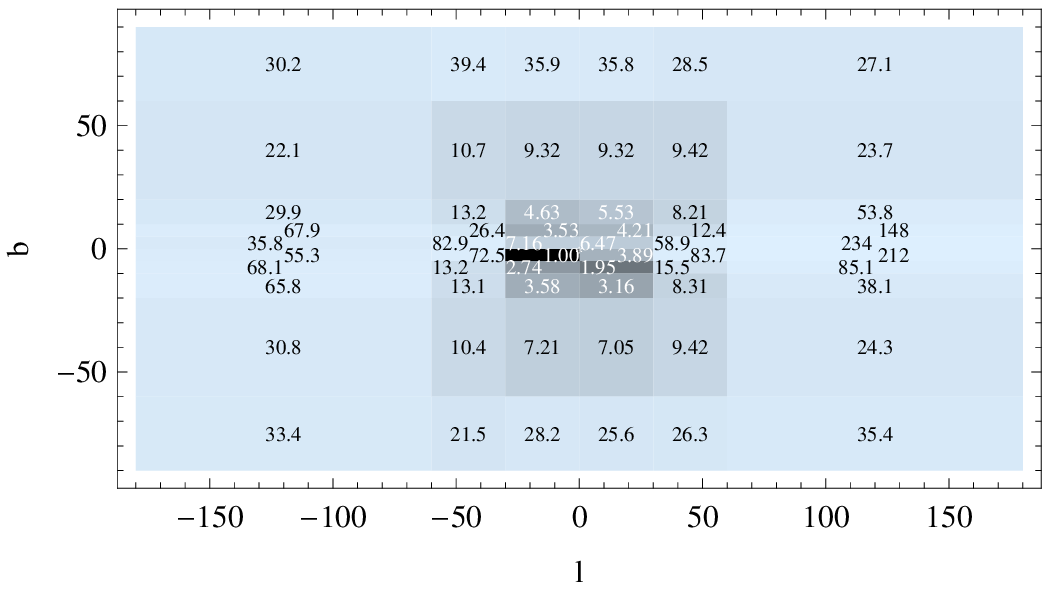}
\includegraphics[scale=0.78]{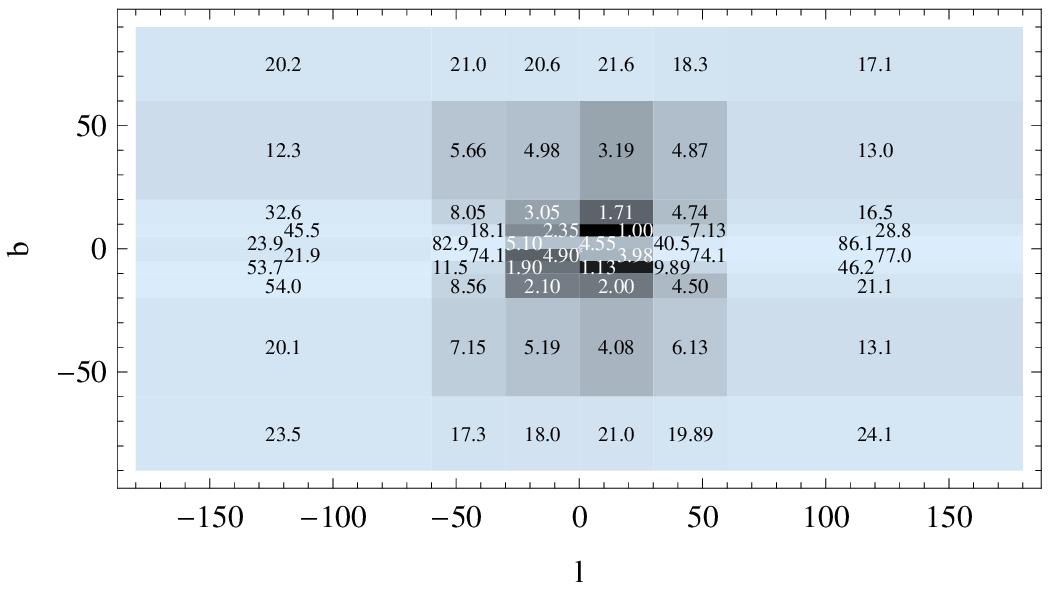} \\
\hspace*{-1.2cm}
\includegraphics[scale=0.78]{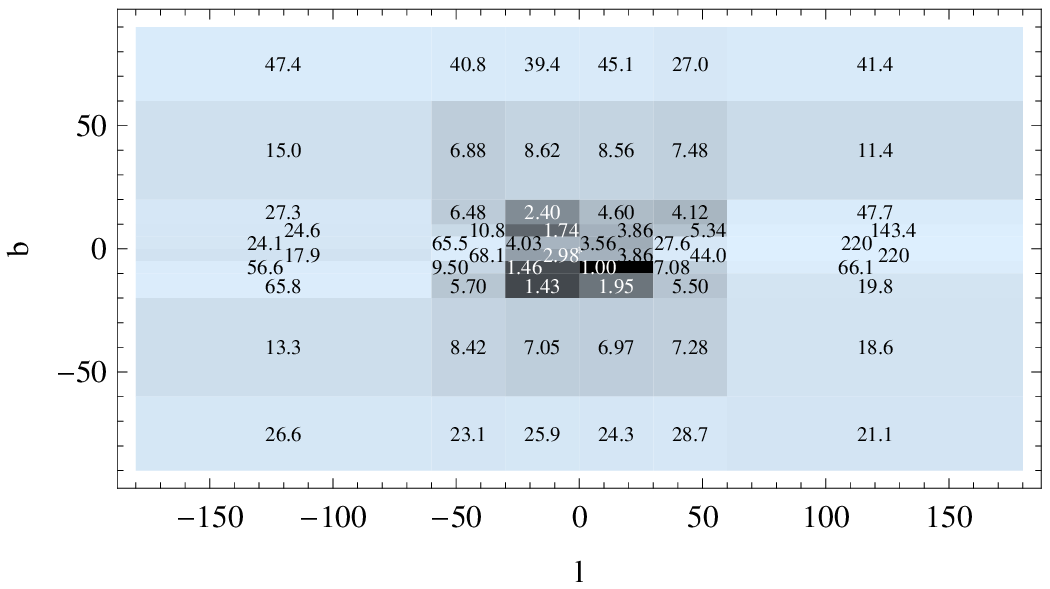}
\includegraphics[scale=0.78]{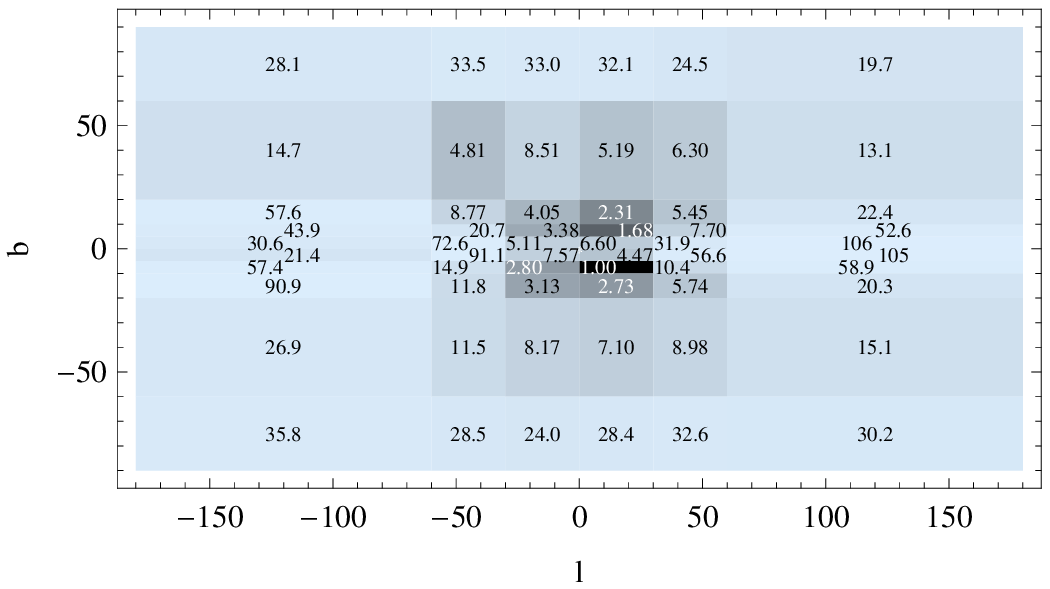}\\
\caption{Relative strength of 3$\sigma$ upper limits on DM annihilation cross section for different channels and masses and for a given DM profile. 
Darker regions give stronger limits. 
Numbers give the ratio of the 3$\sigma$ upper limit from each window to the lowest 3$\sigma$ upper limit among 60 windows under study, 
$ \sigma v ^{3\sigma}/ \sigma v ^{3\sigma}_{min}$. 
\textit{Top left:} DM particles with $m_{\chi}=10~\GeV$ annihilating into $b\bar{b}$. 
The window with $-5^{\circ} < b < 0^{\circ}$, $-30^{\circ} < l < 0^{\circ}$ gives the tightest 3$\sigma$ upper limit on annihilation cross section with $ \sigma v ^{3\sigma}_{min} = 1.08 \times 10^{-27}$ cm$^{3}$s$^{-1}$. 
\textit{Top right:} the same as \textit{left panel} with the total gas contribution free within a factor of 2.
The tightest 3$\sigma$ annihilation cross section is $\sigma v ^{3\sigma}_{min} = 2.49 \times 10^{-27}$ cm$^{3}$s$^{-1}$.
\textit{Middle left}: DM particles with $m_{\chi}=100~\GeV$  annihilating into $W^{+}W^{-}$. 
The tightest 3$\sigma$ limit is from the window of $5^{\circ} < b < 10^{\circ}$, $0^{\circ} < l < 30^{\circ}$ and is equal to $\sigma v ^{3\sigma}_{min} = 1.11 \times 10^{-25}$ cm$^{3}$s$^{-1}$.
\textit{Middle right:} the same as \textit{left panel} with "free" total gas.
The tightest 3$\sigma$ limit is $\sigma v ^{3\sigma}_{min} = 9.3 \times 10^{-26}$ cm$^{3}$s$^{-1}$.
\textit{Bottom left}: DM particles with $m_{\chi}=1.6~ \TeV$ annihilating into a pair of intermediate light bosons $\phi$ which then decay to  $e^{+}e^{-}$,  $\mu^{+}\mu^{-}$ and  $\pi^{+}\pi^{-}$ at a ratio of 1:1:2. 
The tightest 3$\sigma$ limit is from the window of $-10^{\circ} < b < -5^{\circ}$, $0^{\circ} < l < 30^{\circ}$ and is equal to $\sigma v ^{3\sigma}_{min} = 8.9 \times 10^{-25}$ cm$^{3}$s$^{-1}$.
\textit{Bottom right:} the same as \textit{left panel} with "free" total gas.
The tightest 3$\sigma$ limit is $\sigma v ^{3\sigma}_{min} = 7.1 \times 10^{-25}$ cm$^{3}$s$^{-1}$.}
\label{fig.gammaLimitsFullSky}
\end{figure}

Adding $\gamma$-ray fluxes generated by DM annihilation to the background spectra, we calculate the WIMPs' annihilation cross sections which best fit the diffuse $\gamma$-ray spectra in our windows of study. 
The nuisance parameter associated with exposure uncertainties $\beta$ is also taken as a free parameter. 
We then derive 3$\sigma$ upper limits on annihilation cross section and the value of $\beta$ parameter for each annihilation channel and mass corresponding to $\Delta \chi^2(\langle \sigma v \rangle, \beta)= \chi^2(\langle \sigma v \rangle, \beta)-\chi^2_{min}=10.27$. 
In left panels of Fig.~\ref{fig.gammaLimitsFullSky} the relative strengths of 3$\sigma$ upper limits on $\langle \sigma v \rangle$ at each region of the sky are shown for DM models with $m_{\chi}$ = 10 GeV annihilating into $b \bar{b}$ (top), with $m_{\chi}$ = 100 GeV annihilating into $W^{+}W^{-}$ (middle) and with $m_{\chi}$ = 1.6 TeV (bottom) annihilating into light intermediate bosons $\chi \chi$ $\longrightarrow$ $\phi \phi$, which subsequently decay to the kinematically allowed light SM particles giving hard injection spectrum leptons \cite{Cholis:2008vb, Cholis:2008qq}. We take the case where each $\phi$ $\longrightarrow$ $e^{+} e^{-}$, $\mu^{+} \mu^{-}$ and $\pi^{+} \pi^{-}$ at a relative branching ratio of 1:1:2  which is in preferred by the leptonic data after the measurement of the positron fraction with \textit{AMS-02} \cite{Cholis:2013psa}. These type of models can also explain the relative large annihilation rates needed to fit the leptonic data, via Sommerfeld enhancement \cite{ArkaniHamed:2008qn}.
The limits are normalized to the tightest 3$\sigma$ upper limit among the 60 windows to show the relative strength between these windows. 
The produced spectra by annihilation channels of $b\bar{b}$ and $W^+W^-$ are dominated by the prompt component, thus they probe well the distribution of DM in the Galaxy.
The spectra for the model annihilating to leptons and charged pions, receive a significant contribution from the inverse Compton component, thus they are influenced also by assumptions on radiation field and the propagation of cosmic rays in the Galaxy.
To account for the main type of  uncertainties on DM limits, originating from uncertainties on the background assumptions, we also derive 3$\sigma$ upper limits when the normalization of the total gas contribution is free.
The suppression or enhancement of the total gas contribution is allowed to be within a factor of 2 with respect to the case of the fixed gas contribution. 
Those normalized limits are shown in right panels of Fig.~\ref{fig.gammaLimitsFullSky} for the same channels and masses as left panels.
The 10, 100 and 1600 GeV masses are probed by different energy ranges of the $\gamma$-ray spectra.  
The allowed freedom leads to weaker constraints for 10 and 100 GeV DM while tighter for the 1.6 TeV DM particles in more regions of the sky. 

\begin{figure}
\hspace{-1.2cm}
\includegraphics[scale=0.78]{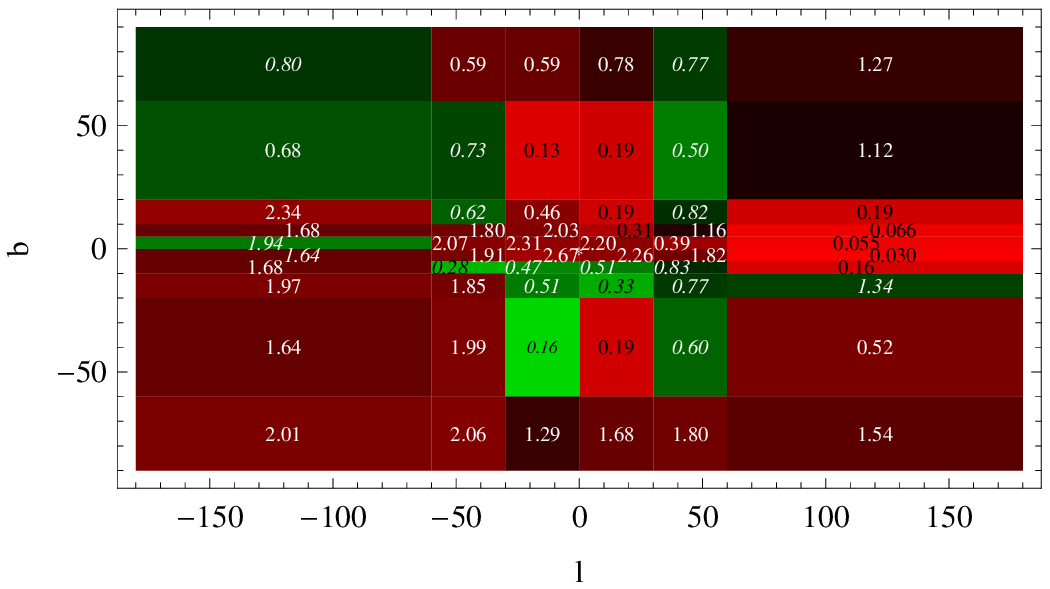}
\includegraphics[scale=0.78]{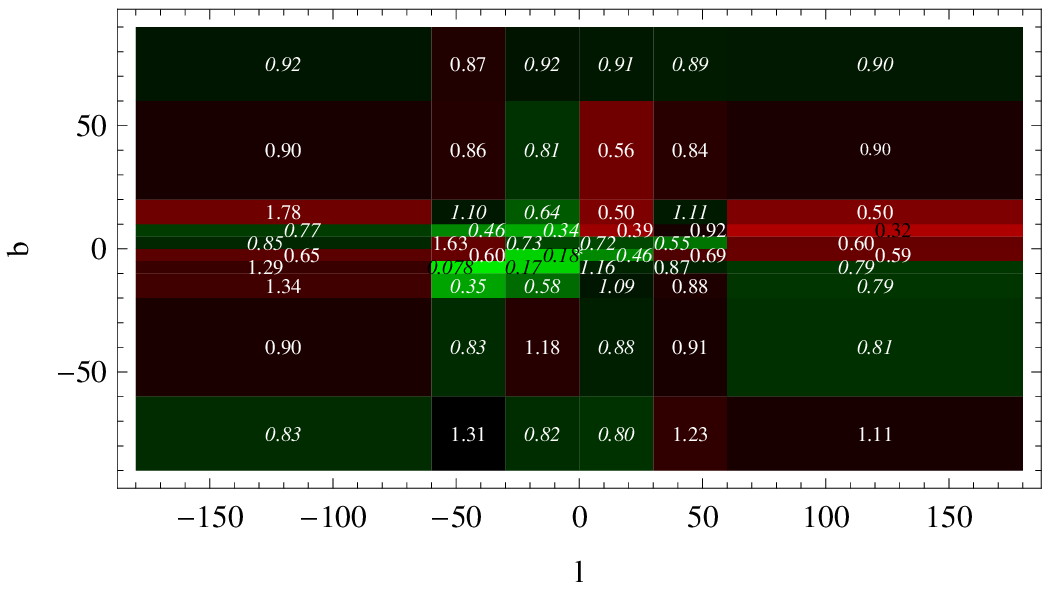} \\
\hspace*{2.9cm}
\includegraphics[scale=0.78]{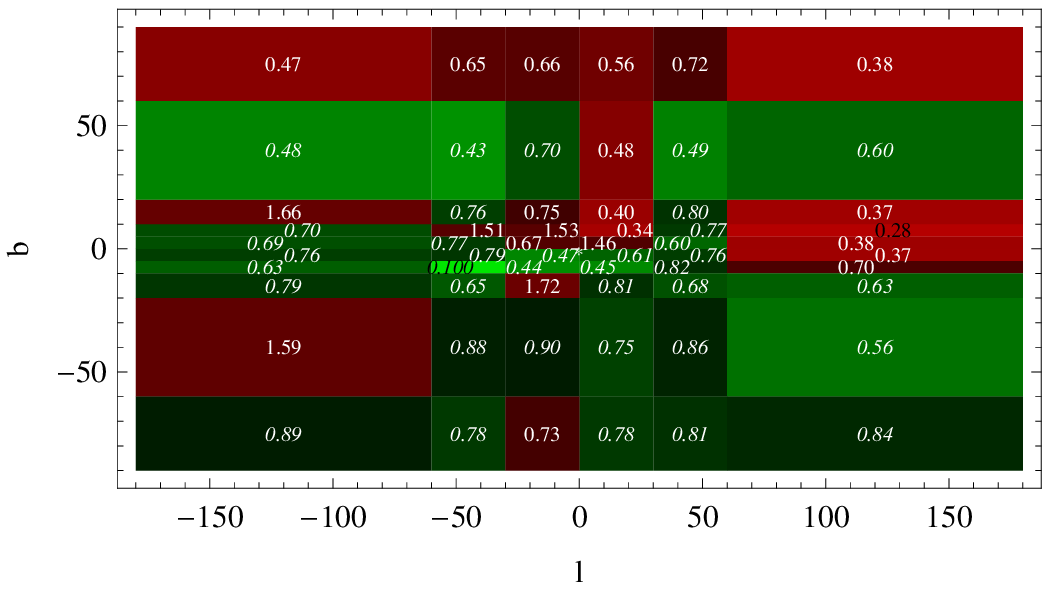}
\caption{The impact of astrophysical uncertainties in deriving 3$\sigma$ upper limits on the DM annihilation cross section for three different channels and masses and a given DM Einasto profile. We compare the  3$\sigma$ upper limit derived under our reference assumptions for ISM gas and radiation field (model $A$) to the 3$\sigma$ upper limits derived under varied assumptions on  either the ISM gas normalization, the XCO radial profile, the radiation field metallicity gradient and the radiation field spacial distribution (model $B_{i}$ with $i$:1-4) (see text for more details). For each angular window, we calculate the four ratios of $\sigma v ^{3\sigma} _{B_{i}}$/$\sigma v ^{3\sigma} _{A}$ and present the value of the ratio that deviated the most from 1. This allows us to check the robustness of the   3$\sigma$ upper limits, where more robust limits come from windows that have the presented ratio being closer to 1.  This test also allows us to check, which type of astrophysical assumptions -not related directly to DM-  impact the most, the limits on DM. 
Darker regions give more robust limits. Red regions refer to the cases where the main uncertainty on the DM limits, comes from uncertainties on, either the ISM gas normalization, or the XCO radial profile (the ratio value is written in normal fonds).  Green regions  refer  to the cases where the main uncertainty on the DM limit, comes from uncertainties on either the the radiation field metallicity gradient, or its spacial distribution (the ratio value is written in \textit{italics}).
\textit{Top left:} DM particles with $m_{\chi}=10~\GeV$ annihilating into $b\bar{b}$. 
\textit{Top right:} particles with $m_{\chi}=100~\GeV$  annihilating into $W^{+}W^{-}$. 
\textit{Bottom}: particles with $m_{\chi}=1.6~\TeV$  annihilating to intermediate light bosons $\phi$ which subsequently decay to $e^{+}e^{-}$, $\mu^{+}\mu^{-}$, $\pi^{+}\pi^{-}$ at a relative ratio of 1:1:2.}
\label{fig.gammaLimitsAstroUncer}
\end{figure}

In addition in Fig.~\ref{fig.gammaLimitsAstroUncer} we show the impact on the 3$\sigma$ upper limits on DM annihilation cross sections, from uncertainties
on the astrophysical assumptions. We keep the DM profile fixed (whose impact on the DM limits has been studied already in \cite{Cirelli:2009dv, Papucci:2009gd}) and instead concentrate on the impact of the ISM gas distribution and the interstellar radiation field energy density. The specific choices on these properties impact both the $\gamma$-ray background contribution and also the non-prompt DM $\gamma$-ray contributions. These contributions are the bremsstrahlung emission and ICS emission by CR $e^{\pm}$ produced in DM annihilations.  In our test, we compare the  3$\sigma$ upper limit derived under our reference assumptions for ISM gas and radiation field which we define as model $A$; to the 3$\sigma$ upper limits derived under four variations on our assumptions. Variation 1: we let the ISM gas normalization be free within a factor of 2 (as in Fig.~\ref{fig.gammaLimitsFullSky} right panels), variation 2: we use the XCO radial profile of \cite{Israel:2000mx}, variation 3:  
we vary the radiation field metallicity gradient and variation 4: we vary the radiation field spacial distribution (see appendices~\ref{sec:XCOfactor} and~\ref{sec:ISRF} for further details). We refer to these alternative models as $B_{i}$ with $i:1-4$.
Then for each of the 60 angular windows we calculate the four ratios of $\sigma v ^{3\sigma} _{B_{i}}$/$\sigma v ^{3\sigma} _{A}$ and present  in Fig.~\ref{fig.gammaLimitsAstroUncer} the value of the ratio that deviated the most from 1. This is a probe for the robustness of the   3$\sigma$ upper limits given that all these models have already been checked with CR and $\gamma$-ray measurements.  The more robust limits come from windows that have the presented ratio being closer to 1 (darker regions). The color code which is either red or green, indicates which type of astrophysical assumptions impact the most, the limits on DM. 
Red regions refer to the cases where the main uncertainty on the DM limits, comes from uncertainties on, either the ISM gas normalization, or the XCO radial profile while green regions  refer  to the cases where the main uncertainty on the DM limit, comes from uncertainties on either the the radiation field metallicity gradient, or its' spacial distribution. For heavier DM models and especially for the lepto-philic 1.6 TeV case the exact assumptions on the ISRF matter in most windows  while for the lighter 10 GeV case to $b\bar{b}$ the gas assumptions matter more. Also the heavier 100 GeV to Ws or 1.6 TeV XDM cases tend to provide robust limits in more windows on the sky compared to the 10 GeV to $b\bar{b}$.  We also note that depending on the DM model under study, different windows provide the more/less robust limits.  
  
For all three cases of DM models depicted in Figs.~\ref{fig.gammaLimitsFullSky} and~\ref{fig.gammaLimitsAstroUncer}, the tightest limits come from regions closer to the GC. From Fig.~\ref{fig.gammaLimitsAstroUncer} we show that all the DM limits in the windows with $\mid l \mid< 30^{\circ}$ and $\mid b \mid < 10^{\circ}$ (inner galaxy) have an uncertainty by typically a factor of $\sim 2$ due to the uncertainties in the ISM properties withe the uncertainties on the 1.6 TeV DM case being the smaller ones.

In Figs.~\ref{fig:chi2FullSky4cases}-\ref{fig.gammaLimitsAstroUncer} we used as reference, a diffusion co-efficient that scales as $D(z) \propto exp\{\mid z \mid /z_{d}\}$, with $z_{d} = 4$ kpc (see eq.~\ref{eq:Diffusion}).  Very thin diffusion zones ($z_{d} < 1$ kpc), suppress the diffuse ICS component and enhance the impact that the ISM gas distribution uncertainties  have on the derived DM limits. In addition, the quality of the fits to the diffuse $\gamma$-ray data is worse for propagation models of $z_{d} < 1$ kpc whose propagation properties (diffusion co-efficient energy scaling and normalization, re-acceleration, convection) have been fitted to the available CR data \cite{Cholis:2011un}. On the contrary, thick diffusion zones ($z_{d} \ge 8$ kpc) do not affect much the quality of neither the CR or the $\gamma$-ray fits, or the derived DM limits from the cases shown for the $z_{d}=4$ kpc.  
Strong convective winds perpendicular to the galactic disk suppress all the diffuse background $\gamma$-ray emission closer to the galactic disk and are not strongly supported from the $\gamma$-ray data at lower latitudes~\cite{Cholis:2011un}; we have thus ignored the possible impact of large scale galactic convective winds in the DM limits. 

We find that by excluding the $|b|<1^{\circ}$ region we can reduce the large uncertainties on both the galactic disk ISRF and the ISM gas column densities, thus deriving the tightest limits from $1^{\circ}< |b| < 9^{\circ}$ and $|l| < 8^{\circ}$ in most cases.
At $|b| > 10^{\circ}$ and $|l|< 30^{\circ}$ the contribution from the bubbles is dominant, thus any limit will be contaminated by their presence.
Since we do not include bubbles in our background, the limits in the relevant windows ($|b| < 60^{\circ}$, $|l|< 30^{\circ}$) are conservative. 
At intermediate latitudes, the most constraining region is the one with $9^{\circ}< |b| < 25^{\circ}$ and $|l| < 8^{\circ}$.
At high latitudes, we choose a large region with $|b| > 60^{\circ}$ and $0^{\circ} <l< 360^{\circ}$ for extracting limits to allow for minimum statistical errors.

\begin{figure}
\centering
\includegraphics[scale=.39]{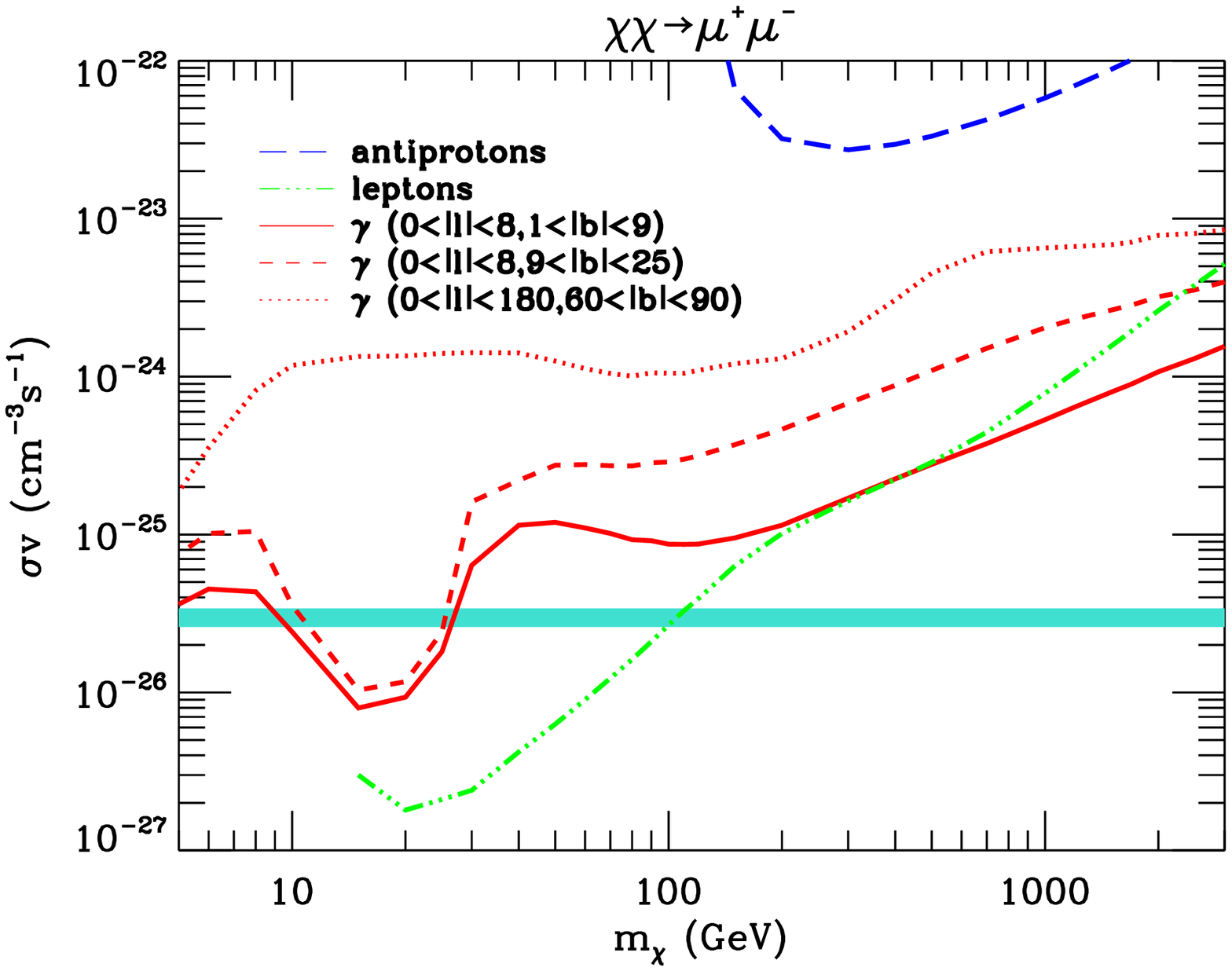}
\hspace{-0.7cm}
\includegraphics[scale=.39]{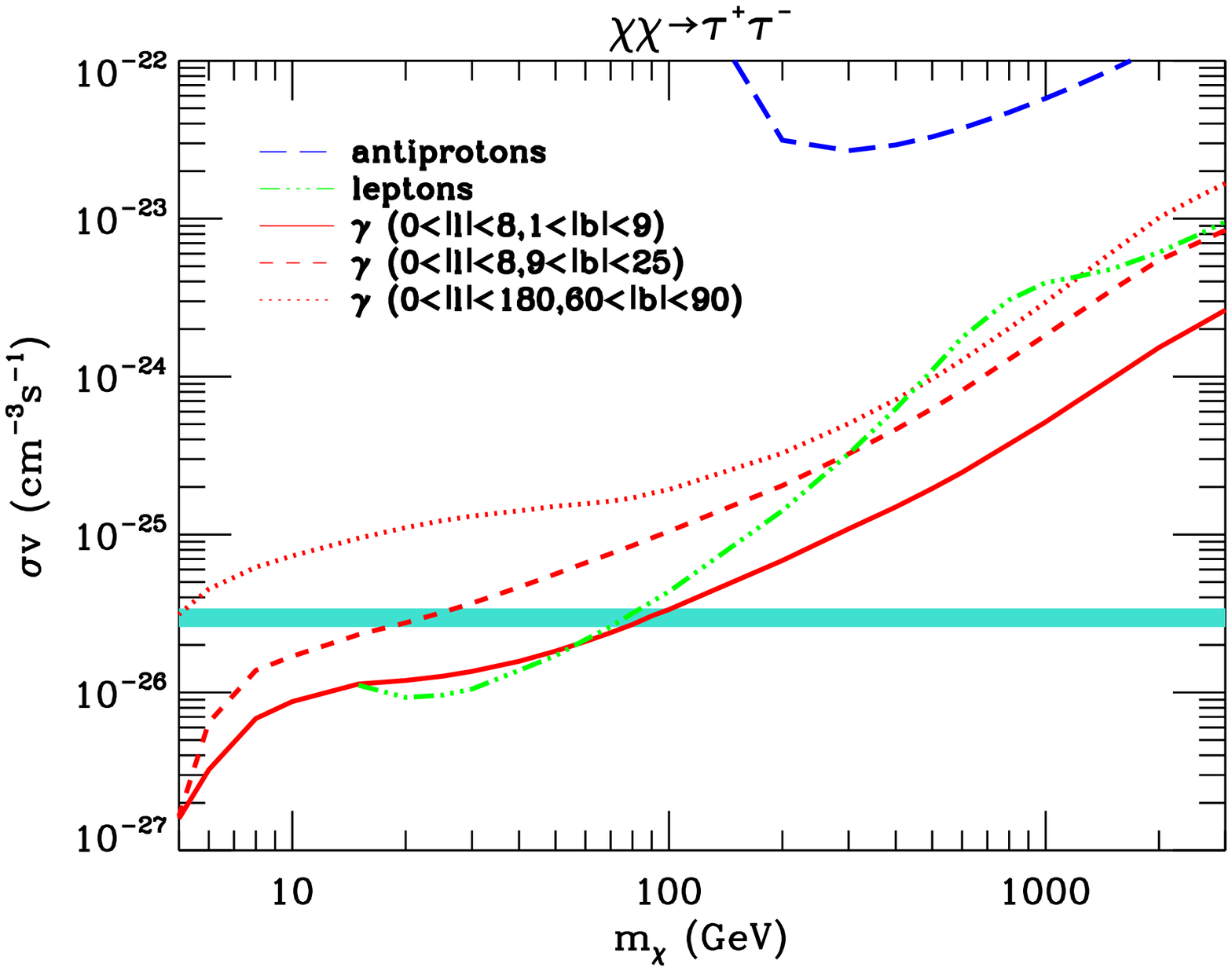} \\
\includegraphics[scale=.39]{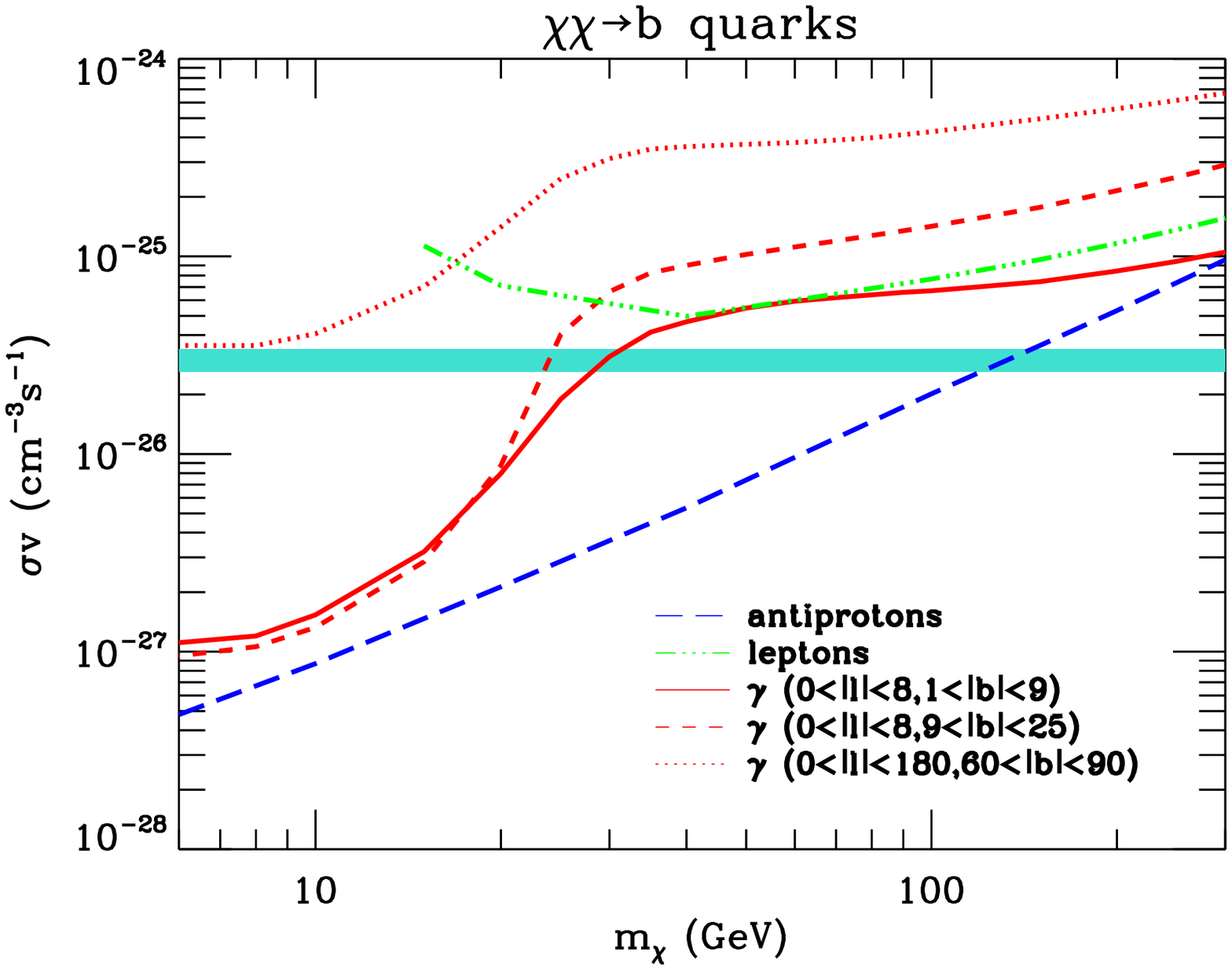}
\hspace{-0.7cm}
\includegraphics[scale=.39]{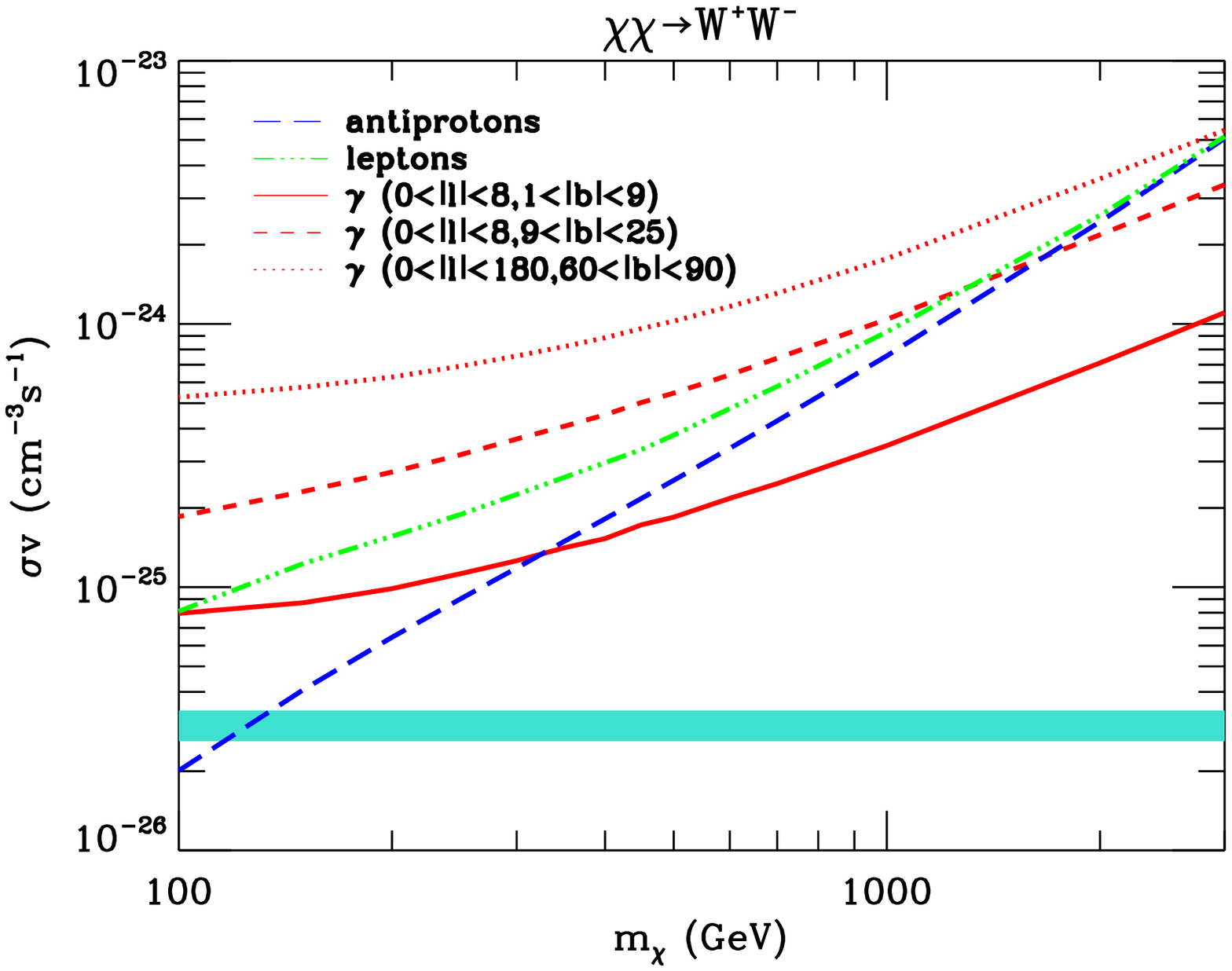}\\
\includegraphics[scale=.39]{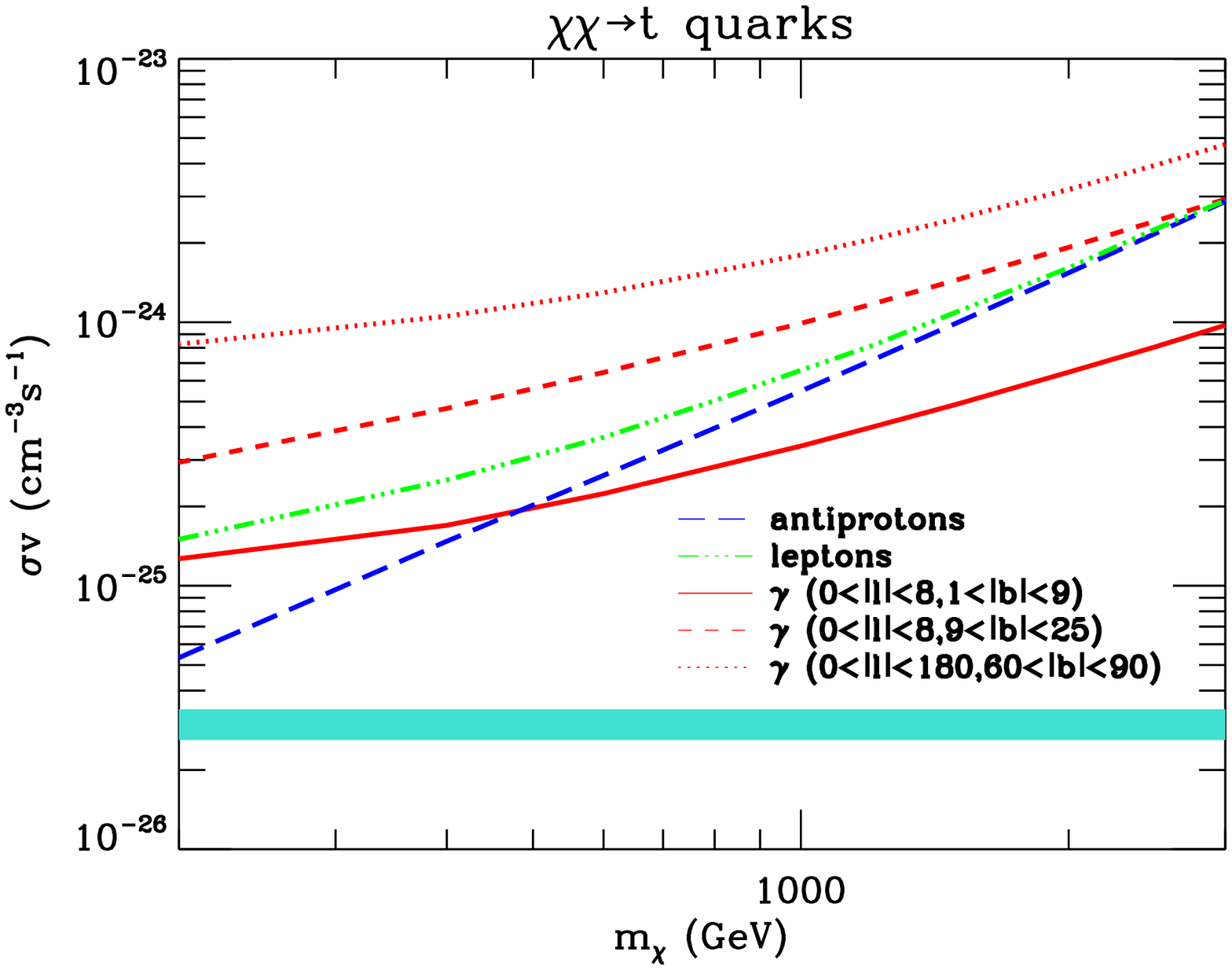}
\caption{3$\sigma$ upper limits on WIMPs annihilation cross section versus $m_{\chi}$.
The annihilation channels are $\mu^+\mu^-$ ({\emph top left}), $\tau^+\tau^-$ ({\emph top right}), b$\bar{b}$ ({\emph middle left}), $W^+W^-$ ({\emph middle right}) and t$\bar{t}$ ({\emph bottom}). 
The lines represent limits from $\gamma$-rays in $|l|<8^{\circ}, 1^{\circ}<|b|<9^{\circ}$ ({\emph dotted green}), $\gamma$-rays in $|l|<8^{\circ}, 9^{\circ}<|b|<25^{\circ}$ ({\emph dashed green}), $\gamma$-rays in $0^{\circ}<l<360^{\circ}, |b|>60^{\circ}$ ({\emph dotted dashed green}), antiprotons ({\emph red}) and leptons ({\emph blue}). Our limits from leptons stop at 15 GeV since in our analysis we ignore leptonic data at lower energies. The ISM gas normalization is kept to be free within a factor of 2 from the reference distribution case (see text for more details). We include all diffuse $\gamma$-ray components of DM origin (prompt, ICS, bremsstrahlung).}
\label{fig.bound}
\end{figure}

For the remaining part of our discussion we will only show limits from the region of $1^{\circ}< |b|< 9^{\circ}$, $|l|<8^{\circ}$, the region of $9^{\circ} <|b| < 25^{\circ}$, $|l|< 8^{\circ}$ and the region of $|b| > 60^{\circ}$.
In Fig.~\ref{fig.bound}, those limits are shown versus the mass of DM for five simplified annihilation channels: $\chi \chi \longrightarrow \mu^{+} \mu^{-}$,
$\chi \chi \longrightarrow \tau^{+} \tau^{-}$, $\chi \chi \longrightarrow b \bar{b}$, $\chi \chi \longrightarrow W^{+} W^{-}$ and  $\chi \chi \longrightarrow t \bar{t}$.   
In spite of the conservative choices, the obtained limits are stringent and compatible with bounds derived by other analysis \cite{Ackermann:2010rg, Ackermann:2011wa, Ackermann:2012rg}.
For the cases of the $\mu^+\mu^-$, $\tau^+\tau^-$  and $b \bar{b}$ annihilation channels, the thermal relic range of the annihilation cross section is excluded by the low latitude limits for $m_{\chi} \le 30, 100$ and 30 GeV respectively. For annihilation to muons the thermal relic cross-section is excluded by the \textit{AMS}-02 positron fraction data for masses less than 100 GeV (see also \cite{Bergstrom:2013jra}). 
We also note that the limits from $b\bar{b}$ annihilation channel at intermediate latitudes ($9^{\circ}< |b| < 25^{\circ}$) are slightly tighter than those at low latitudes ($1^{\circ}< |b|< 9^{\circ}$) for $m_{\chi} < 30$ GeV, with the limits from antiprotons being the most competitive. 
Finally for the annihilation channels to $W^{+}W^{-}$ and to the $t \bar{t}$ quarks, the $\gamma$-ray limits from the lower latitude region are stronger than the limits derived from CR leptons at all masses up to 3 TeV and stronger than the limits from CR anti-protons for masses heavier than $\sim$ 500 GeV.

\begin{figure}
\centering
\includegraphics[scale=.39]{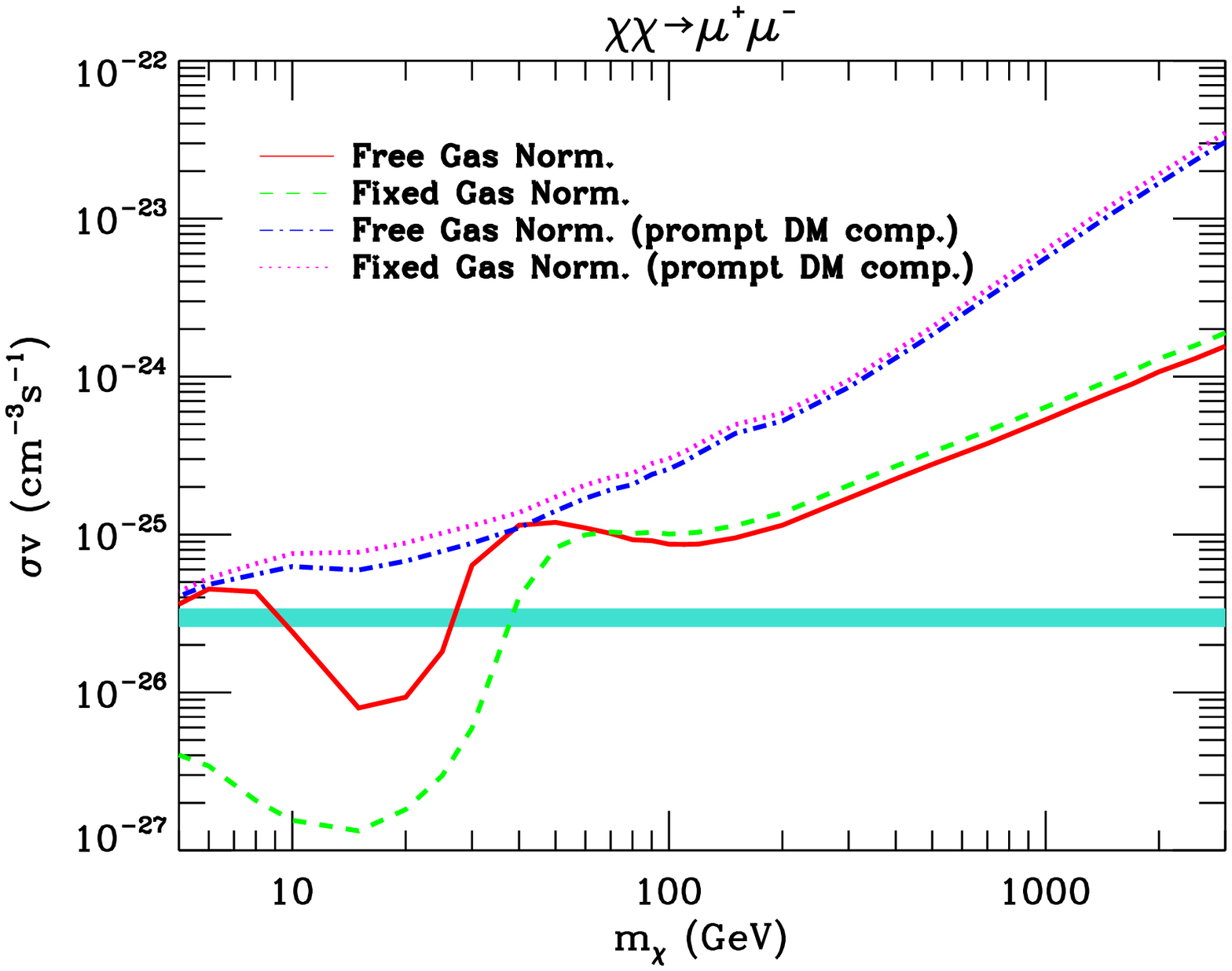}
\hspace{-0.7cm}
\includegraphics[scale=.39]{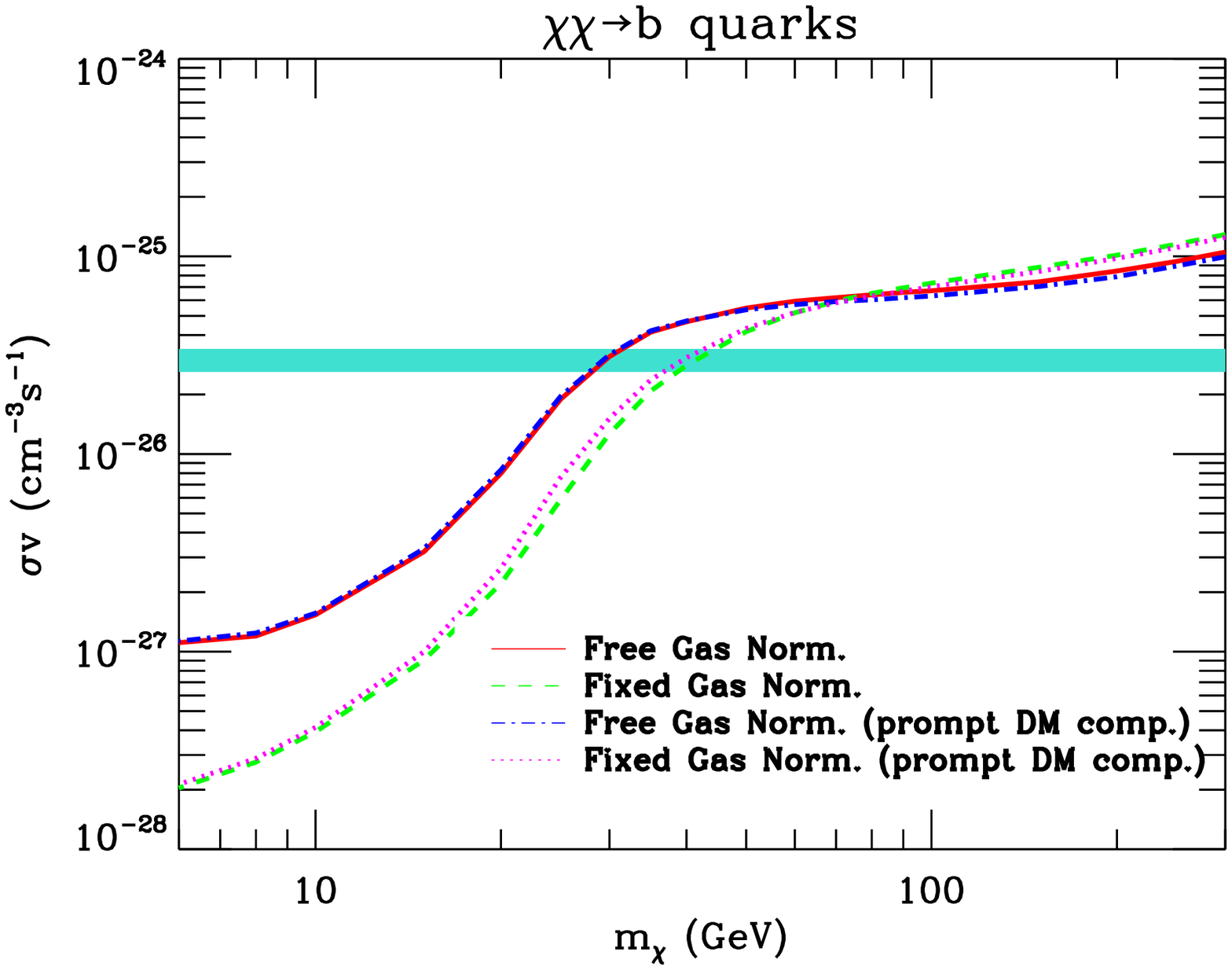} 
\caption{3$\sigma$ upper limits on WIMPs annihilation cross section versus $m_{\chi}$, including the impact of having the ISM gas normalization fixed to the reference distribution, or free within a factor of 2 for the chosen window of $|l|<8^{\circ}, 1^{\circ}<|b|<9^{\circ}$. We also show limits from that same window for the case where we ignore the ICS and bremsstrahlung diffuse $\gamma$-ray components. 
The annihilation channels are $\mu^+\mu^-$ ({\emph left}) and $b \bar{b}$ ({\emph right}). 
The lines represent limits for the case of free ISM gas normalization with all DM diffuse components as is the case presented in Fig.~\ref{fig.bound} ({\emph solid red}),
fixed ISM gas normalization with all DM diffuse components ({\emph dashed green}), free ISM gas normalization with just the prompt DM diffuse component ({\emph dotted dashed blue}) and for the case of fixed ISM gas normalization with just the prompt DM diffuse component ({\emph dotted magenta}).}
\label{fig.bound_prompt}
\end{figure}

In deriving the 3$\sigma$ limits, we allow the DM to contribute in the best fit to the data, with respect to which the 3$\sigma$ limits are defined. 
In Fig.~\ref{fig.bound_prompt} we show both limits with only the prompt DM diffuse $\gamma$-ray component and limits with all the DM originated diffuse 
$\gamma$-ray components (prompt, ICS, bremsstrahlung). We also show the limits when keeping the ISM gas normalization fixed to the reference galactic  
distribution, or having it free within a factor of 2 as is the case in Fig.~\ref{fig.bound}. Allowing for a fee gas normalization can lead to either weaker or stronger 
limits on DM, since the best fit value of the DM contribution also changes.  

For the case of direct annihilation to muons, we get that including the ICS (dominant) and bremsstrahlung (subdominant) components, can affect the derived limits
by up to a factor of 10. Moreover for the muon channel, the DM diffuse ICS component is significant in its amplitude and does not peak close to 
the energy scale of the  $m_{\chi}$ (as does the prompt: FSR,VIB). Thus, including it changes significantly the spectral DM $\gamma$-ray signature and
as a consequence the derived limits in a non-trivial manner (see the red solid line vs the blue dashed-dotted line, or the green dashed vs the magenta dotted line in 
Fig.~\ref{fig.bound_prompt}, left panel). 
For the annihilation to $b$-quarks, the ICS and bremsstrahlung components are strongly subdominant, compared to the prompt and thus including them 
affects weakly the derived limits. The only significant uncertainty (other than the DM distribution) in deriving limits is that of the background, that at the lightest mass
range, can lead to an uncertainty in the limits of up to a factor of 5.


\subsection{Limits from Antiprotons}
\label{subsec:pbarLimits}

Despite the number of experiments devoted to study the antiproton signal, the $\bar{p}$ spectrum has not revealed deviations from standard astrophysical backgrounds that require any exotic contribution. 
However, the balloon campaigns by the BESS detector~\citep{Matsunaga:1998,Maeno:2001,Asaoka:2002} and the more recent measurements by the \textit{PAMELA} satellite~\citep{Adriani:2010} have provided fairly good-precision antiproton data at energies up to about 180 GeV which allowed to derive good constraints on a variety of DM scenarios~\citep{Bergstrom:1999,Donato:2009,Cholis:2011,Garny:2011,Evoli:2011id}.
Indeed, if hadron production in WIMP pair annihilation is not forbidden either by kinematics or by some symmetry enforcing WIMPs to be coupled with leptons only, the ratio between the DM signal and the background from standard astrophysical sources is usually much larger in the antiproton channel with respect to all other indirect detection methods. 
A second aspect that makes antiprotons appealing in constraining DM contribution regards the fact that the theoretical prediction for the background component is fairly under control: the production of secondary antiprotons from the interaction of primary cosmic rays with the interstellar medium and, subsequently, their propagation in the Galaxy have to be modeled in close analogy to secondary versus primary cosmic ray nuclei, such as B/C. 
Once a given phenomenological model is tuned to reproduce the latter, the spread in predictions for the antiproton flux is modest~\citep{Evoli:2011id}. 

In the present analysis, we compare the WIMP annihilation cross section constraints obtained by studying the diffuse emission with those which can be derived from the antiproton spectrum provided by the \textit{PAMELA} experiment.  
For the $b\bar{b}$ annihilation channel, the antiproton limits are the most constraining ones as shown in Fig.~\ref{fig.bound}.
In $t\bar{t}$ and $W^+W^-$ annihilation channels, the extracted limits are competitive with those from low latitude diffuse $\gamma$-rays for DM masses below about 300 GeV.
The lepto-philic annihilation channels can also lead to the production of antiprotons through the radiative emission of electroweak gauge bosons if the DM mass is heavy enough \cite{Ciafaloni:2010ti}.  
In these cases, the antiproton limits are weaker than limits derived from other detection methods and sharply rise when the DM mass is light. 


\subsection{Limits from Leptons}
\label{subsec:Leptons}

The spectra of cosmic ray $e^{\pm}$s at high energies show a hardening with respect to the prediction of standard astrophysical sources \cite{Adriani:2010ib, Adriani:2008zr, FermiLAT:2011ab, 2013PhRvL.110n1102A, Ackermann:2010ij}. In fact the rise of the positron fraction first observed from \textit{PAMELA}  \cite{Adriani:2008zr} and confirmed by \textit{AMS}-02
\cite{2013PhRvL.110n1102A}, provides the most compelling evidence for hard positron population not predicted by standard calculations on the secondary cosmic ray positron production in the ISM. 
One possible interpretation is that the observed excess is connected with WIMPs annihilation and/or decay; 
however, there are other explanations related to astrophysical sources, such as local pulsars. For a review of the status before \textit{AMS}-02 data see \cite{Serpico:2011wg} while for the most recent analysis on the physical implications of the  \textit{AMS}-02 positron fraction see \cite{Yuan:2013eja, Linden:2013mqa, Cholis:2013psa, Yin:2013vaa, Gaggero:2013rya}.
Indeed, a combination of sources could be responsible for the observed features in leptons spectra.
In \cite{Cholis:2011un}, we fitted those spectra by including only the contribution from the distribution of pulsars following \cite{Malyshev:2009tw}. 
Here, we consider the combined contribution of WIMPs annihilation to leptons and of pulsars in such a manner as to allow for maximal DM contribution in both deriving a best fit and a $3\sigma$ upper limit values to the annihilation cross-sections. We use the spectrum of positron fraction measured by \textit{PAMELA} \cite{Adriani:2010ib, Adriani:2008zr} and \textit{AMS}-02 \cite{2013PhRvL.110n1102A}, the spectrum of electrons measured by \textit{PAMELA} \cite{PAMELA:2011xv} as well as the spectrum of $e^-+e^+$ measured by \textit{Fermi}-LAT \cite{Ackermann:2010ij}, MAGIC \cite{BorlaTridon:2011dk} and  H.E.S.S \cite{Aharonian:2008aa,Aharonian:2009ah}. 
To avoid uncertainties regarding the solar modulation and the calculation of the secondary production rate, we ignore data points below 10 GeV. 
The averaged spectral properties of pulsars (see \cite{Cholis:2011un} for details) and WIMPs annihilation cross section are fitted to those data sets.
Then 3$\sigma$ upper limits on $\langle\sigma v\rangle$ are derived by lowering the pulsar contribution to allow the maximum possible contribution from DM.  
Those limits are shown in Fig.~\ref{fig.bound}. 
Since we only consider data points above 10 GeV, no limit can be extracted from DM particles lighter than about 10 GeV.
The limits from leptons are less constraining than those from low latitude $\gamma$-rays for all studied annihilation channels except for the $\mu^+\mu^-$ annihilation channel where for masses bellow 1 TeV, they provide as tight limits as the $\gamma$-ray limits from our innermost region of study.

\section{Gamma-ray Limits and the Dark Matter Profile}
\label{sec:DMprofile}

\begin{figure}
\centering
\includegraphics[scale=.55]{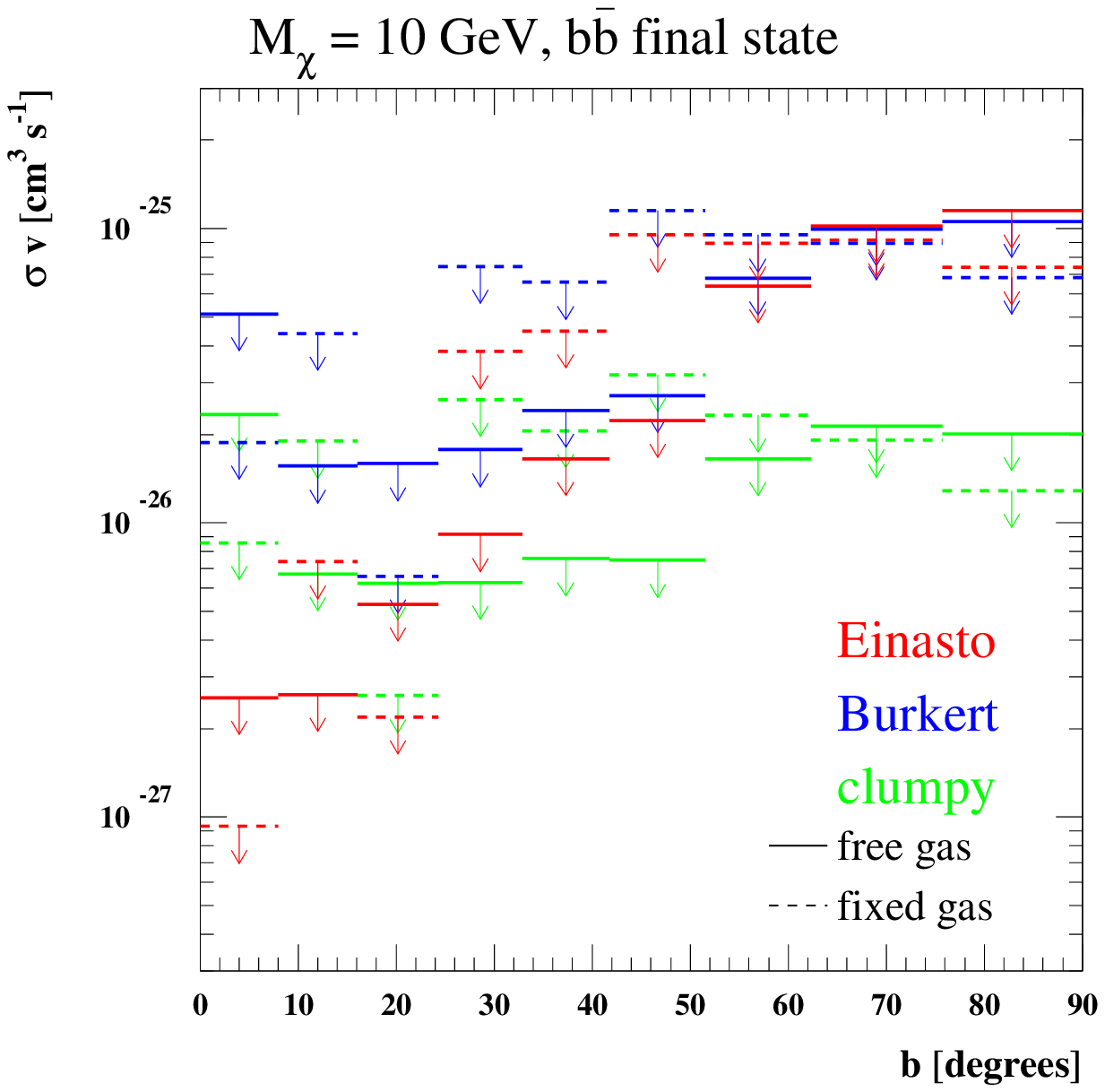}
\hspace{-0.7cm}
\includegraphics[scale=.55]{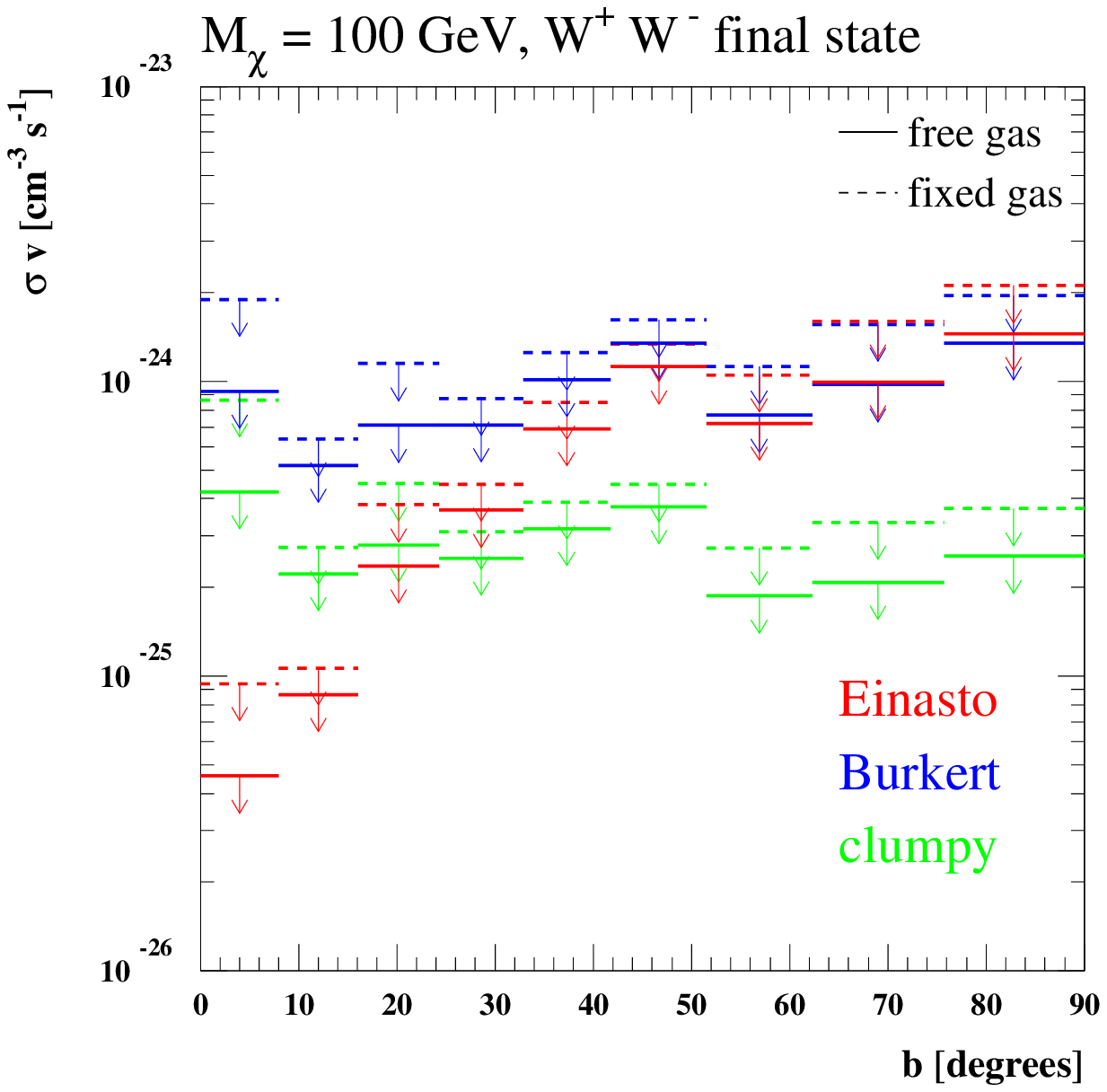} \\
\caption{Limits on the annihilation cross section for two sample WIMP  
models from the \textit{Fermi} diffuse $\gamma$-ray spectrum  in a few  
latitude windows  (longitude range
appropriately chosen to keep the angular window fixed, see the text  
for details) and
considering an Einasto profile, a Burkert profile and a case in which  
substructures dominate the annihilation yield ("clumpy"). Results are  
shown in our reference gas model ("fixed" gas) as well
as allowing for its rescaling ("free" gas).}
\label{fig:profile1}
\end{figure}

We have so far focussed the discussion on a single model for the DM
density profile, the Einasto profile introduced in  
Eq.~(\ref{eq:Einasto}).  This model is motivated
by results of numerical N-body simulations of hierarchical clustering  
of CDM
cosmologies, see e.g.~\cite{Navarro:2004, Graham:2006ae}, however is  
still controversial
whether it should be applied in this form to the Milky Way. For instance, the  
central enhancement
in the DM density seen in the simulations, could be washed out  
as a back-reaction
of a baryon infall scenario with large exchange of angular momentum  
between the gas
and the DM particles, see, e.g.~\cite{ElZant:2003rp}, turning the  
cuspy profile into a
profile with a constant density core, such as the Burkert  
profile~\cite{Burkert:1995yz}:
\begin{equation}
   \rho_{\chi}(r) = \frac{\rho_{B}}{(a+r)(a^2+r^2)}\,.
      \label{eq:Burkert}
\end{equation}
Such a profile is also phenomenologically motivated, since it is in  
better agreement with the
gentle rise in the rotation curve at small radii, which seems to be  
observed for many external
galaxies; especially in the case of low-mass dark-matter-dominated low  
surface brightness  and
dwarf galaxies, see, e.g., \cite{Gentile:2006hv}, 
but also suggested by observations towards spiral galaxies \cite{Gentile:2004}. Since the $\gamma$-ray  
  WIMP signal
scales with the integral along the line of sight of the square of the  
density, the limits we derive
depend critically on the assumptions made for the DM distribution.  
In Fig.~\ref{fig:profile1} we
consider two sample cases of WIMP annihilation channel and mass and  
extract the limit
on the annihilation cross section, considering separately 9 latitude  
windows, starting from the
GC with a  window encompassing $0^{\circ}<|b|<8^{\circ}$  
and $|l|<8^{\circ}$, and
moving towards $|b|=90^{\circ}$. We vary the longitude range to keep  
fixed the subtended
solid angle. Results are shown for the reference Einasto profile and  
for a Burkert profile
with the same local halo density normalization  of
$\rho_\chi(R_\odot)=0.4$~GeV~cm$^{-3}$
and a core radius $a=10$~kpc~\cite{Catena:2009mf}. We also show the  
case in which
annihilations are dominantly taking place in DM substructures,  
distributed according
to the same Burkert profile and a "boost factor" (in this case the product  
between the effective
density contrast times the fraction of DM in substructures),  
that is taken to be equal to 10. The latter case is
labeled as "clumpy" in the plot. The analysis is performed in the  
setups in which we
use our reference gas model (label "fixed gas" in the plot) or allow  
free rescaling of the gas density
(label "free gas" in the plot), see the discussion above.  One can  
clearly see that going from
the Einasto to the Burkert and further more to the clumpy profile, the  
model assumptions of the
flux at lower latitudes get more critical.

\begin{figure}
\centering
\includegraphics[scale=.55]{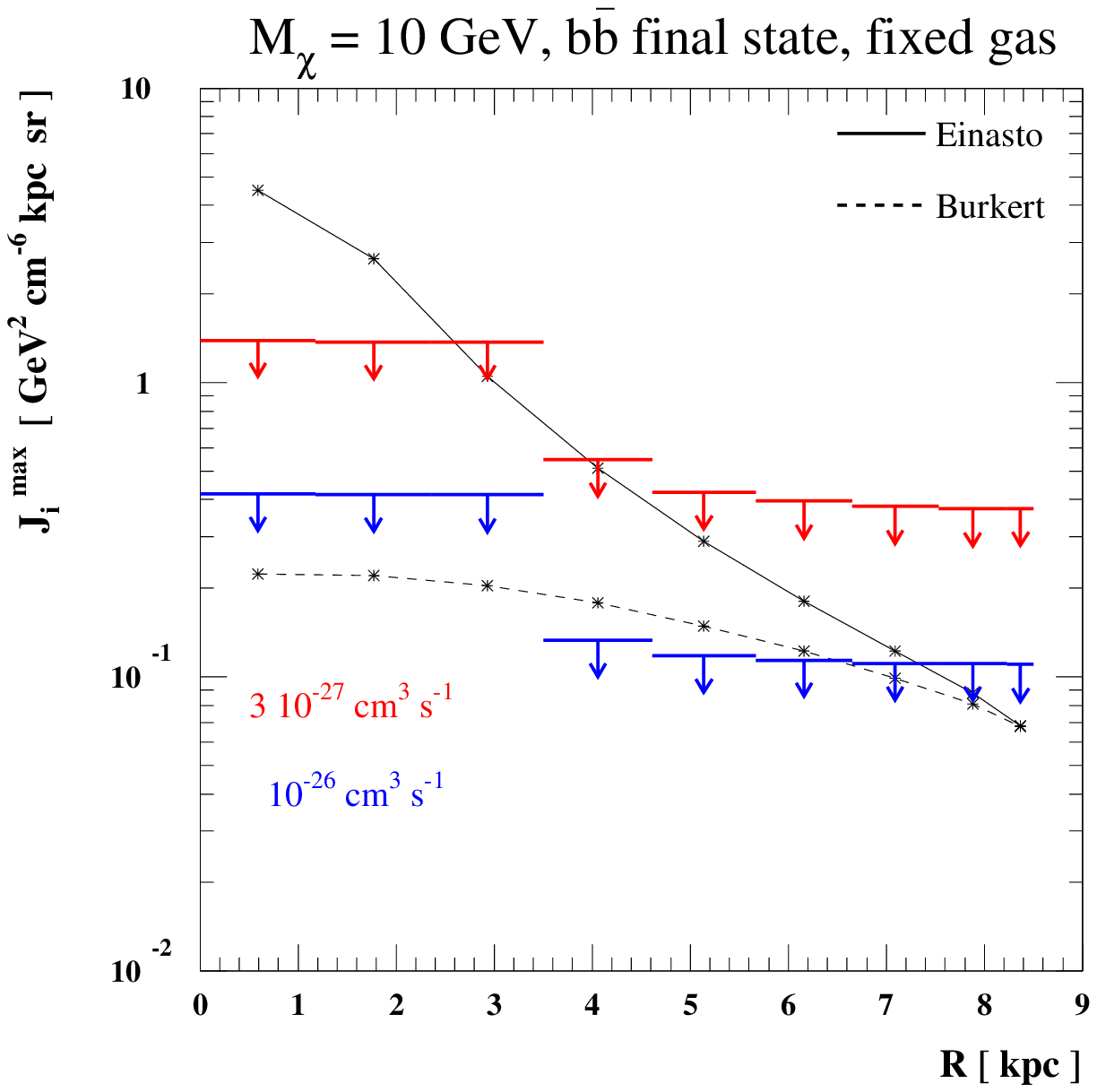}
\hspace{-0.7cm}
\includegraphics[scale=.55]{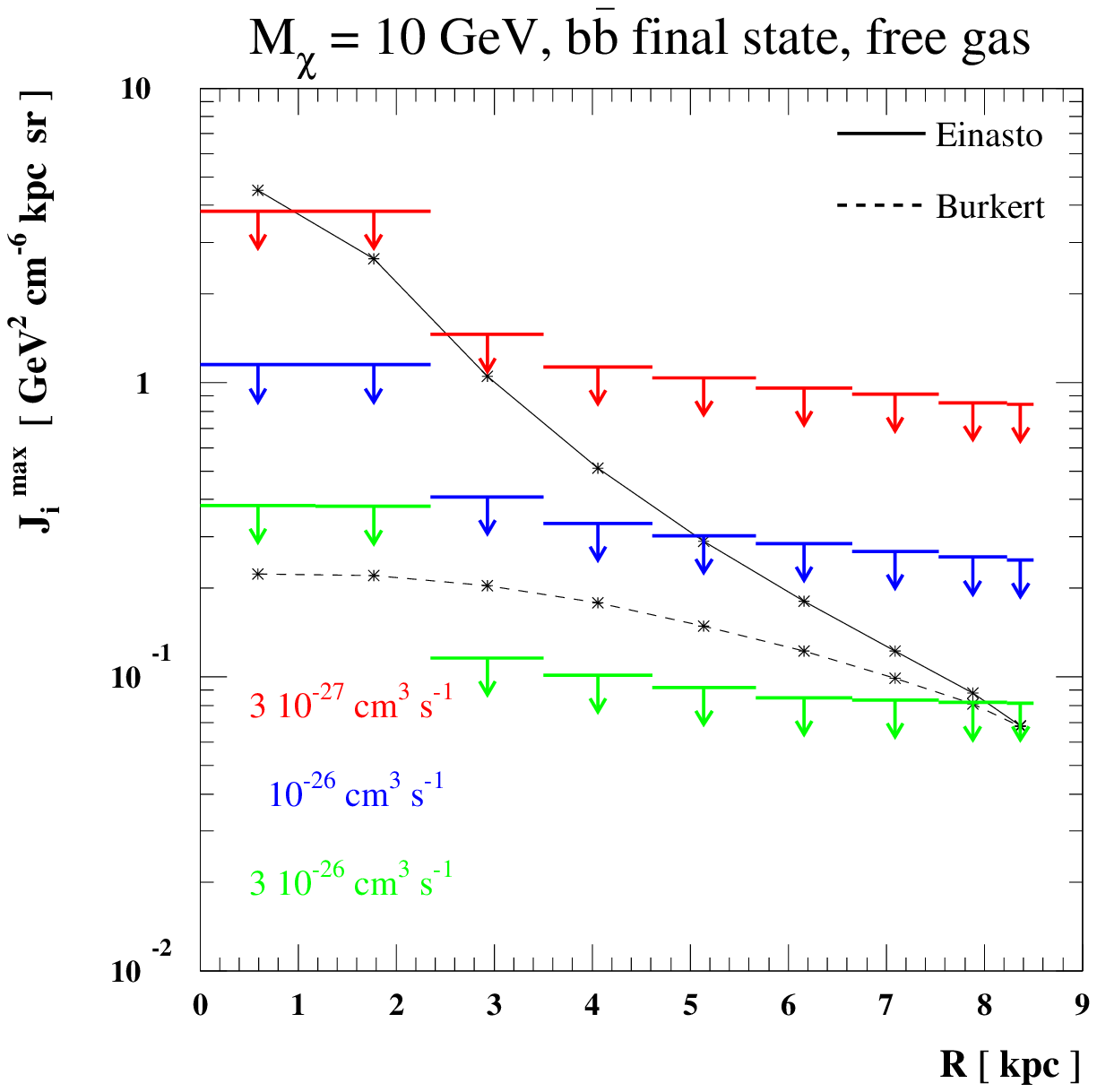} \\
\caption{Limits from different angular windows on the DM density line of sight integration factors $J_i$. We show results 
for given assumptions on the DM mass (10 GeV), annihilation channel (to $b\bar{b}$) and annihilation cross-sections ( $\sigma v $ = $3 \times 10^{-27}$, $1\times 10^{-26}$, $3 \times 10^{-26}$ cm$^{3}$s$^{-1}$). We also show the impact of assumptions made on the ISM gas distribution, \textit{left}: reference, \textit{right}: gas normalization free within a factor of 2.}
\label{fig:profile2}
\end{figure}

In Fig. ~\ref{fig:profile2} we turn the perspective around and for  
given WIMP model (namely
specifying its mass, annihilation channel and annihilation rate) we  
discuss what limits can
be extracted on the DM profile. To that extent, rather than  
specifying the profile via
a functional form and a few parameters as done so far, we define it  
parametrically at a set of
radii $r_i$ corresponding to the tangential points of the  
same latitude bins considered in
the previous figure. We still keep fixed the value for local halo  
density and assume that the
DM profile at larger radii follows the Einasto profile introduced above.
In the inner Galaxy the DM profile is constructed as a log-log interpolation between the  
values $\rho_i$
at the chosen radius $r_i$ (when varying $\rho_i$ we only impose that  
profile is monotonically
decreasing). The analysis is performed scanning the parameter space  
defined by the values
$\rho_i$, computing the line of sight integration in each of the  
considered angular windows
and comparing against the data to find whether the given configuration  
is allowed.
For all surviving models, we consider the bin encompassing the  
GC and compute
the line of sight integration factors $J_i$ obtained by imposing that  
the density profile is constant
below  $r_i$. In the figure we plot the maximum of  $J_i$ in our scan  
and compare it to the
analogous quantity for the preferred parametric profile: this gives a  
feeling for how close to
the GC one needs to trust the extrapolation of a  
parametric profile to claim a given
upper limit.

While the method leading to Fig.~\ref{fig:profile2} gives results that are strongly model dependent (on the background, the DM mass, annihilation channel and cross-section), 
it provides a global picture of the compatibility of specific model assumptions to the \textit{Fermi} $\gamma$-ray data. 
Given the fact that the \textit{Fermi}-LAT has been collecting data for almost 5 years, such constraints on the integration factors $J_i$ (for a given particle physics DM model), can not be evaded if the background and its uncertainties are properly modeled.   

\section{Comparison to other Indirect DM searches}
\label{sec:multiwavelength}

Cosmic ray $e^{\pm}$ which are produced by DM annihilations, depending on the DM mass and intermediate products of annihilation, have different injection spectra.
Those $e^{\pm}$ loose energy not only via bremsstrahlung radiation and inverse Compton scattering, which yields $\gamma$-rays, but also via synchrotron radiation. 
The synchrotron emission can be observed at microwave wavelengths toward regions with strong galactic magnetic fields and large DM densities.
Such a case can be the extended region toward the GC, where the \textit{WMAP} haze \cite{Finkbeiner:2003im} has been observed \cite{Finkbeiner:2004us, Hooper:2007kb, Dobler:2007wv}, or regions with strong localized magnetic fields traced by filamentary structures in the inner Galaxy \cite{Linden:2011au, Hooper:2011ti}.
The analysis of \textit{Planck}~data \cite{Planck:2012fb}, has shown that at latitudes $|b|<30^{\circ}$ the microwave haze morphology correlates well with the \textit{Fermi} bubbles.  
The hard spectrum and the extended nature of the haze make any possible DM contribution concentrated at the lower latitudes and subdominant unless large annihilation rates are imposed  \cite{Cholis:2008vb, Linden:2010eu}.  
At frequencies below the \textit{WMAP} and the \textit{Planck} data bands, the limits on DM models with $1<m_{\chi}<100$ GeV have been derived in the galactic halo \cite{Fornengo:2011iq, Mambrini:2012ue}. These limits are comparable to our limits from the regions with $\mid b \mid \leq 25^{\circ}$ apart from the case of DM lighter than 20 GeV annihilating to $b$-quarks  where our limits of Fig.~\ref{fig.bound} become more competitive. 

The CMB temperature and polarization power spectra measured by \textit{WMAP} can be used to constrain DM models and especially lepto-philic annihilation channels suggested to explain the excess in leptonic cosmic ray spectra \cite{Galli:2009zc, Slatyer:2009yq, Galli:2011rz, Evoli:2012qh, Lopez-Honorez:2013cua, Galli:2013dna}.
The \textit{Planck} data are expected to probe TeV scale lepto-philic DM models \cite{Galli:2009zc, Slatyer:2009yq}. Yet the same class of DM models with masses less than 300 GeV,  already phase their tightest limits from the absence of strong features in the \textit{AMS} positron fraction data \cite{Bergstrom:2013jra}. 
Our $\gamma$-ray limits from the inner region of the Galaxy, shown in Fig.~\ref{fig.bound}, are stronger than the current CMB limits while our lepton limits for the muon channel are at worse a factor of 2 weaker to those of \cite{Bergstrom:2013jra} and for the tau channel similar or stronger by a factor of $\sim 2$ to those of 
\cite{Bergstrom:2013jra}. 

Dwarf spheroidal galaxies (dSphs) being DM dominated objects provide a relatively clean set of targets for DM detection with $\gamma$-rays \cite{Evans:2003sc, Colafrancesco:2006he, Strigari:2006rd, Bovy:2009zs, Scott:2009jn} and up to now have provided the tightest limits on DM annihilation rate \cite{Ackermann:2011wa, Cholis:2012am} in annihilation channels with a strong prompt $\gamma$-ray contribution. 
While these targets are expected to have suppressed background $\gamma$-ray fluxes, the foreground $\gamma$-rays along the line of sight impose large uncertainties ($O(10)$) in the residual $\gamma$-ray spectra.
Therefore, the limits from dSphs strongly depend on the assumptions for these foregrounds \cite{Cholis:2012am}.
Additionally, the assumptions on their DM profiles may have a strong impact (factor of $\sim$ 3-10) on the limits derived from them \cite{Evans:2003sc,Charbonnier:2011ft,GeringerSameth:2011iw}.
Since the air Cherenkov telescope arrays \cite{Aliu:2012ga,Aleksic:2011jx} have a much better angular resolution, they will be able to separate the signal from  annihilating DM with mass heavier than 500 GeV from backgrounds toward both dwarf spheroidal galaxies and galaxy clusters. Our $\gamma$-ray limits from the inner part of the Galaxy and from leptons are for the $\chi \chi \longrightarrow \mu^{\pm}$ channel, at least an order of magnitude stronger than the most optimistic DM dwarf spheroidal limits presented in \cite{Ackermann:2011wa}; a factor of few stronger for the $\chi \chi \longrightarrow \tau^{\pm}$ case and for the $\chi \chi \longrightarrow b\bar{b}$ channel the limits from antiprotons are stronger bellow 300 GeV by a factor of typically a few (for reference propagation assumptions). We note that our $\gamma$-ray limits on the  $\chi \chi \longrightarrow b\bar{b}$ channel  from the inner galactic region are not stronger than those of  \cite{Ackermann:2011wa} down to $\simeq$30 GeV and get stronger only for the lower than 30 GeV masses. For DM annihilating to $W^{\pm}$ gauge bosons our $\gamma$-ray limits are similar to those presented in \cite{Ackermann:2011wa}.

An alternative indirect DM probe comes from the closest galaxy clusters which have a typical extension radius of a few degrees in the sky. 
In such small windows very few photons above 100 GeV are observed.
In addition, the $\gamma$-ray radiation from the isotropic and the diffuse galactic components and from uncorrelated point sources, which lay along the same 
line of sight, dominate any signal at $\gamma$-rays from these targets. 
Furthermore, galaxy clusters have intergalactic atomic and molecular gas which produce $\gamma$-rays of non DM origin, 
via $\pi^{0}$ decays and bremsstrahlung radiation.
The emissivity of $\pi^{0}$ and bremsstrahlung backgrounds is proportional to the product of the cosmic ray proton or electron densities (produced in galaxies) with the atomic and molecular gas densities that exist both inside the galaxies and in the intergalactic medium. 
There is also another contribution to $\gamma$-rays from inverse Compton scattering which may be the most difficult to evaluate because of uncertainties in the radiation field in galaxy clusters.
The exact assumptions on the radiation field and gas densities can considerably affect the calculations on the background contribution. 
Furthermore, the assumptions on distribution of DM substructures and the evolution of the DM profile in galaxy clusters can significantly change (factors of up to $10^{3}$) the DM limits from these targets (for recent analysis see \cite{Pinzke:2011ek, Ando:2012vu, Han:2012au}). 
The analysis of \cite{Han:2012au} and \cite{Hektor:2012kc} have been suggested possible signals of DM annihilation from $\gamma$-ray observations toward nearby galaxy clusters, while \cite{Aharonian:2009bc, Aleksic:2009ir, Ackermann:2010rg, Dugger:2010ys, Zimmer:2011vy, Huang:2011xr} have seen no evident $\gamma$-ray excess, yet their limits are systematically weaker than these from dwarf spheroidal galaxies and of our tighter limits shown in Fig.~\ref{fig.bound}. 

Searches for $\gamma$-ray lines in the $\textit{Fermi}$ data, have recently provided both excesses and general limits \cite{Vertongen:2011mu, Bringmann:2012vr,Weniger:2012tx, Su:2012ft, Rajaraman:2012db,Tempel:2012ey, Fermi-LAT:2013uma}. Yet, the exact interpretation of these excesses/limits on DM is highly model dependent (see for instance \cite{Buchmuller:2012rc, Cohen:2012me, Buckley:2012ws, Fan:2012gr, Asano:2012zv, Weiner:2012gm}). Thus we do not make any comparison between our results and this probe of indirect DM searches. 

The isotropic diffuse $\gamma$-ray emission at high latitudes, can be employed to set limits on signals from the extragalactic DM  structures and from unresolved galactic DM substructures, which are within the Milky Way virial radius \cite{Abdo:2010dk, Hutsi:2010ai, Calore:2011bt, Singal:2011yi}. 
A different approach to extract a DM signal is to search for the small angular scale fluctuations of the isotropic $\gamma$-ray background \cite{Hensley:2009gh, Cuoco:2010jb}.
Thanks to the high sensitivity of the \textit{Fermi}-LAT measurements in a wide energy range, the diffuse $\gamma$-ray power spectrum has been analyzed in four energy bins \cite{Cuoco:2011ng, Ackermann:2012uf}. 
This has allowed constraining the anisotropies of the DM induced emission \cite{Fornasa:2012gu}.
Those constraints strongly depend on the distribution of galactic and extragalactic DM (sub)halos, which is calculated by cosmological simulations \cite{Diemand:2006ik, Diemand:2007qr, BoylanKolchin:2009nc}, and their uncertainties are dominated by the abundance of low-mass (sub)halos.
The contribution of extragalactic DM structures to the isotropic $\gamma$-ray background can be affected by uncertainties in the evolution of the intergalactic radiation field \cite{Aharonian:2005gh,Franceschini:2008tp,Gilmore:2011ks,Cavadini:2011ig}.
However, these uncertainties are less significant than those of low mass (sub)halo population. 
In our limits from the inner part of the Galaxy, we avoid the inner  $1^{\circ}$ in latitude, where the DM profile matters significantly; while we have also shown in Fig.~\ref{fig.gammaLimitsAstroUncer} the general impact of the other astrophysical uncertainties. Currently our limits are stronger by a factor of $\sim$ 10 and are significantly more robust than those from anisotropies in the diffuse $\gamma$-ray background \cite{Fornasa:2012gu, Gomez-Vargas:2013cna}. Yet, improvements in that search probe are expected to be important, as more data are being collected and better understanding of the relevant backgrounds  is being reached \cite{SiegalGaskins:2010mp, Cuoco:2012yf}.

\section{Summary and Conclusions}
\label{sec:conclusions}

In this work, we study the properties of annihilating dark matter using the  \textit{Fermi}-LAT $\gamma$-ray spectral data.
Since the bulk of the $\gamma$-rays is expected to be of astrophysical origin in most directions on the galactic sky, it has been necessary to first evaluate a reference diffuse $\gamma$-ray background, including also a discussion on its uncertainties and their impact in deriving limits on DM via $\gamma$-ray observations.
We model the diffuse galactic $\gamma$-ray components produced by the $\pi^0$ decay, inverse Compton scattering and bremsstrahlung emission.
To that end, we determine the properties of cosmic ray galactic diffusion based on results of \cite{Cholis:2011un} and then fit the parameters governing the propagation of cosmic rays within the Galaxy against the local flux of cosmic rays. 
To calculate the $\pi^0$ and bremsstrahlung components, we employ the most recent models for the distribution of the atomic and molecular hydrogen gas while including as well the contribution of the dark gas (see section~\ref{sec:background}). 
In addition to the mentioned components, there are contributions from point and extended sources, the isotropic extragalactic background radiation and from specific extended diffuse features, i.e. the \textit{Fermi} bubbles, the Loop I and the northern arc which are modeled in \cite{Su:2010qj}.
Having constructed the diffuse $\gamma$-ray background, we then examine its consistency with the \textit{Fermi} data in the 1-200 GeV energy range and in 60 angular windows covering the entire sky.   
Our reference model has a generally good agreement with the \textit{Fermi} diffuse $\gamma$-ray data in the sky regions under study, as shown in Fig.~\ref{fig:chi2FullSky4cases}.
We apply alternative models for the galactic distribution of the molecular hydrogen gas, motivated by different assumptions on the radial dependence of $X_{CO}$. We also test the  distribution of the interstellar radiation field for different assumptions on the metallicity gradient on the galactic disk and the stellar population on the Disk and the Bulge. The level of agreement between the \textit{Fermi} data and the prediction of the model changes are shown in Figs.~\ref{fig.gamma_XCO} and \ref{fig.gamma_ISRF}.   

Our reference diffuse background model, is used to search for the possible contribution of DM to the diffuse $\gamma$-ray flux.
To get conservative constraints, we replace the spectrum of the extragalactic background \cite{Abdo:2010nz} with that of the minimal non-DM extragalactic $\gamma$-ray background which is about 40\% of the EGBR flux between 100 MeV and 100 GeV measured by \cite{Abdo:2010nz} (see discussion in Appendix~\ref{sec:ISRF}). 
Since the exact normalization of the fluxes of the \textit{Fermi} bubbles, Loop I and the northern arc is uncertain, we ignore their contribution in extracting conservative limits. 
The relative strength in placing limits on DM from $\gamma$-ray observations at different patches on the galactic sky is shown in Fig.~\ref{fig.gammaLimitsFullSky}
for three characteristically different models of DM ranging between 10 GeV and 1.6 TeV. In Fig.~\ref{fig.gammaLimitsAstroUncer} we show for the same 
windows on the galactic sky, the impact that the uncertainties in the gas and the radiation field distribution have on the derived DM limits for the same three reference DM models. DM models with a dominant prompt component receive their tightest constraints from the inner few degrees, but suffer from uncertainties in the derived limits that are up to a factor of 4 due to uncertainties in the gas distribution (in the selected windows of Figs.~\ref{fig.gammaLimitsAstroUncer}). TeV scale 
DM models annihilating dominantly to leptons receive their tightest limits from regions above the galactic center, with these constraints having an uncertainty by a factor of up to $\sim$2 due to uncertainties in the radiation field. 

The limits on WIMPs annihilation rate for various annihilation channels, which are shown in Fig.~\ref{fig.bound}, are stringent and compatible with bounds derived by other analyses \cite{Ackermann:2010rg, Ackermann:2011wa, Ackermann:2012rg}.
We find that for DM models with mass $\lessapprox 100$ GeV annihilating into $\tau^+\tau^-$ and mass $\lessapprox 30$ GeV annihilating into $\mu^+\mu^-$, $b\bar{b}$, the thermal relic value of the annihilation cross section is excluded by the \textit{Fermi} spectral data in $1^{\circ}< |b| < 9^{\circ}$ and $|l| < 8^{\circ}$ angular window.  
Considering the flux of antiprotons, we can exclude DM models with masses $\lessapprox 150$ GeV and large branching ratios to quarks and gauge bosons.
Limits from the \textit{AMS}-02 cosmic ray positron fraction are also constraining for channels producing hard spectrum leptons.   

In most cases our most stringent $\gamma$-ray limits come from our window of $1^{\circ}< |b|< 9^{\circ}$, $|l|<8^{\circ}$, i.e. we
do not need to resort to the inner $1^{\circ}$ around the GC where both backgrounds and the DM halo profile suffer from their greatest uncertainties. 
In addition we note that models that have already been excluded from XENON100, typically have large couplings to quarks and as a result would also give large CR antiproton fluxes.
As a result, the measured spectrum of CR antiprotons while very strong in setting limits  when compared to only other indirect DM probes, it is less effective  
when combined with direct detection results (within for instance the MSSM) than $\gamma$-rays are, which provide a less degenerate DM search probe.

In addition in Fig.~\ref{fig:profile1} we study the impact that assumptions on the DM distribution have, over the DM annihilation cross-section limits vs latitude, 
for given choices of DM mass and annihilation products. While the assumptions on the cuspiness  of the DM profile impact the limits by more than a factor of 2 
only in the inner 30 degrees, the assumptions on the significance of DM clumps can have an impact on the derived limits by more than a factor of 2 in most galactic latitudes.
Finally we turn the argument around and also present a method in deriving constraints on the DM distribution profile from the diffuse $\gamma$-ray 
spectra for fixed DM particle properties i.e. its mass and annihilation products (see Fig.~\ref{fig:profile2}). 
 
\acknowledgments

The authors acknowledge the use of the \textit{Fermi}-LAT data and parts of the ``FermiTools'' publicly available at \texttt{http://fermi.gsfc.nasa.gov/ssc/}. 
We also acknowledge using HEALPix \cite{Gorski:2004by}.
We would like to thank Alessandro Cuoco, Dan Hooper, Hani  Nurbiantoro Santosa, Meng Su and Gabrijela Zaharijas for the valuable discussions we have shared.
The work of MT is supported by the German Science Foundation (DFG) under the Collaborative Research Center (SFB) 676 ÒParticles, Strings and the Early UniverseÓ.
MT also thanks ICTP-SAIFR for its hospitality during the final stages of this work.
CE~acknowledges support from the Helmholtz Alliance for Astro-particle Physics funded by the Initiative and Networking Fund of the Helmholtz Association.
This work has been supported in part by the US Department of Energy (IC).
  
\begin{appendix}

\section{Impact of $X_{CO}$ factor on Diffuse $\gamma$-ray Background}
\label{sec:XCOfactor}

\begin{figure}
\centering
\includegraphics[scale=.39]{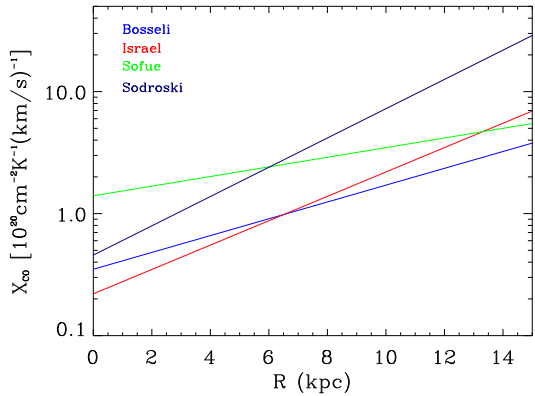}
\caption{Radial profile of the examined models for $X_{CO}$ in this analysis. The models are provided by Bosseli et al. \cite{Boselli:2001wj} (\textit{blue line}), Israel \cite{Israel:1997wn, Israel:2000mx} (\textit{red}), Sofue et al.\cite{1996PASJ...48..275A, Nakanishi:2006zf} (\textit{green}) and Sodroski et al. \cite{1995ApJ...452..262S} (\textit{dark blue}).}
\label{fig.XCO}
\end{figure}

The spatial variation of $X_{CO}$ in our Galaxy influences the distribution of molecular hydrogen gas and the estimate of diffuse $\gamma$-ray components which are produced by $\pi^0$ decay and bremsstrahlung emission.
To investigate this effect, we examine different phenomenological assumptions on radial dependence of $X_{CO}$.  
The profile of these models is shown in Fig.~\ref{fig.XCO}.
The models labelled by ``Bosseli" \cite{Boselli:2001wj}, ``Israel" \cite{Israel:1997wn, Israel:2000mx} and ``Sofue" \cite{1996PASJ...48..275A, Nakanishi:2006zf} have been derived by combining the analysis of the $X_{CO}$ dependence on the ultraviolet radiation field and the metallicity for a sample of nearby galaxies with the measurement of the galactic metallicity gradient.
The model labelled by ``Sodroski" \cite{1995ApJ...452..262S} has been obtained by the study of the \textit{COBE} diffuse infrared background experiment (DIRBE) and several virial analysis of giant molecular cloud complexes in the galactic disk.  

\begin{figure}
\hspace{-0.5cm}
\includegraphics[scale=.78]{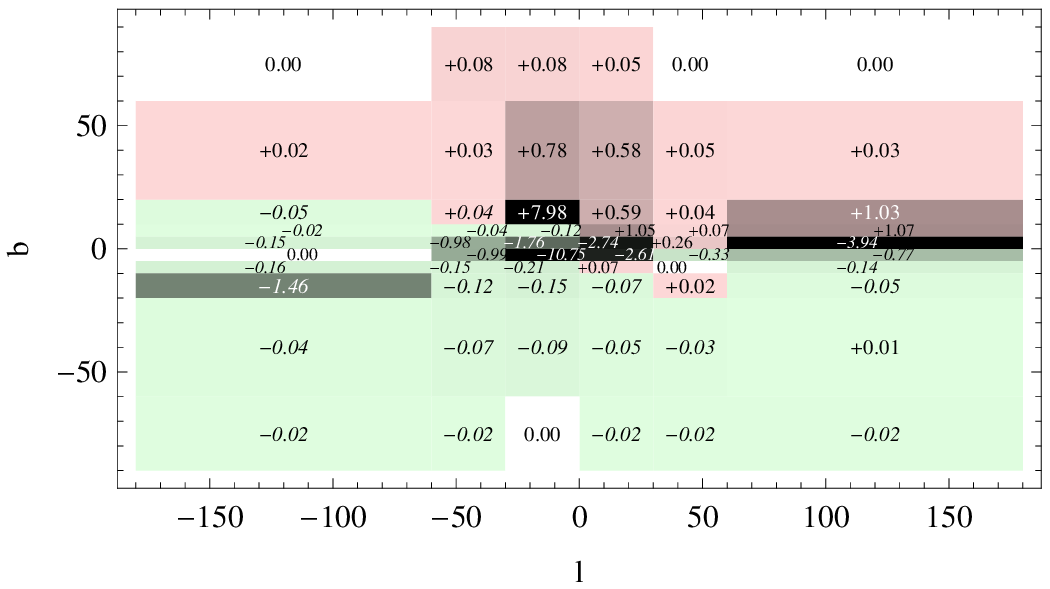}
\includegraphics[scale=.78]{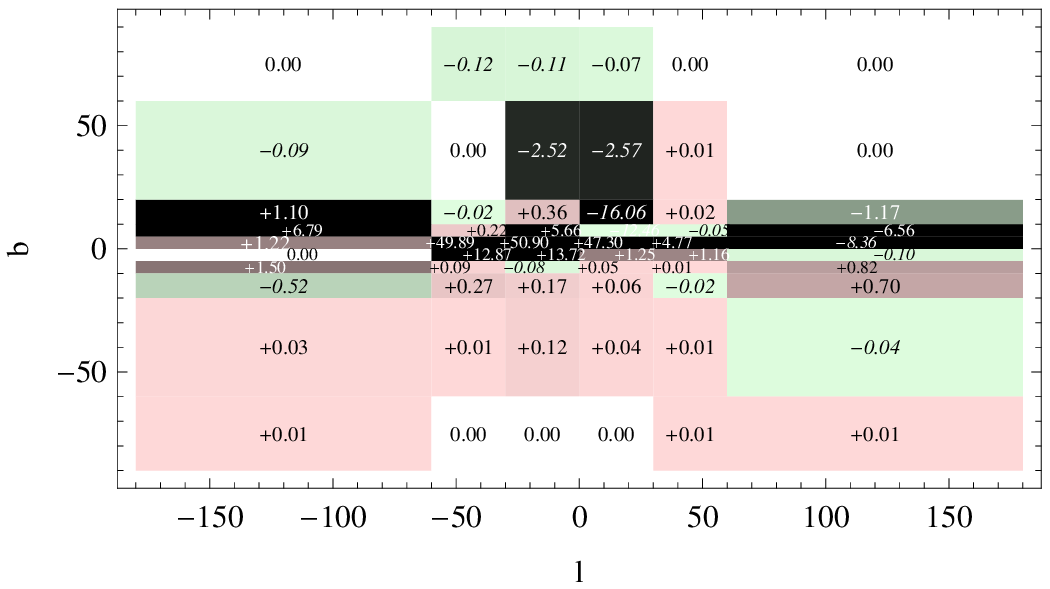} 
\caption{Difference between the reduced $\chi^2$ of the spectrum predicted by $X_{CO}$ profiles labelled by ``Israel" (\textit{left}) and  ``Sofue" (\textit{right}) and that by our reference $X_{CO}$ profile ``Bosseli"  in sky regions under study.}
\label{fig.gamma_XCO}
\end{figure}

For each assumption on the radial distribution of $X_{CO}$, we fit the propagation parameters against the local spectra of cosmic rays and then evaluate the spectra of diffuse $\gamma$-rays. 
The level of agreement between the \textit{Fermi} data and the predicted diffuse $\gamma$-ray spectra by each model is then compared with that by the reference model, labelled by ``Bosseli".  
In Fig.~\ref{fig.gamma_XCO}, the difference between the reduced $\chi^2$ of the spectrum predicted by the profiles labelled by ``Israel" (Fig.~\ref{fig.gamma_XCO}, left panel) and ``Sofue" (Fig.~\ref{fig.gamma_XCO}, right panel) and that by our reference profile (Fig.~\ref{fig:chi2FullSky4cases}, top right) is shown in all windows under study.
The ``Israel" profile, which is slightly steeper than our reference profile, provides better results at $|b|<5^{\circ}$.
At higher latitudes the difference between the predictions of the ``Israel" and the reference profiles is very small except for the region with $-30^{\circ}<l<0^{\circ}~\&~ 10^{\circ}<b<20^{\circ}$ which has rather larger reduced $\chi^2$ with respect to the reference model. 
The predicted spectra by the ``Sofue" profile in the inner Galaxy, $-50^{\circ}<l<50^{\circ}~\&~ |b|<5^{\circ}$, are in poor agreement with the \textit{Fermi} data.
This profile results in smaller reduced $\chi^2$ in the north where the \textit{Fermi} bubbles/haze, Loop I and the northern arc are expected to contribute.
In the rest of windows, the differences are negligible apart from the windows with $60^{\circ}<l<180^{\circ}~\&~ |b|<5^{\circ}$ and $-180^{\circ}<l<-60^{\circ}~\&~5^{\circ}< |b|<10^{\circ}$ which have, respectively smaller and larger reduced $\chi^2$ than the reference model.
The diffuse $\gamma$-ray spectra obtained by the ``Sodroski" profile are in serious tension with the \textit{Fermi} data at $|b|<10^{\circ}$ while they agree well with the data at higher latitudes.
The differences between the reduced $\chi^2$'s of the spectra provided by the reference and the ``Sodroski" profiles, which are not shown here, are significantly large at low latitudes especially in the northern hemisphere windows.

Among the tested radial profiles, the predicted diffuse $\gamma$-ray spectra by the reference and the ``Israel" profiles are in relatively good agreement with the \textit{Fermi} data in all regions under study. 
The ``Israel" profile is slightly preferred, since it gives better results at low latitudes, while the ``Sofue" and ``Sodroski" profiles are disfavored. 

\section{Impact of Interstellar Radiation Field on the Diffuse $\gamma$-ray Background}
\label{sec:ISRF}

The stellar population in our Galaxy has four spatially distinct components; the thin and the thick galactic disks, the galactic bulge or bar and the stellar halo.

The thin disk has continuous star formation and high metallicity, while the thick disk with the age of $\sim 1$ Gyr has very suppressed star formation and lower metallicity than the thin disk.
The density of the thin and the thick components exponentially decreases outward the galactic disk as follows, 
\footnote{While $d\rho /dz$ is not continuos at $z \longrightarrow 0$, the overall function is still justified at small $z$ because the interstellar gas concentrated to the disk modifies the gravitational field \cite{2008gady.book.....B}.}
\be
\rho_{disk} = \rho_{disk} (R,0) e^{-\mid z \mid/z_{D}}.
\ee
The scale heights $z_D$ of the thin and the thick disks are, respectively, $\simeq 0.3$ kpc and $\simeq 1$ kpc.
The mid plane density of the thick disk is about 50 times smaller than that of the thin disk \cite{2008gady.book.....B}.
Thus, the disk which is the combination of the thin and the thick disks is dominated by the thin component.
The disk stars have a variety of ages because of the continuous star formation.
The disk comprises the major part of the galactic luminosity with $L = 2.5 \pm 1 \times 10^{10} L_{\odot}$ and the remaining contribution, $\simeq 5 \pm 2 \times 10^9 L_{\odot}$, is provided by the bulge component \cite{2008gady.book.....B}.
From surface brightness measurements of disk galaxies the intensity is expected to be described by,
\be
I \propto e^{-R/R_{D}}, 
\ee     
where $R_{D} \simeq 2-3$ kpc.
Since the disk is the dominant component which provides the luminosity, we assume that the contribution of both the thin and the thick disks to the ISRF is spatially described by,
\be
\rho_{disk_i}(R,z) = \rho_{disk_i} e^{-R/R_{D}} e^{-\mid z \mid/z_{D}},
\ee  
where $R_{D} = 3$ kpc and $z_{D}$ is the same as the scale height of the distribution of stars.
The relative normalization of the thin to the thick disks is taken to be 50 to 1 on the disk plane. 

The bulge luminosity is $\simeq$ 1/5 of the disk luminosity. 
The bulge stars, unlike the disk stars, originate from the early formation stages of the Galaxy and compose a more evolved distribution.
This results in significantly higher stellar density in the bulge with respect to the disk.
The morphology of the bulge is triaxial (bar-like) with axis ratios of 3:1:1 \cite{Dwek:1995xu, Babusiaux:2005zi, Zoccali:2009hu}.
We use the 2-D model of  \cite{Porter:2005qx} as reference.
The profile of the bulge can be described by,
\be
\rho_{bulge}(R,z) \sim exp \left[ -\frac{1}{2} \left( \frac{R^2}{R_{0}^{2}} + \frac{z^2}{z_{0}^{2}}\right)\right],
\ee  
where $R_{0}=0.9$ kpc and $z_{0}=0.5$ kpc from fitting to the \textit{COBE} DIRBE near infrared luminosity data \cite{Dwek:1995xu}.

Finally, the profile of the very old stellar halo which has low metallicity and contains $\simeq 1 \%$ of the total stars is described by,
\be
\rho_{halo}(r) \sim \left(\frac{1}{r+r_{c}}\right)^{3},
\ee
where $r_{c}=2$ kpc \cite{Malyshev:2010xc, Juric:2005zr}.

\begin{figure}
\hspace{-0.5cm}
\includegraphics[scale=.78]{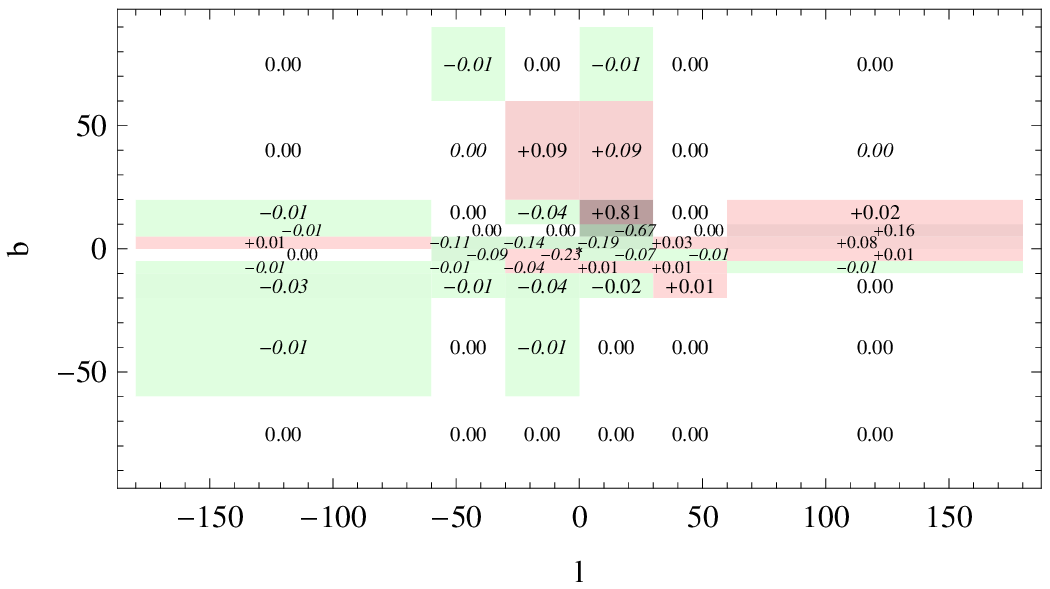}
\includegraphics[scale=.78]{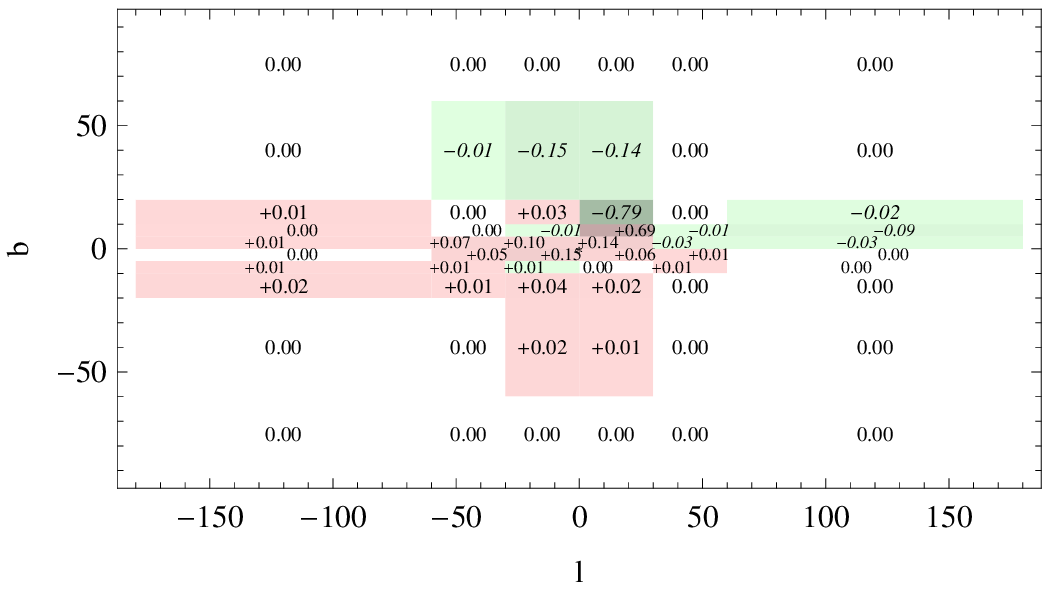} \\
\hspace*{-0.5cm}
\includegraphics[scale=.78]{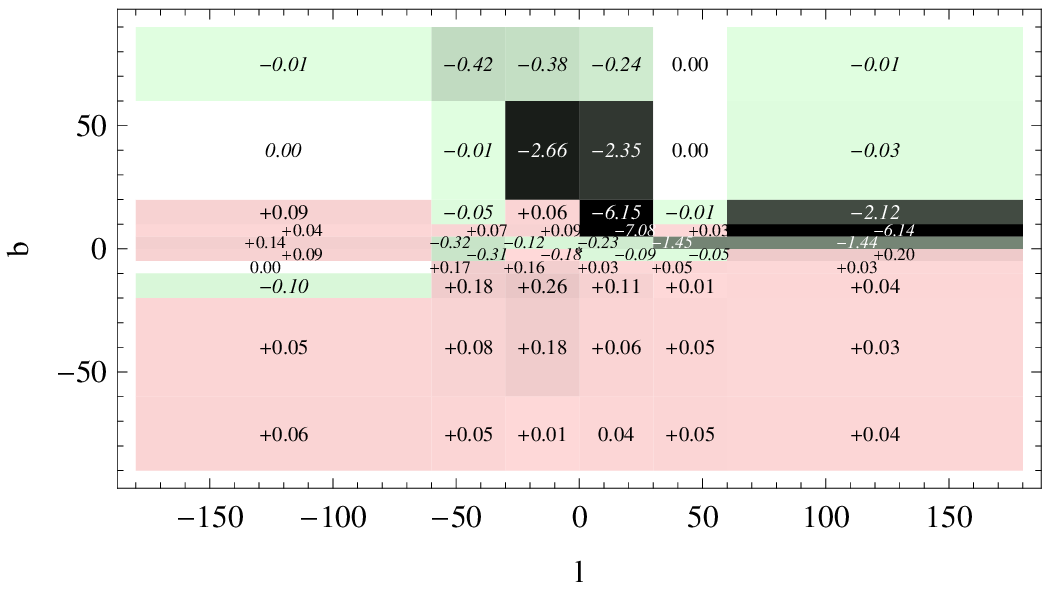}
\includegraphics[scale=.78]{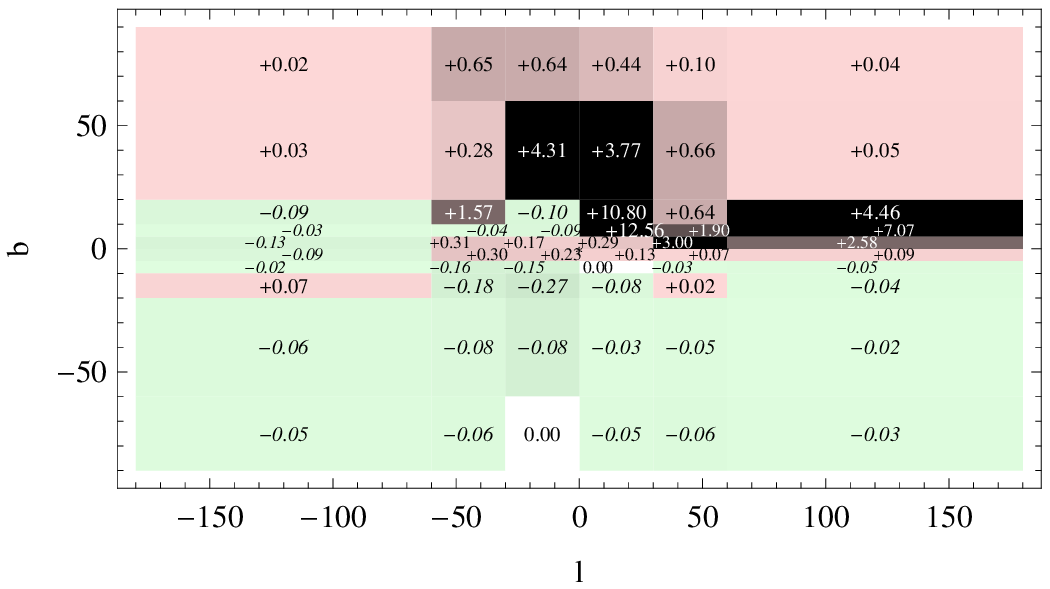} \\
\hspace*{-0.5cm}
\includegraphics[scale=.78]{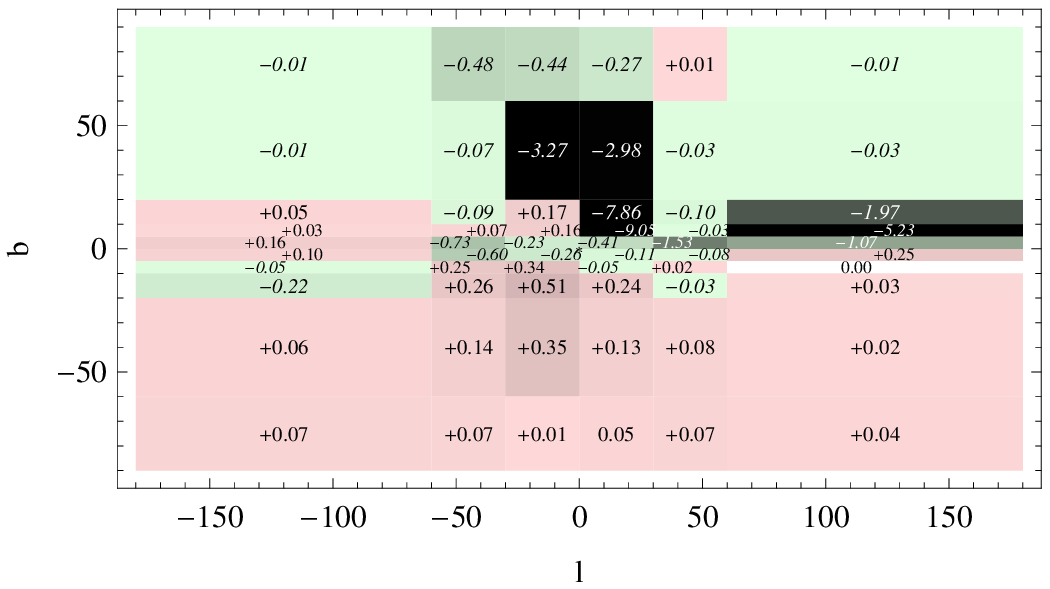}
\includegraphics[scale=.78]{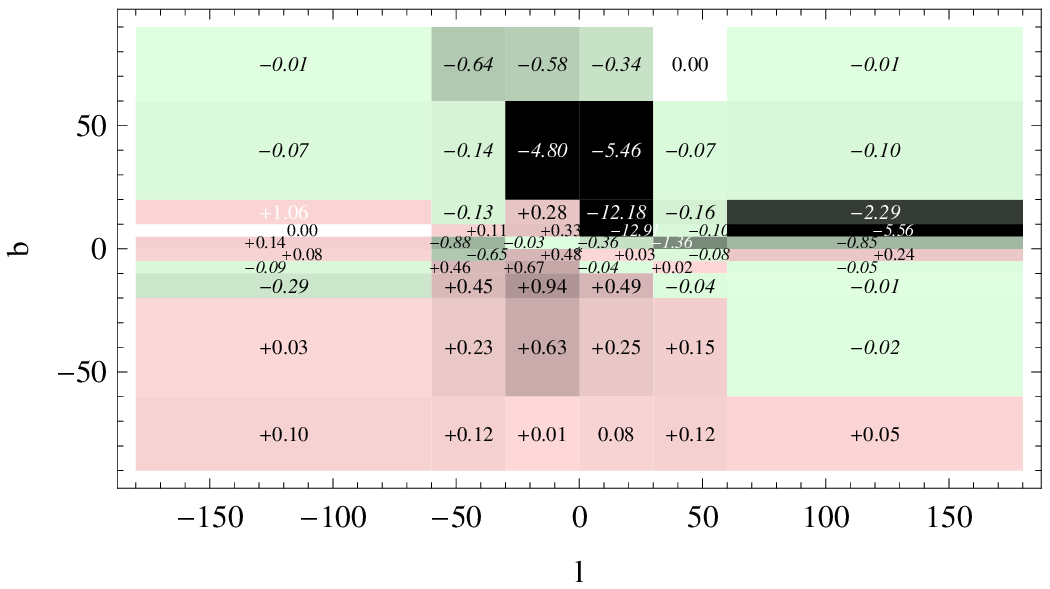}
\caption{Difference between the reduced $\chi^2$ of the spectrum predicted by different ISRF assumptions and our reference ISRF. 
\textit{Top row}: varying the amplitude of the Bulge stellar component; maximum bulge (\textit{left}), minimum bulge (\textit{right}).
\textit{Middle row}: varying the amplitude of the Disk stellar component; maximum disk (\textit{left}), minimum disk (\textit{right}).
\textit{Botom row}: varying the metallicity gradient; maximum gradient (\textit{left}), minimum gradient (\textit{right}).}
\label{fig.gamma_ISRF}
\end{figure}

A more accurate description includes many different sub-populations of stars \cite{Strong:1998fr, Porter:2005qx, Moskalenko:2005ng, Porter:2006tb, Porter:2008ve}.
As reference, we use the model of \cite{Porter:2005qx} which takes the various subcomponents into account.
Apart from the reference model, we examine different assumptions on the major stellar components to investigate the effect of the ISRF uncertainties on the estimated $\gamma$-ray spectra.
We keep the spatial distribution and morphology of the disk and bulge contributions fixed, since they are well measured, and test different values of luminosity which is more uncertain.
The difference between the reduced $\chi^2$'s obtained by different examined ISRF models and the reference model are shown in Fig.~\ref{fig.gamma_ISRF} for  all windows under study.  
In models which are labelled as``maximum (minimum) bulge", we increase (decrease) the bulge luminosity from 1/6 to 1/3 (1/20) and keep the disk contribution fixed.
In models labelled by ``maximum (minimum) disk", the luminosity of the disk is increased (suppressed) by $50 \%$ while the bulge component is kept fixed. 
The ratio of the disk to the bulge luminosity is 8:1 for the``maximum disk" and 5:2 for the``minimum disk".

The radiation emitted by stars is scattered or absorbed by the interstellar dust grains and re-emitted at far infrared.
Therefore, the distribution of dust which is correlated with the assumptions on the metallicity gradient significantly affects both the spatial profile and the spectrum of the ISRF \cite{Porter:2008ve}.
Based on results of \cite{Porter:2008ve}, the maximum (minimum) metallicity gradient is 0.07 dex kpc$^{-1}$ (0 dex kpc$^{-1}$). 
These two extreme cases which are labelled by ``maximum (minimum) gradient" are examined as well. 

The impact of uncertainties of the bulge properties on the diffuse $\gamma$-ray spectra, is subdominant compared to those of the galactic disk.
The ``maximum disk" model provides better agreement to the $\gamma$-ray data than the ``minimum disk" and the reference model. 
While both variations of the metallicity gradient lead to better agreement with the data than the reference model,  the ``minimum gradient" gives the best fit to the data.

\section{Minimal non-DM Contribution to Extragalactic Background Radiation}
\label{sec:EGBR}

The extragalactic background radiation (EGBR), which is measured by the \textit{Fermi} collaboration \cite{Abdo:2010nz, Collaboration:2010gqa}, is described by a single power-law $dN_{\gamma}/dE \sim E^{-2.41}$ in the energy range of 100 MeV to 100 GeV.
The total flux of $\gamma$-rays in this energy range is 1.09 $\times 10^{-5}$ ph $\textrm{cm}^{-2} \textrm{s}^{-1} \textrm{sr}^{-1}$. 
DM annihilation in the main halo and in substructures can contribute to the diffuse $\gamma$-ray spectrum even at high latitudes.
Thus, it is involved in the isotropic $\gamma$-ray spectrum. 
In addition, DM annihilation in early proto-halos at high redshifts can contribute to both EGBR and its power spectrum \cite{Ackermann:2012uf}.

In this paper, we aim at deriving conservative upper limits on the DM annihilation rate in the DM halo.
Therefore, we will consider the minimal contribution from non-DM sources of $\gamma$-rays to the EGBR spectrum. 
Among the known sources that can contribute to the EGBR, we consider the contribution of BL Lacertae-like objects (BL Lacs), flat spectrum radio quasars (FSRQs),
millisecond pulsars (MSPs), unresolved star-forming galaxies (SFG), Fanarof-Riley I and II radio galaxies (FRI, FRII) and the contribution from ultra high energy cosmic rays (UHECRs) which produce $\gamma$-rays through interactions with the cosmic microwave background and the interstellar radiation field (UHECR CMB). 
We also include the possible contribution to the EGBR from undetected/unidentified gamma-ray bursts (GRBs), from starburst and luminous infrared galaxies (SBG), the contribution from cascades which are produced by interaction of ultra high energy cosmic ray protons with the inter-cluster medium (UHEpr ICM) and finally the contribution from electrons which are accelerated by gravitationally induced shock waves in the intergalactic medium and up-scatter low energy photons to $\gamma$-rays (IGS).
 
BL Lacs and FSRQs are two populations of Blazars, the active galactic nuclei (AGN) with a relativistic jet pointing straight at us.
They are common among the detected point sources and their unresolved members are expected to contribute to the EGBR. 
Their contribution to the EGBR can be determined from identified point sources which belong to those categories.
BL Lacs are objects with strong radio emission, rapid and strong variability in their optical luminosity and non-thermal optical spectrum. 
Following the assumptions of \cite{Collaboration:2010gqa}, the differential spectrum of the unresolved BL Lacs is described by
\be
\frac{dN_{\gamma}}{dE}_{BL~Lac} = 3.9 \times 10^{-8} E_{\gamma}^{-2.23} \textrm{GeV}^{-1}\textrm{cm}^{-2}\textrm{s}^{-1}\textrm{sr}^{-1},
\label{eq:BL_Lac}
\ee  
where $E_{\gamma}$ is in GeV and the normalization is chosen in such a way that the spectrum of BL Lacs is 1$\sigma$ below the estimate of \cite{Collaboration:2010gqa}.
The contribution of BL Lacs to the EGBR flux is $5.4 \times 10^{-7}$ ph $\textrm{cm}^{-2}\textrm{s}^{-1}\textrm{sr}^{-1}$ in energy range of 10 MeV $< E_{\gamma} <$10 GeV. 
FSRQs are energetic flat spectrum quasi-stellar radio sources.
Their spectrum is slightly softer than that of the BL Lacs as it follows:
\be
\frac{dN_{\gamma}}{dE}_{FSRQ} = 3.1 \times 10^{-8} E_{\gamma}^{-2.45} \textrm{GeV}^{-1}\textrm{cm}^{-2}\textrm{s}^{-1}\textrm{sr}^{-1}.
\label{eq:FSRQ}
\ee
Taking 1$\sigma$ low values, the contribution to the EGBR is $6.1 \times 10^{-7}$ ph $\textrm{cm}^{-2}\textrm{s}^{-1}\textrm{sr}^{-1}$ which, in terms of photon number, is rather similar to that of BL Lacs. 
FSQRs are the dominant population below 0.5 GeV.

Although the millisecond pulsars are galactic faint $\gamma$-ray sources, they are expected to rather significantly contribute to $\gamma$-rays at high latitudes. 
Their minimum contribution at latitudes $|b|> 40^{\circ}$ has been estimated to be 
$8.0 \times 10^{-7}$ ph $\textrm{cm}^{-2}\textrm{s}^{-1}\textrm{sr}^{-1}$ for 
100 MeV $< E_{\gamma} <$ 10 GeV \cite{FaucherGiguere:2009df}. In addition their contribution to the EGBR can be constrained by their contribution to the diffuse $\gamma$-ray  power spectrum at high latitudes \cite{SiegalGaskins:2010mp}.
Using the spectral properties of millisecond pulsars measured by \cite{Abdo:2009ax}, we model their differential spectrum at $|b|>40^{\circ}$ as 
\be
\frac{dN_{\gamma}}{dE}_{MSP} = 1.8 \times 10^{-7} E_{\gamma}^{-1.5} 
exp\left(-\frac{E_{\gamma}}{1.9}\right)
\textrm{GeV}^{-1}\textrm{cm}^{-2}\textrm{s}^{-1}\textrm{sr}^{-1}.
\label{eq:MSP}
\ee

Unresolved star-forming galaxies constitute at least about 5\% of the EGBR \cite{Makiya:2010zt}. 
The evaluation of their contribution to the EGBR has great uncertainties; so SFGs may even account for $50$\% of the flux \cite{Fields:2010bw}.
Assuming that their $\gamma$-ray spectrum is mainly of hadronic origin, i.e. from the decay of $\pi^{0}$s which are produced by inelastic pp collisions in galaxies,
their spectrum is a power-law above 1 GeV and peaks at $\simeq 0.5$ GeV.
For the minimal contribution of SFGs we take
\be
\frac{dN_{\gamma}}{dE}_{SFG} = 1.3 \times 10^{-7} E_{\gamma}^{-2.75}
\textrm{GeV}^{-1}\textrm{cm}^{-2}\textrm{s}^{-1}\textrm{sr}^{-1} \; \textrm{for} 
\; E_{\gamma} > 1 \textrm{GeV},
\label{eq:SFG}
\ee
which results in $5.5 \times 10^{-7}$ ph $\textrm{cm}^{-2}\textrm{s}^{-1}\textrm{sr}^{-1}$ for $100$ MeV $<E_{\gamma} < 100$ GeV.

Fanarof-Riley radio galaxies of type I (FRI) and type II (FRII) are misaligned AGN populations of, respectively, BL Lacs and FSQRs.
FRI galaxies have decelerating subsonic jets and edge-darkened lobes, while FRII galaxies have relativistic supersonic jets and edge-brightened radio lobes.
Their $\gamma$-ray spectral energy distributions can be explained by synchrotron-self-Compton emission models.
However, low statistics of both classes in $\gamma$-rays makes the modeling of the spectral energy distribution very uncertain. 
Following the assumptions of \cite{Inoue:2011bm}, the differential spectrum of Fanarof-Riley galaxies at $E_{\gamma} > 5$ MeV is given by a power law with the index of 2.39. 
Including the attenuation of $\gamma$-rays, it is expressed by
\be
\frac{dN_{\gamma}}{dE}_{FR} = 5.7 \times 10^{-6} E_{\gamma}^{-2.39}
exp\left[ - \left(\frac{E_{\gamma}}{50}\right)\right]
\textrm{GeV}^{-1}\textrm{cm}^{-2}\textrm{s}^{-1}\textrm{sr}^{-1}.
\label{eq:FRI_FRII}
\ee
These sources give $1.0 \times 10^{-6}$ ph $\textrm{cm}^{-2}\textrm{s}^{-1}\textrm{sr}^{-1}$ for 100 MeV $< E_{\gamma} < $ 100 GeV even by conservative estimates. 
Therefore, they constitute at least 10 \% of the EGBR.

Cascades produced by ultra high energy cosmic ray interactions with the CMB give a hard $\gamma$-ray spectrum. 
Their contribution to the EGBR has been estimated in \cite{Berezinsky:2010xa} (see also \cite{Calore:2011bt}) assuming that the UHECRs are proton dominated and have a power law spectrum with the index of $\simeq 2$ up to $E_{p} = 10^{21}$ eV. 
The spectrum has been evaluated by Monte Carlo simulations on the development of the cascades 
$p + \gamma \longrightarrow \pi^{0}$ and $p + \gamma \longrightarrow \pi^{\pm}$ up to redshifts of $z=2$.
Adopting the minimum estimate of \cite{Berezinsky:2010xa}, it can be approximated as
\be
\frac{dN_{\gamma}}{dE}_{UHECR} = 4.8 \times 10^{-9} E_{\gamma}^{-1.8} 
exp\left[ - \left(\frac{E_{\gamma}}{100}\right)^{0.35}\right]
\textrm{GeV}^{-1}\textrm{cm}^{-2}\textrm{s}^{-1}\textrm{sr}^{-1} \; \textrm{for}
\; E_{\gamma} > 1 \textrm{GeV}.
\label{eq:UHECR_CMB}
\ee
Their contribution to the total number of extragalactic $\gamma$-ray photons is minimal equal to $3.3 \times 10^{-8}$ ph $\textrm{cm}^{-2}\textrm{s}^{-1}\textrm{sr}^{-1}$ for  
100 MeV $<E_{\gamma}<$ 100 GeV.
It can only become important at the highest energies of the observed spectrum.

Long duration GRBs amount for about 1\% of the EGBR based on the estimates from the \textit{EGRET} extragalactic background \cite{Le:2008au}. 
Their differential spectrum is described as 
\be
\frac{dN_{\gamma}}{dE}_{GRB} = 8.9 \times 10^{-9} E_{\gamma}^{-2.1}
\textrm{GeV}^{-1}\textrm{cm}^{-2}\textrm{s}^{-1}\textrm{sr}^{-1}
\label{eq:GRB}
\ee
where the attenuation has to be added for $\gamma$-rays above $\sim$ 50 GeV.
GRBs give the total number of $1.0 \times 10^{-7}$ ph $\textrm{cm}^{-2}\textrm{s}^{-1}\textrm{sr}^{-1}$ for $E_{\gamma}>$ 100 MeV.

In starburst galaxies the star formation rate and the interstellar medium density are higher than those in the Milky Way. 
The contribution of SBGs to the EGBR is mostly via inverse Compton scattering of electrons and positrons; thus it is more important at high energies. 
Following \cite{Thompson:2006qd}, the differential spectrum of SBGs is expressed by
\be
\frac{dN_{\gamma}}{dE}_{SBG} = 0.3 \times 10^{-7} E_{\gamma}^{-2.4}
\textrm{GeV}^{-1}\textrm{cm}^{-2}\textrm{s}^{-1}\textrm{sr}^{-1},
\label{eq:SBG}
\ee
which gives $5.4 \times 10^{-7}$ ph $\textrm{cm}^{-2}\textrm{s}^{-1}\textrm{sr}^{-1}$ for $E_{\gamma} >$ 100 MeV and constitutes $\simeq 5$\% of the EGBR. 

Ultra high energy cosmic ray protons produce $\pi^{0}$ and secondary $e^{\pm}$ cascades through interaction with the inter-cluster medium.
The subsequent decay of $\pi^{0}$s and the inverse Compton scattering of $e^{\pm}$s lead to the production of $\gamma$-rays. 
Since the estimate of the inverse Compton component depends on the exact assumptions on the inter-cluster radiation field, we ignore it in our conservative approach.
The contribution of $\pi^{0}$ component above about 1 GeV can be approximated by 
\be
\frac{dN_{\gamma}}{dE}_{UHEp \; ICM} = 3.1 \times 10^{-9} E_{\gamma}^{-2.75}
\textrm{GeV}^{-1}\textrm{cm}^{-2}\textrm{s}^{-1}\textrm{sr}^{-1}.
\label{eq:UHEp_ICM}
\ee
The chosen normalization results in $1.0 \times 10^{-7}$ ph $\textrm{cm}^{-2}\textrm{s}^{-1}\textrm{sr}^{-1}$ for energies above 100 MeV \cite{Pfrommer:2007sz, Blasi:2007pm}.

Finally, gravitationally induced shock waves in the intergalactic medium can accelerate electrons and protons. 
The inverse Compton scattering of those electrons, which have a very hard spectrum $\sim E^{-2}$, gives rise to $\gamma$-rays.
However, the contribution of the accelerated protons to $\gamma$-rays is negligible because of their long timescale for energy loss.
While these $\gamma$-rays are not extragalactic, they mimic the EGBR because of their contribution at high latitudes.
Following \cite{Keshet:2002sw} (see also \cite{Gabici:2002fg}) we take
\begin{equation}
\frac{dN_{\gamma}}{dE}_{IGS} = 0.87 \times 10^{-10} \times\left\{\begin{array}{ll} 
& \left(\frac{E_{\gamma}}{10}\right)^{-2.04} \; \textrm{for} 
\; E_{\gamma} < 10 \textrm{GeV} \\
& \left(\frac{E_{\gamma}}{10}\right)^{-2.13} \; \textrm{for} \; E_{\gamma} > 10 \textrm{GeV}
\end{array}\right\} \textrm{GeV}^{-1}\textrm{cm}^{-2}\textrm{s}^{-1}\textrm{sr}^{-1}.
\label{eq:IGS}
\end{equation}

\begin{figure}
\centering
\includegraphics[width=4.80in,angle=0]{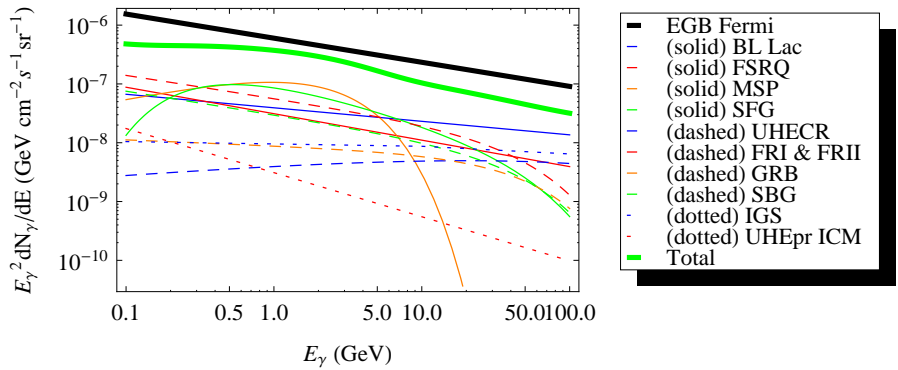}
\caption{Extragalactic $\gamma$-ray background measured by \textit{Fermi} (EGB Fermi) (\textit{solid black}) and the minimal contribution of various known sources. 
The contributions are from 
\textit{solid blue}: BL Lacs, 
\textit{solid red}: FSRQs,  
\textit{solid orange}: millisecond pulsars (MSP), 
\textit{solid green }: star-forming galaxies (SFG), 
\textit{dashed blue}: interactions of ultra high energy cosmic rays with the CMB and the galactic radiation field (UHECR), 
\textit{dashed red}: Fanarov Riley I and II galaxies (FRI \& FRII),
\textit{dashed orange} gamma ray bursts (GRB),
\textit{dashed green} starburst galaxies,
\textit{dotted blue}: gravitationally induced shock waves in the intergalactic medium (IGS) and from 
\textit{dotted red}:  interactions of ultra high energy cosmic ray protons in the inter-cluster medium (UHEpr ICM). 
The \textit{thick solid green} is the total minimal contribution to the EGBR (Total).}
\label{fig:EGBR}
\end{figure}

We include the attenuation of $\gamma$-rays for those sources which are expected to be important at high redshifts.
The spectra of the mentioned sources are shown in Fig.~\ref{fig:EGBR}.
The combination of these spectra results in a total flux of $4.3 \times 10^{-6}$ ph
$\textrm{cm}^{-2}\textrm{s}^{-1}\textrm{sr}^{-1}$ between 100 MeV and 100 GeV or $\simeq 40$\% of the total EGBR measured by \cite{Abdo:2010nz}.

\end{appendix}

\bibliography{DMGammaRays}

\begin{thebibliography}{190}
\expandafter\ifx\csname natexlab\endcsname\relax\def\natexlab#1{#1}\fi
\expandafter\ifx\csname bibnamefont\endcsname\relax
  \def\bibnamefont#1{#1}\fi
\expandafter\ifx\csname bibfnamefont\endcsname\relax
  \def\bibfnamefont#1{#1}\fi
\expandafter\ifx\csname citenamefont\endcsname\relax
  \def\citenamefont#1{#1}\fi
\expandafter\ifx\csname url\endcsname\relax
  \def\url#1{\texttt{#1}}\fi
\expandafter\ifx\csname urlprefix\endcsname\relax\def\urlprefix{URL }\fi
\providecommand{\bibinfo}[2]{#2}
\providecommand{\eprint}[2][]{\url{#2}}

\bibitem[{\citenamefont{Hinshaw et~al.}(2012)}]{Hinshaw:2012aka}
\bibinfo{author}{\bibfnamefont{G.}~\bibnamefont{Hinshaw}} \bibnamefont{et~al.}
  (\bibinfo{collaboration}{WMAP Collaboration}) (\bibinfo{year}{2012}),
  \eprint{1212.5226}.

\bibitem[{\citenamefont{Ade et~al.}(2013)}]{Ade:2013zuv}
\bibinfo{author}{\bibfnamefont{P.}~\bibnamefont{Ade}} \bibnamefont{et~al.}
  (\bibinfo{collaboration}{Planck Collaboration}) (\bibinfo{year}{2013}),
  \eprint{1303.5076}.

\bibitem[{\citenamefont{Aprile et~al.}(2013)}]{Aprile:2013doa}
\bibinfo{author}{\bibfnamefont{E.}~\bibnamefont{Aprile}} \bibnamefont{et~al.}
  (\bibinfo{collaboration}{XENON100 Collaboration}) (\bibinfo{year}{2013}),
  \eprint{1301.6620}.

\bibitem[{\citenamefont{Ahmed et~al.}(2010)}]{Ahmed:2009zw}
\bibinfo{author}{\bibfnamefont{Z.}~\bibnamefont{Ahmed}} \bibnamefont{et~al.}
  (\bibinfo{collaboration}{CDMS-II Collaboration}), \bibinfo{journal}{Science}
  \textbf{\bibinfo{volume}{327}}, \bibinfo{pages}{1619} (\bibinfo{year}{2010}),
  \eprint{0912.3592}.

\bibitem[{\citenamefont{Agnese et~al.}(2013)}]{Agnese:2013rvf}
\bibinfo{author}{\bibfnamefont{R.}~\bibnamefont{Agnese}} \bibnamefont{et~al.}
  (\bibinfo{collaboration}{CDMS Collaboration}),
  \bibinfo{journal}{Phys.Rev.Lett.}  (\bibinfo{year}{2013}),
  \eprint{1304.4279}.

\bibitem[{\citenamefont{Aalseth et~al.}(2013)}]{Aalseth:2012if}
\bibinfo{author}{\bibfnamefont{C.}~\bibnamefont{Aalseth}} \bibnamefont{et~al.}
  (\bibinfo{collaboration}{CoGeNT Collaboration}), \bibinfo{journal}{Physical
  Review D 88,} \textbf{\bibinfo{volume}{012002}} (\bibinfo{year}{2013}),
  \eprint{1208.5737}.

\bibitem[{\citenamefont{Bernabei et~al.}(2010)}]{Bernabei:2010mq}
\bibinfo{author}{\bibfnamefont{R.}~\bibnamefont{Bernabei}} \bibnamefont{et~al.}
  (\bibinfo{collaboration}{DAMA Collaboration, LIBRA Collaboration}),
  \bibinfo{journal}{Eur.Phys.J.} \textbf{\bibinfo{volume}{C67}},
  \bibinfo{pages}{39} (\bibinfo{year}{2010}), \eprint{1002.1028}.

\bibitem[{\citenamefont{Angloher et~al.}(2012)\citenamefont{Angloher, Bauer,
  Bavykina, Bento, Bucci et~al.}}]{Angloher:2011uu}
\bibinfo{author}{\bibfnamefont{G.}~\bibnamefont{Angloher}},
  \bibinfo{author}{\bibfnamefont{M.}~\bibnamefont{Bauer}},
  \bibinfo{author}{\bibfnamefont{I.}~\bibnamefont{Bavykina}},
  \bibinfo{author}{\bibfnamefont{A.}~\bibnamefont{Bento}},
  \bibinfo{author}{\bibfnamefont{C.}~\bibnamefont{Bucci}},
  \bibnamefont{et~al.}, \bibinfo{journal}{Eur.Phys.J.}
  \textbf{\bibinfo{volume}{C72}}, \bibinfo{pages}{1971} (\bibinfo{year}{2012}),
  \eprint{1109.0702}.

\bibitem[{\citenamefont{Akerib et~al.}(2013)}]{Akerib:2012ys}
\bibinfo{author}{\bibfnamefont{D.}~\bibnamefont{Akerib}} \bibnamefont{et~al.}
  (\bibinfo{collaboration}{LUX Collaboration}),
  \bibinfo{journal}{Nucl.Instrum.Meth.} \textbf{\bibinfo{volume}{A704}},
  \bibinfo{pages}{111} (\bibinfo{year}{2013}), \eprint{1211.3788}.

\bibitem[{\citenamefont{Behnke et~al.}(2011)\citenamefont{Behnke, Behnke,
  Brice, Broemmelsiek, Collar et~al.}}]{Behnke:2010xt}
\bibinfo{author}{\bibfnamefont{E.}~\bibnamefont{Behnke}},
  \bibinfo{author}{\bibfnamefont{J.}~\bibnamefont{Behnke}},
  \bibinfo{author}{\bibfnamefont{S.}~\bibnamefont{Brice}},
  \bibinfo{author}{\bibfnamefont{D.}~\bibnamefont{Broemmelsiek}},
  \bibinfo{author}{\bibfnamefont{J.}~\bibnamefont{Collar}},
  \bibnamefont{et~al.}, \bibinfo{journal}{Phys.Rev.Lett.}
  \textbf{\bibinfo{volume}{106}}, \bibinfo{pages}{021303}
  (\bibinfo{year}{2011}), \eprint{1008.3518}.

\bibitem[{\citenamefont{Armengaud et~al.}(2012)}]{Armengaud:2012pfa}
\bibinfo{author}{\bibfnamefont{E.}~\bibnamefont{Armengaud}}
  \bibnamefont{et~al.} (\bibinfo{collaboration}{EDELWEISS Collaboration}),
  \bibinfo{journal}{Phys.Rev.} \textbf{\bibinfo{volume}{D86}},
  \bibinfo{pages}{051701} (\bibinfo{year}{2012}), \eprint{1207.1815}.

\bibitem[{\citenamefont{Aaltonen et~al.}(2012)}]{Aaltonen:2012jb}
\bibinfo{author}{\bibfnamefont{T.}~\bibnamefont{Aaltonen}} \bibnamefont{et~al.}
  (\bibinfo{collaboration}{CDF Collaboration}),
  \bibinfo{journal}{Phys.Rev.Lett.} \textbf{\bibinfo{volume}{108}},
  \bibinfo{pages}{211804} (\bibinfo{year}{2012}), \eprint{1203.0742}.

\bibitem[{\citenamefont{Fox et~al.}(2012)\citenamefont{Fox, Harnik, Primulando,
  and Yu}}]{Fox:2012ee}
\bibinfo{author}{\bibfnamefont{P.~J.} \bibnamefont{Fox}},
  \bibinfo{author}{\bibfnamefont{R.}~\bibnamefont{Harnik}},
  \bibinfo{author}{\bibfnamefont{R.}~\bibnamefont{Primulando}},
  \bibnamefont{and} \bibinfo{author}{\bibfnamefont{C.-T.} \bibnamefont{Yu}},
  \bibinfo{journal}{Phys.Rev.} \textbf{\bibinfo{volume}{D86}},
  \bibinfo{pages}{015010} (\bibinfo{year}{2012}), \eprint{1203.1662}.

\bibitem[{\citenamefont{Chatrchyan et~al.}(2012)}]{Chatrchyan:2012tea}
\bibinfo{author}{\bibfnamefont{S.}~\bibnamefont{Chatrchyan}}
  \bibnamefont{et~al.} (\bibinfo{collaboration}{CMS Collaboration}),
  \bibinfo{journal}{Phys.Rev.Lett.} \textbf{\bibinfo{volume}{108}},
  \bibinfo{pages}{261803} (\bibinfo{year}{2012}), \eprint{1204.0821}.

\bibitem[{\citenamefont{Bell et~al.}(2012)\citenamefont{Bell, Dent, Galea,
  Jacques, Krauss et~al.}}]{Bell:2012rg}
\bibinfo{author}{\bibfnamefont{N.~F.} \bibnamefont{Bell}},
  \bibinfo{author}{\bibfnamefont{J.~B.} \bibnamefont{Dent}},
  \bibinfo{author}{\bibfnamefont{A.~J.} \bibnamefont{Galea}},
  \bibinfo{author}{\bibfnamefont{T.~D.} \bibnamefont{Jacques}},
  \bibinfo{author}{\bibfnamefont{L.~M.} \bibnamefont{Krauss}},
  \bibnamefont{et~al.}, \bibinfo{journal}{Phys.Rev.}
  \textbf{\bibinfo{volume}{D86}}, \bibinfo{pages}{096011}
  (\bibinfo{year}{2012}), \eprint{1209.0231}.

\bibitem[{\citenamefont{Aad et~al.}(2013)}]{ATLAS:2012ky}
\bibinfo{author}{\bibfnamefont{G.}~\bibnamefont{Aad}} \bibnamefont{et~al.}
  (\bibinfo{collaboration}{ATLAS Collaboration}), \bibinfo{journal}{JHEP}
  \textbf{\bibinfo{volume}{1304}}, \bibinfo{pages}{075} (\bibinfo{year}{2013}),
  \eprint{1210.4491}.

\bibitem[{\citenamefont{Dutta et~al.}(2013)\citenamefont{Dutta, Kamon, Kolev,
  Sinha, Wang et~al.}}]{Dutta:2013sta}
\bibinfo{author}{\bibfnamefont{B.}~\bibnamefont{Dutta}},
  \bibinfo{author}{\bibfnamefont{T.}~\bibnamefont{Kamon}},
  \bibinfo{author}{\bibfnamefont{N.}~\bibnamefont{Kolev}},
  \bibinfo{author}{\bibfnamefont{K.}~\bibnamefont{Sinha}},
  \bibinfo{author}{\bibfnamefont{K.}~\bibnamefont{Wang}}, \bibnamefont{et~al.}
  (\bibinfo{year}{2013}), \eprint{1302.3231}.

\bibitem[{\citenamefont{Lin et~al.}(2013)\citenamefont{Lin, Kolb, and
  Wang}}]{Lin:2013sca}
\bibinfo{author}{\bibfnamefont{T.}~\bibnamefont{Lin}},
  \bibinfo{author}{\bibfnamefont{E.~W.} \bibnamefont{Kolb}}, \bibnamefont{and}
  \bibinfo{author}{\bibfnamefont{L.-T.} \bibnamefont{Wang}}
  (\bibinfo{year}{2013}), \eprint{1303.6638}.

\bibitem[{\citenamefont{Cirelli}(2012)}]{Cirelli:2012tf}
\bibinfo{author}{\bibfnamefont{M.}~\bibnamefont{Cirelli}},
  \bibinfo{journal}{Pramana} \textbf{\bibinfo{volume}{79}},
  \bibinfo{pages}{1021} (\bibinfo{year}{2012}), \eprint{1202.1454}.

\bibitem[{\citenamefont{Serpico}(2012)}]{Serpico:2011wg}
\bibinfo{author}{\bibfnamefont{P.~D.} \bibnamefont{Serpico}},
  \bibinfo{journal}{Astropart.Phys.} \textbf{\bibinfo{volume}{39-40}},
  \bibinfo{pages}{2} (\bibinfo{year}{2012}), \eprint{1108.4827}.

\bibitem[{\citenamefont{Bringmann and Weniger}(2012)}]{Bringmann:2012ez}
\bibinfo{author}{\bibfnamefont{T.}~\bibnamefont{Bringmann}} \bibnamefont{and}
  \bibinfo{author}{\bibfnamefont{C.}~\bibnamefont{Weniger}},
  \bibinfo{journal}{Phys.Dark Univ.} \textbf{\bibinfo{volume}{1}},
  \bibinfo{pages}{194} (\bibinfo{year}{2012}), \eprint{1208.5481}.

\bibitem[{\citenamefont{Daw et~al.}(2012)\citenamefont{Daw, Fox, Gauvreau,
  Ghag, Harmon et~al.}}]{Daw:2010ud}
\bibinfo{author}{\bibfnamefont{E.}~\bibnamefont{Daw}},
  \bibinfo{author}{\bibfnamefont{J.}~\bibnamefont{Fox}},
  \bibinfo{author}{\bibfnamefont{J.}~\bibnamefont{Gauvreau}},
  \bibinfo{author}{\bibfnamefont{C.}~\bibnamefont{Ghag}},
  \bibinfo{author}{\bibfnamefont{L.}~\bibnamefont{Harmon}},
  \bibnamefont{et~al.}, \bibinfo{journal}{Astropart.Phys.}
  \textbf{\bibinfo{volume}{35}}, \bibinfo{pages}{397} (\bibinfo{year}{2012}),
  \eprint{1010.3027}.

\bibitem[{\citenamefont{Bertone~(ed.)}(2010)}]{Bertone:2010zz}
\bibinfo{author}{\bibfnamefont{G.}~\bibnamefont{Bertone~(ed.)}}
  (\bibinfo{year}{2010}).

\bibitem[{\citenamefont{\url{http://dragon.hepforge.org}}()}]{DRAGONweb}
\bibinfo{author}{\bibnamefont{\url{http://dragon.hepforge.org}}}.

\bibitem[{\citenamefont{Evoli et~al.}(2008)\citenamefont{Evoli, Gaggero,
  Grasso, and Maccione}}]{Evoli:2008dv}
\bibinfo{author}{\bibfnamefont{C.}~\bibnamefont{Evoli}},
  \bibinfo{author}{\bibfnamefont{D.}~\bibnamefont{Gaggero}},
  \bibinfo{author}{\bibfnamefont{D.}~\bibnamefont{Grasso}}, \bibnamefont{and}
  \bibinfo{author}{\bibfnamefont{L.}~\bibnamefont{Maccione}},
  \bibinfo{journal}{JCAP} \textbf{\bibinfo{volume}{0810}}, \bibinfo{pages}{018}
  (\bibinfo{year}{2008}), \eprint{0807.4730}.

\bibitem[{\citenamefont{Collaboration}(2012{\natexlab{a}})}]{FermiLAT:2012aa}
\bibinfo{author}{\bibfnamefont{T.~F.-L.} \bibnamefont{Collaboration}}
  (\bibinfo{collaboration}{The Fermi-LAT Collaboration}),
  \bibinfo{journal}{Astrophys.J.} \textbf{\bibinfo{volume}{750}},
  \bibinfo{pages}{3} (\bibinfo{year}{2012}{\natexlab{a}}), \eprint{1202.4039}.

\bibitem[{\citenamefont{Timur et~al.}(2011)\citenamefont{Timur, Armand, Pohl,
  and Salati}}]{Timur:2011vv}
\bibinfo{author}{\bibfnamefont{D.}~\bibnamefont{Timur}},
  \bibinfo{author}{\bibfnamefont{F.}~\bibnamefont{Armand}},
  \bibinfo{author}{\bibfnamefont{M.}~\bibnamefont{Pohl}}, \bibnamefont{and}
  \bibinfo{author}{\bibfnamefont{P.}~\bibnamefont{Salati}},
  \bibinfo{journal}{Astron.Astrophys.} \textbf{\bibinfo{volume}{531}},
  \bibinfo{pages}{A37} (\bibinfo{year}{2011}), \eprint{1102.0744}.

\bibitem[{\citenamefont{Cholis et~al.}(2012)\citenamefont{Cholis, Tavakoli,
  Evoli, Maccione, and Ullio}}]{Cholis:2011un}
\bibinfo{author}{\bibfnamefont{I.}~\bibnamefont{Cholis}},
  \bibinfo{author}{\bibfnamefont{M.}~\bibnamefont{Tavakoli}},
  \bibinfo{author}{\bibfnamefont{C.}~\bibnamefont{Evoli}},
  \bibinfo{author}{\bibfnamefont{L.}~\bibnamefont{Maccione}}, \bibnamefont{and}
  \bibinfo{author}{\bibfnamefont{P.}~\bibnamefont{Ullio}},
  \bibinfo{journal}{JCAP} \textbf{\bibinfo{volume}{1205}}, \bibinfo{pages}{004}
  (\bibinfo{year}{2012}), \eprint{1106.5073}.

\bibitem[{\citenamefont{Tavakoli et~al.}(2011)\citenamefont{Tavakoli, Cholis,
  Evoli, and Ullio}}]{Tavakoli:2011wz}
\bibinfo{author}{\bibfnamefont{M.}~\bibnamefont{Tavakoli}},
  \bibinfo{author}{\bibfnamefont{I.}~\bibnamefont{Cholis}},
  \bibinfo{author}{\bibfnamefont{C.}~\bibnamefont{Evoli}}, \bibnamefont{and}
  \bibinfo{author}{\bibfnamefont{P.}~\bibnamefont{Ullio}}
  (\bibinfo{year}{2011}), \eprint{1110.5922}.

\bibitem[{\citenamefont{Ackermann
  et~al.}(2012{\natexlab{a}})}]{Ackermann:2012rg}
\bibinfo{author}{\bibfnamefont{M.}~\bibnamefont{Ackermann}}
  \bibnamefont{et~al.} (\bibinfo{collaboration}{LAT collaboration}),
  \bibinfo{journal}{Astrophys.J.} \textbf{\bibinfo{volume}{761}},
  \bibinfo{pages}{91} (\bibinfo{year}{2012}{\natexlab{a}}), \eprint{1205.6474}.

\bibitem[{\citenamefont{Cirelli et~al.}(2013)\citenamefont{Cirelli, Serpico,
  and Zaharijas}}]{Cirelli:2013mqa}
\bibinfo{author}{\bibfnamefont{M.}~\bibnamefont{Cirelli}},
  \bibinfo{author}{\bibfnamefont{P.~D.} \bibnamefont{Serpico}},
  \bibnamefont{and} \bibinfo{author}{\bibfnamefont{G.}~\bibnamefont{Zaharijas}}
  (\bibinfo{year}{2013}), \eprint{1307.7152}.

\bibitem[{\citenamefont{Tavakoli}(2012)}]{Tavakoli:2012jx}
\bibinfo{author}{\bibfnamefont{M.}~\bibnamefont{Tavakoli}}
  (\bibinfo{year}{2012}), \eprint{1207.6150}.

\bibitem[{\citenamefont{Pohl et~al.}(2008)\citenamefont{Pohl, Englmaier, and
  Bissantz}}]{Pohl:2007dz}
\bibinfo{author}{\bibfnamefont{M.}~\bibnamefont{Pohl}},
  \bibinfo{author}{\bibfnamefont{P.}~\bibnamefont{Englmaier}},
  \bibnamefont{and} \bibinfo{author}{\bibfnamefont{N.}~\bibnamefont{Bissantz}},
  \bibinfo{journal}{Astrophys.J.} \textbf{\bibinfo{volume}{677}},
  \bibinfo{pages}{283} (\bibinfo{year}{2008}), \eprint{0712.4264}.

\bibitem[{\citenamefont{Adriani et~al.}(2011{\natexlab{a}})}]{Adriani:2011cu}
\bibinfo{author}{\bibfnamefont{O.}~\bibnamefont{Adriani}} \bibnamefont{et~al.}
  (\bibinfo{collaboration}{PAMELA Collaboration}), \bibinfo{journal}{Science}
  \textbf{\bibinfo{volume}{332}}, \bibinfo{pages}{69}
  (\bibinfo{year}{2011}{\natexlab{a}}), \eprint{1103.4055}.

\bibitem[{\citenamefont{Yoon et~al.}(2011)}]{Yoon:2011zz}
\bibinfo{author}{\bibfnamefont{Y.~S.} \bibnamefont{Yoon}} \bibnamefont{et~al.},
  \bibinfo{journal}{Astrophys. J.} \textbf{\bibinfo{volume}{728}},
  \bibinfo{pages}{122} (\bibinfo{year}{2011}), \eprint{1102.2575}.

\bibitem[{\citenamefont{Ackermann
  et~al.}(2010{\natexlab{a}})}]{Ackermann:2010ij}
\bibinfo{author}{\bibfnamefont{M.}~\bibnamefont{Ackermann}}
  \bibnamefont{et~al.} (\bibinfo{collaboration}{Fermi LAT}),
  \bibinfo{journal}{Phys. Rev.} \textbf{\bibinfo{volume}{D82}},
  \bibinfo{pages}{092004} (\bibinfo{year}{2010}{\natexlab{a}}),
  \eprint{1008.3999}.

\bibitem[{\citenamefont{Borla~Tridon et~al.}(2011)\citenamefont{Borla~Tridon,
  Colin, Cossio, Doro, and Scalzotto}}]{BorlaTridon:2011dk}
\bibinfo{author}{\bibfnamefont{D.}~\bibnamefont{Borla~Tridon}},
  \bibinfo{author}{\bibfnamefont{P.}~\bibnamefont{Colin}},
  \bibinfo{author}{\bibfnamefont{L.}~\bibnamefont{Cossio}},
  \bibinfo{author}{\bibfnamefont{M.}~\bibnamefont{Doro}}, \bibnamefont{and}
  \bibinfo{author}{\bibfnamefont{V.}~\bibnamefont{Scalzotto}}
  (\bibinfo{collaboration}{MAGIC Collaboration}) (\bibinfo{year}{2011}),
  \eprint{1110.4008}.

\bibitem[{\citenamefont{Aharonian et~al.}(2008)}]{Aharonian:2008aa}
\bibinfo{author}{\bibfnamefont{F.}~\bibnamefont{Aharonian}}
  \bibnamefont{et~al.} (\bibinfo{collaboration}{H.E.S.S. Collaboration}),
  \bibinfo{journal}{Phys.Rev.Lett.} \textbf{\bibinfo{volume}{101}},
  \bibinfo{pages}{261104} (\bibinfo{year}{2008}), \eprint{0811.3894}.

\bibitem[{\citenamefont{Aharonian et~al.}(2009)}]{Aharonian:2009ah}
\bibinfo{author}{\bibfnamefont{F.}~\bibnamefont{Aharonian}}
  \bibnamefont{et~al.} (\bibinfo{collaboration}{H.E.S.S. Collaboration}),
  \bibinfo{journal}{Astron.Astrophys.} \textbf{\bibinfo{volume}{508}},
  \bibinfo{pages}{561} (\bibinfo{year}{2009}), \eprint{0905.0105}.

\bibitem[{\citenamefont{Adriani et~al.}(2011{\natexlab{b}})}]{PAMELA:2011xv}
\bibinfo{author}{\bibfnamefont{O.}~\bibnamefont{Adriani}} \bibnamefont{et~al.}
  (\bibinfo{collaboration}{PAMELA Collaboration}),
  \bibinfo{journal}{Phys.Rev.Lett.} \textbf{\bibinfo{volume}{106}},
  \bibinfo{pages}{201101} (\bibinfo{year}{2011}{\natexlab{b}}),
  \eprint{1103.2880}.

\bibitem[{\citenamefont{Adriani et~al.}(2010{\natexlab{a}})}]{Adriani:2010ib}
\bibinfo{author}{\bibfnamefont{O.}~\bibnamefont{Adriani}} \bibnamefont{et~al.},
  \bibinfo{journal}{Astropart. Phys.} \textbf{\bibinfo{volume}{34}},
  \bibinfo{pages}{1} (\bibinfo{year}{2010}{\natexlab{a}}), \eprint{1001.3522}.

\bibitem[{\citenamefont{Adriani et~al.}(2009)}]{Adriani:2008zr}
\bibinfo{author}{\bibfnamefont{O.}~\bibnamefont{Adriani}} \bibnamefont{et~al.}
  (\bibinfo{collaboration}{PAMELA}), \bibinfo{journal}{Nature}
  \textbf{\bibinfo{volume}{458}}, \bibinfo{pages}{607} (\bibinfo{year}{2009}),
  \eprint{0810.4995}.

\bibitem[{\citenamefont{{Aguilar} et~al.}(2013)\citenamefont{{Aguilar},
  {Alberti}, {Alpat}, {Alvino}, {Ambrosi}, {Andeen}, {Anderhub}, {Arruda},
  {Azzarello}, {Bachlechner} et~al.}}]{2013PhRvL.110n1102A}
\bibinfo{author}{\bibfnamefont{M.}~\bibnamefont{{Aguilar}}},
  \bibinfo{author}{\bibfnamefont{G.}~\bibnamefont{{Alberti}}},
  \bibinfo{author}{\bibfnamefont{B.}~\bibnamefont{{Alpat}}},
  \bibinfo{author}{\bibfnamefont{A.}~\bibnamefont{{Alvino}}},
  \bibinfo{author}{\bibfnamefont{G.}~\bibnamefont{{Ambrosi}}},
  \bibinfo{author}{\bibfnamefont{K.}~\bibnamefont{{Andeen}}},
  \bibinfo{author}{\bibfnamefont{H.}~\bibnamefont{{Anderhub}}},
  \bibinfo{author}{\bibfnamefont{L.}~\bibnamefont{{Arruda}}},
  \bibinfo{author}{\bibfnamefont{P.}~\bibnamefont{{Azzarello}}},
  \bibinfo{author}{\bibfnamefont{A.}~\bibnamefont{{Bachlechner}}},
  \bibnamefont{et~al.}, \bibinfo{journal}{Physical Review Letters}
  \textbf{\bibinfo{volume}{110}}, \bibinfo{eid}{141102} (\bibinfo{year}{2013}).

\bibitem[{\citenamefont{Adriani et~al.}(2010{\natexlab{b}})}]{Adriani:2010rc}
\bibinfo{author}{\bibfnamefont{O.}~\bibnamefont{Adriani}} \bibnamefont{et~al.}
  (\bibinfo{collaboration}{PAMELA Collaboration}),
  \bibinfo{journal}{Phys.Rev.Lett.} \textbf{\bibinfo{volume}{105}},
  \bibinfo{pages}{121101} (\bibinfo{year}{2010}{\natexlab{b}}),
  \eprint{1007.0821}.

\bibitem[{\citenamefont{Kalberla et~al.}(2005)\citenamefont{Kalberla, Burton,
  Hartmann, Arnal, Bajaja et~al.}}]{Kalberla:2005ts}
\bibinfo{author}{\bibfnamefont{P.~M.} \bibnamefont{Kalberla}},
  \bibinfo{author}{\bibfnamefont{W.}~\bibnamefont{Burton}},
  \bibinfo{author}{\bibfnamefont{D.}~\bibnamefont{Hartmann}},
  \bibinfo{author}{\bibfnamefont{E.}~\bibnamefont{Arnal}},
  \bibinfo{author}{\bibfnamefont{E.}~\bibnamefont{Bajaja}},
  \bibnamefont{et~al.}, \bibinfo{journal}{Astron.Astrophys.}
  \textbf{\bibinfo{volume}{440}}, \bibinfo{pages}{775} (\bibinfo{year}{2005}),
  \eprint{astro-ph/0504140}.

\bibitem[{\citenamefont{Dame et~al.}(2001)\citenamefont{Dame, Hartmann, and
  Thaddeus}}]{Dame:2000sp}
\bibinfo{author}{\bibfnamefont{T.}~\bibnamefont{Dame}},
  \bibinfo{author}{\bibfnamefont{D.}~\bibnamefont{Hartmann}}, \bibnamefont{and}
  \bibinfo{author}{\bibfnamefont{P.}~\bibnamefont{Thaddeus}},
  \bibinfo{journal}{Astrophys.J.} \textbf{\bibinfo{volume}{547}},
  \bibinfo{pages}{792} (\bibinfo{year}{2001}), \eprint{astro-ph/0009217}.

\bibitem[{\citenamefont{Bissantz et~al.}(2003)\citenamefont{Bissantz,
  Englmaier, and Gerhard}}]{Bissantz:2002ge}
\bibinfo{author}{\bibfnamefont{N.}~\bibnamefont{Bissantz}},
  \bibinfo{author}{\bibfnamefont{P.}~\bibnamefont{Englmaier}},
  \bibnamefont{and} \bibinfo{author}{\bibfnamefont{O.}~\bibnamefont{Gerhard}},
  \bibinfo{journal}{Mon.Not.Roy.Astron.Soc.} \textbf{\bibinfo{volume}{340}},
  \bibinfo{pages}{949} (\bibinfo{year}{2003}), \eprint{astro-ph/0212516}.

\bibitem[{\citenamefont{Shetty et~al.}(2010)\citenamefont{Shetty, Glover,
  Dullemond, and Klessen}}]{Shetty:2010dm}
\bibinfo{author}{\bibfnamefont{R.}~\bibnamefont{Shetty}},
  \bibinfo{author}{\bibfnamefont{S.~C.} \bibnamefont{Glover}},
  \bibinfo{author}{\bibfnamefont{C.~P.} \bibnamefont{Dullemond}},
  \bibnamefont{and} \bibinfo{author}{\bibfnamefont{R.~S.}
  \bibnamefont{Klessen}} (\bibinfo{year}{2010}), \eprint{1011.2019}.

\bibitem[{\citenamefont{Glover and Mac~Low}(2010)}]{Glover:2010uz}
\bibinfo{author}{\bibfnamefont{S.}~\bibnamefont{Glover}} \bibnamefont{and}
  \bibinfo{author}{\bibfnamefont{M.-M.} \bibnamefont{Mac~Low}}
  (\bibinfo{year}{2010}), \eprint{1003.1340}.

\bibitem[{\citenamefont{Nakagawa et~al.}(2005)\citenamefont{Nakagawa, Onishi,
  Mizuno, and Fukui}}]{Nakagawa:2005tw}
\bibinfo{author}{\bibfnamefont{M.}~\bibnamefont{Nakagawa}},
  \bibinfo{author}{\bibfnamefont{T.}~\bibnamefont{Onishi}},
  \bibinfo{author}{\bibfnamefont{A.}~\bibnamefont{Mizuno}}, \bibnamefont{and}
  \bibinfo{author}{\bibfnamefont{Y.}~\bibnamefont{Fukui}},
  \bibinfo{journal}{Publ.Astron.Soc.Jap.}  (\bibinfo{year}{2005}),
  \eprint{astro-ph/0510473}.

\bibitem[{\citenamefont{Strong et~al.}(2004)\citenamefont{Strong, Moskalenko,
  Reimer, Digel, and Diehl}}]{Strong:2004td}
\bibinfo{author}{\bibfnamefont{A.}~\bibnamefont{Strong}},
  \bibinfo{author}{\bibfnamefont{I.}~\bibnamefont{Moskalenko}},
  \bibinfo{author}{\bibfnamefont{O.}~\bibnamefont{Reimer}},
  \bibinfo{author}{\bibfnamefont{S.}~\bibnamefont{Digel}}, \bibnamefont{and}
  \bibinfo{author}{\bibfnamefont{R.}~\bibnamefont{Diehl}},
  \bibinfo{journal}{Astron.Astrophys.} \textbf{\bibinfo{volume}{422}},
  \bibinfo{pages}{L47} (\bibinfo{year}{2004}), \eprint{astro-ph/0405275}.

\bibitem[{\citenamefont{Boselli et~al.}(2002)\citenamefont{Boselli, Lequeux,
  and Gavazzi}}]{Boselli:2001wj}
\bibinfo{author}{\bibfnamefont{A.}~\bibnamefont{Boselli}},
  \bibinfo{author}{\bibfnamefont{J.}~\bibnamefont{Lequeux}}, \bibnamefont{and}
  \bibinfo{author}{\bibfnamefont{G.}~\bibnamefont{Gavazzi}},
  \bibinfo{journal}{Astron.Astrophys.} \textbf{\bibinfo{volume}{384}},
  \bibinfo{pages}{33} (\bibinfo{year}{2002}), \eprint{astro-ph/0112275}.

\bibitem[{\citenamefont{Israel}(1997)}]{Israel:1997wn}
\bibinfo{author}{\bibfnamefont{F.}~\bibnamefont{Israel}},
  \bibinfo{journal}{Astron.Astrophys.} \textbf{\bibinfo{volume}{328}},
  \bibinfo{pages}{471} (\bibinfo{year}{1997}), \eprint{astro-ph/9709194}.

\bibitem[{\citenamefont{Israel}(2000)}]{Israel:2000mx}
\bibinfo{author}{\bibfnamefont{F.}~\bibnamefont{Israel}}
  (\bibinfo{year}{2000}), \eprint{astro-ph/0001250}.

\bibitem[{\citenamefont{Nakanishi and Sofue}(2006)}]{Nakanishi:2006zf}
\bibinfo{author}{\bibfnamefont{H.}~\bibnamefont{Nakanishi}} \bibnamefont{and}
  \bibinfo{author}{\bibfnamefont{Y.}~\bibnamefont{Sofue}},
  \bibinfo{journal}{Publ.Astron.Soc.Jap.}  (\bibinfo{year}{2006}),
  \eprint{astro-ph/0610769}.

\bibitem[{\citenamefont{{Arimoto} et~al.}(1996)\citenamefont{{Arimoto},
  {Sofue}, and {Tsujimoto}}}]{1996PASJ...48..275A}
\bibinfo{author}{\bibfnamefont{N.}~\bibnamefont{{Arimoto}}},
  \bibinfo{author}{\bibfnamefont{Y.}~\bibnamefont{{Sofue}}}, \bibnamefont{and}
  \bibinfo{author}{\bibfnamefont{T.}~\bibnamefont{{Tsujimoto}}},
  \bibinfo{journal}{\pasj} \textbf{\bibinfo{volume}{48}}, \bibinfo{pages}{275}
  (\bibinfo{year}{1996}).

\bibitem[{\citenamefont{{Sodroski} et~al.}(1995)\citenamefont{{Sodroski},
  {Odegard}, {Dwek}, {Hauser}, {Franz}, {Freedman}, {Kelsall}, {Wall},
  {Berriman}, {Odenwald} et~al.}}]{1995ApJ...452..262S}
\bibinfo{author}{\bibfnamefont{T.~J.} \bibnamefont{{Sodroski}}},
  \bibinfo{author}{\bibfnamefont{N.}~\bibnamefont{{Odegard}}},
  \bibinfo{author}{\bibfnamefont{E.}~\bibnamefont{{Dwek}}},
  \bibinfo{author}{\bibfnamefont{M.~G.} \bibnamefont{{Hauser}}},
  \bibinfo{author}{\bibfnamefont{B.~A.} \bibnamefont{{Franz}}},
  \bibinfo{author}{\bibfnamefont{I.}~\bibnamefont{{Freedman}}},
  \bibinfo{author}{\bibfnamefont{T.}~\bibnamefont{{Kelsall}}},
  \bibinfo{author}{\bibfnamefont{W.~F.} \bibnamefont{{Wall}}},
  \bibinfo{author}{\bibfnamefont{G.~B.} \bibnamefont{{Berriman}}},
  \bibinfo{author}{\bibfnamefont{S.~F.} \bibnamefont{{Odenwald}}},
  \bibnamefont{et~al.}, \bibinfo{journal}{\apj} \textbf{\bibinfo{volume}{452}},
  \bibinfo{pages}{262} (\bibinfo{year}{1995}).

\bibitem[{\citenamefont{{Grenier} et~al.}(2005)\citenamefont{{Grenier},
  {Casandjian}, and {Terrier}}}]{2005Sci...307.1292G}
\bibinfo{author}{\bibfnamefont{I.~A.} \bibnamefont{{Grenier}}},
  \bibinfo{author}{\bibfnamefont{J.-M.} \bibnamefont{{Casandjian}}},
  \bibnamefont{and}
  \bibinfo{author}{\bibfnamefont{R.}~\bibnamefont{{Terrier}}},
  \bibinfo{journal}{Science} \textbf{\bibinfo{volume}{307}},
  \bibinfo{pages}{1292} (\bibinfo{year}{2005}).

\bibitem[{\citenamefont{{Schlegel} et~al.}(1998)\citenamefont{{Schlegel},
  {Finkbeiner}, and {Davis}}}]{1998ApJ...500..525S}
\bibinfo{author}{\bibfnamefont{D.~J.} \bibnamefont{{Schlegel}}},
  \bibinfo{author}{\bibfnamefont{D.~P.} \bibnamefont{{Finkbeiner}}},
  \bibnamefont{and} \bibinfo{author}{\bibfnamefont{M.}~\bibnamefont{{Davis}}},
  \bibinfo{journal}{\apj} \textbf{\bibinfo{volume}{500}}, \bibinfo{pages}{525}
  (\bibinfo{year}{1998}), \eprint{arXiv:astro-ph/9710327}.

\bibitem[{\citenamefont{Dobler et~al.}(2010)\citenamefont{Dobler, Finkbeiner,
  Cholis, Slatyer, and Weiner}}]{Dobler:2009xz}
\bibinfo{author}{\bibfnamefont{G.}~\bibnamefont{Dobler}},
  \bibinfo{author}{\bibfnamefont{D.~P.} \bibnamefont{Finkbeiner}},
  \bibinfo{author}{\bibfnamefont{I.}~\bibnamefont{Cholis}},
  \bibinfo{author}{\bibfnamefont{T.~R.} \bibnamefont{Slatyer}},
  \bibnamefont{and} \bibinfo{author}{\bibfnamefont{N.}~\bibnamefont{Weiner}},
  \bibinfo{journal}{Astrophys.J.} \textbf{\bibinfo{volume}{717}},
  \bibinfo{pages}{825} (\bibinfo{year}{2010}), \eprint{0910.4583}.

\bibitem[{\citenamefont{Collaboration}(2012{\natexlab{b}})}]{Collaboration:2011bm}
\bibinfo{author}{\bibfnamefont{T.~F.-L.} \bibnamefont{Collaboration}},
  \bibinfo{journal}{Astrophys.J.Suppl.} \textbf{\bibinfo{volume}{199}},
  \bibinfo{pages}{31} (\bibinfo{year}{2012}{\natexlab{b}}), \eprint{1108.1435}.

\bibitem[{\citenamefont{Abdo et~al.}(2010{\natexlab{a}})}]{Abdo:2010nz}
\bibinfo{author}{\bibfnamefont{A.~A.} \bibnamefont{Abdo}} \bibnamefont{et~al.}
  (\bibinfo{collaboration}{The Fermi-LAT}), \bibinfo{journal}{Phys. Rev. Lett.}
  \textbf{\bibinfo{volume}{104}}, \bibinfo{pages}{101101}
  (\bibinfo{year}{2010}{\natexlab{a}}), \eprint{1002.3603}.

\bibitem[{\citenamefont{Su et~al.}(2010)\citenamefont{Su, Slatyer, and
  Finkbeiner}}]{Su:2010qj}
\bibinfo{author}{\bibfnamefont{M.}~\bibnamefont{Su}},
  \bibinfo{author}{\bibfnamefont{T.~R.} \bibnamefont{Slatyer}},
  \bibnamefont{and} \bibinfo{author}{\bibfnamefont{D.~P.}
  \bibnamefont{Finkbeiner}}, \bibinfo{journal}{Astrophys. J.}
  \textbf{\bibinfo{volume}{724}}, \bibinfo{pages}{1044} (\bibinfo{year}{2010}),
  \eprint{1005.5480}.

\bibitem[{\citenamefont{Su and Finkbeiner}(2012{\natexlab{a}})}]{Su:2012gu}
\bibinfo{author}{\bibfnamefont{M.}~\bibnamefont{Su}} \bibnamefont{and}
  \bibinfo{author}{\bibfnamefont{D.~P.} \bibnamefont{Finkbeiner}},
  \bibinfo{journal}{Astrophys.J.} \textbf{\bibinfo{volume}{753}},
  \bibinfo{pages}{61} (\bibinfo{year}{2012}{\natexlab{a}}), \eprint{1205.5852}.

\bibitem[{\citenamefont{Porter and Strong}(2005)}]{Porter:2005qx}
\bibinfo{author}{\bibfnamefont{T.~A.} \bibnamefont{Porter}} \bibnamefont{and}
  \bibinfo{author}{\bibfnamefont{A.~W.} \bibnamefont{Strong}}
  (\bibinfo{year}{2005}), \eprint{astro-ph/0507119}.

\bibitem[{\citenamefont{Graham et~al.}(2006{\natexlab{a}})\citenamefont{Graham,
  Merritt, Moore, Diemand, and Terzic}}]{Graham:2005xx}
\bibinfo{author}{\bibfnamefont{A.~W.} \bibnamefont{Graham}},
  \bibinfo{author}{\bibfnamefont{D.}~\bibnamefont{Merritt}},
  \bibinfo{author}{\bibfnamefont{B.}~\bibnamefont{Moore}},
  \bibinfo{author}{\bibfnamefont{J.}~\bibnamefont{Diemand}}, \bibnamefont{and}
  \bibinfo{author}{\bibfnamefont{B.}~\bibnamefont{Terzic}},
  \bibinfo{journal}{Astron.J.} \textbf{\bibinfo{volume}{132}},
  \bibinfo{pages}{2685} (\bibinfo{year}{2006}{\natexlab{a}}),
  \eprint{astro-ph/0509417}.

\bibitem[{\citenamefont{Catena and Ullio}(2010)}]{Catena:2009mf}
\bibinfo{author}{\bibfnamefont{R.}~\bibnamefont{Catena}} \bibnamefont{and}
  \bibinfo{author}{\bibfnamefont{P.}~\bibnamefont{Ullio}},
  \bibinfo{journal}{JCAP} \textbf{\bibinfo{volume}{1008}}, \bibinfo{pages}{004}
  (\bibinfo{year}{2010}), \eprint{0907.0018}.

\bibitem[{\citenamefont{Salucci et~al.}(2010)\citenamefont{Salucci, Nesti,
  Gentile, and Martins}}]{Salucci:2010qr}
\bibinfo{author}{\bibfnamefont{P.}~\bibnamefont{Salucci}},
  \bibinfo{author}{\bibfnamefont{F.}~\bibnamefont{Nesti}},
  \bibinfo{author}{\bibfnamefont{G.}~\bibnamefont{Gentile}}, \bibnamefont{and}
  \bibinfo{author}{\bibfnamefont{C.}~\bibnamefont{Martins}},
  \bibinfo{journal}{Astron.Astrophys.} \textbf{\bibinfo{volume}{523}},
  \bibinfo{pages}{A83} (\bibinfo{year}{2010}), \eprint{1003.3101}.

\bibitem[{\citenamefont{Cholis et~al.}(2009{\natexlab{a}})\citenamefont{Cholis,
  Goodenough, and Weiner}}]{Cholis:2008vb}
\bibinfo{author}{\bibfnamefont{I.}~\bibnamefont{Cholis}},
  \bibinfo{author}{\bibfnamefont{L.}~\bibnamefont{Goodenough}},
  \bibnamefont{and} \bibinfo{author}{\bibfnamefont{N.}~\bibnamefont{Weiner}},
  \bibinfo{journal}{Phys. Rev.} \textbf{\bibinfo{volume}{D79}},
  \bibinfo{pages}{123505} (\bibinfo{year}{2009}{\natexlab{a}}),
  \eprint{0802.2922}.

\bibitem[{\citenamefont{Cholis et~al.}(2009{\natexlab{b}})\citenamefont{Cholis,
  Finkbeiner, Goodenough, and Weiner}}]{Cholis:2008qq}
\bibinfo{author}{\bibfnamefont{I.}~\bibnamefont{Cholis}},
  \bibinfo{author}{\bibfnamefont{D.~P.} \bibnamefont{Finkbeiner}},
  \bibinfo{author}{\bibfnamefont{L.}~\bibnamefont{Goodenough}},
  \bibnamefont{and} \bibinfo{author}{\bibfnamefont{N.}~\bibnamefont{Weiner}},
  \bibinfo{journal}{JCAP} \textbf{\bibinfo{volume}{0912}}, \bibinfo{pages}{007}
  (\bibinfo{year}{2009}{\natexlab{b}}), \eprint{0810.5344}.

\bibitem[{\citenamefont{Cholis and Hooper}(2013)}]{Cholis:2013psa}
\bibinfo{author}{\bibfnamefont{I.}~\bibnamefont{Cholis}} \bibnamefont{and}
  \bibinfo{author}{\bibfnamefont{D.}~\bibnamefont{Hooper}},
  \bibinfo{journal}{Phys. Rev. D 88,} \textbf{\bibinfo{volume}{023013}}
  (\bibinfo{year}{2013}), \eprint{1304.1840}.

\bibitem[{\citenamefont{Arkani-Hamed et~al.}(2009)\citenamefont{Arkani-Hamed,
  Finkbeiner, Slatyer, and Weiner}}]{ArkaniHamed:2008qn}
\bibinfo{author}{\bibfnamefont{N.}~\bibnamefont{Arkani-Hamed}},
  \bibinfo{author}{\bibfnamefont{D.~P.} \bibnamefont{Finkbeiner}},
  \bibinfo{author}{\bibfnamefont{T.~R.} \bibnamefont{Slatyer}},
  \bibnamefont{and} \bibinfo{author}{\bibfnamefont{N.}~\bibnamefont{Weiner}},
  \bibinfo{journal}{Phys.Rev.} \textbf{\bibinfo{volume}{D79}},
  \bibinfo{pages}{015014} (\bibinfo{year}{2009}), \eprint{0810.0713}.

\bibitem[{\citenamefont{Cirelli et~al.}(2010)\citenamefont{Cirelli, Panci, and
  Serpico}}]{Cirelli:2009dv}
\bibinfo{author}{\bibfnamefont{M.}~\bibnamefont{Cirelli}},
  \bibinfo{author}{\bibfnamefont{P.}~\bibnamefont{Panci}}, \bibnamefont{and}
  \bibinfo{author}{\bibfnamefont{P.~D.} \bibnamefont{Serpico}},
  \bibinfo{journal}{Nucl.Phys.} \textbf{\bibinfo{volume}{B840}},
  \bibinfo{pages}{284} (\bibinfo{year}{2010}), \eprint{0912.0663}.

\bibitem[{\citenamefont{Papucci and Strumia}(2010)}]{Papucci:2009gd}
\bibinfo{author}{\bibfnamefont{M.}~\bibnamefont{Papucci}} \bibnamefont{and}
  \bibinfo{author}{\bibfnamefont{A.}~\bibnamefont{Strumia}},
  \bibinfo{journal}{JCAP} \textbf{\bibinfo{volume}{1003}}, \bibinfo{pages}{014}
  (\bibinfo{year}{2010}), \eprint{0912.0742}.

\bibitem[{\citenamefont{Ackermann
  et~al.}(2010{\natexlab{b}})\citenamefont{Ackermann, Ajello, Allafort,
  Baldini, Ballet et~al.}}]{Ackermann:2010rg}
\bibinfo{author}{\bibfnamefont{M.}~\bibnamefont{Ackermann}},
  \bibinfo{author}{\bibfnamefont{M.}~\bibnamefont{Ajello}},
  \bibinfo{author}{\bibfnamefont{A.}~\bibnamefont{Allafort}},
  \bibinfo{author}{\bibfnamefont{L.}~\bibnamefont{Baldini}},
  \bibinfo{author}{\bibfnamefont{J.}~\bibnamefont{Ballet}},
  \bibnamefont{et~al.}, \bibinfo{journal}{JCAP}
  \textbf{\bibinfo{volume}{1005}}, \bibinfo{pages}{025}
  (\bibinfo{year}{2010}{\natexlab{b}}), \eprint{1002.2239}.

\bibitem[{\citenamefont{Ackermann et~al.}(2011)}]{Ackermann:2011wa}
\bibinfo{author}{\bibfnamefont{M.}~\bibnamefont{Ackermann}}
  \bibnamefont{et~al.} (\bibinfo{collaboration}{Fermi-LAT collaboration}),
  \bibinfo{journal}{Phys.Rev.Lett.} \textbf{\bibinfo{volume}{107}},
  \bibinfo{pages}{241302} (\bibinfo{year}{2011}), \eprint{1108.3546}.

\bibitem[{\citenamefont{Bergstrom et~al.}(2013)\citenamefont{Bergstrom,
  Bringmann, Cholis, Hooper, and Weniger}}]{Bergstrom:2013jra}
\bibinfo{author}{\bibfnamefont{L.}~\bibnamefont{Bergstrom}},
  \bibinfo{author}{\bibfnamefont{T.}~\bibnamefont{Bringmann}},
  \bibinfo{author}{\bibfnamefont{I.}~\bibnamefont{Cholis}},
  \bibinfo{author}{\bibfnamefont{D.}~\bibnamefont{Hooper}}, \bibnamefont{and}
  \bibinfo{author}{\bibfnamefont{C.}~\bibnamefont{Weniger}}
  (\bibinfo{year}{2013}), \eprint{1306.3983}.

\bibitem[{\citenamefont{{Matsunaga} et~al.}(1998)\citenamefont{{Matsunaga},
  {Orito}, {Matsumoto}, {Yoshimura}, {Moiseev}, {Anraku}, {Golden}, {Imori},
  {Makida}, {Mitchell} et~al.}}]{Matsunaga:1998}
\bibinfo{author}{\bibfnamefont{H.}~\bibnamefont{{Matsunaga}}},
  \bibinfo{author}{\bibfnamefont{S.}~\bibnamefont{{Orito}}},
  \bibinfo{author}{\bibfnamefont{H.}~\bibnamefont{{Matsumoto}}},
  \bibinfo{author}{\bibfnamefont{K.}~\bibnamefont{{Yoshimura}}},
  \bibinfo{author}{\bibfnamefont{A.}~\bibnamefont{{Moiseev}}},
  \bibinfo{author}{\bibfnamefont{K.}~\bibnamefont{{Anraku}}},
  \bibinfo{author}{\bibfnamefont{R.}~\bibnamefont{{Golden}}},
  \bibinfo{author}{\bibfnamefont{M.}~\bibnamefont{{Imori}}},
  \bibinfo{author}{\bibfnamefont{Y.}~\bibnamefont{{Makida}}},
  \bibinfo{author}{\bibfnamefont{J.}~\bibnamefont{{Mitchell}}},
  \bibnamefont{et~al.}, \bibinfo{journal}{Physical Review Letters}
  \textbf{\bibinfo{volume}{81}}, \bibinfo{pages}{4052} (\bibinfo{year}{1998}),
  \eprint{arXiv:astro-ph/9809326}.

\bibitem[{\citenamefont{{Maeno} et~al.}(2001)\citenamefont{{Maeno}, {Orito},
  {Matsunaga}, {Abe}, {Anraku}, {Asaoka}, {Fujikawa}, {Imori}, {Makida},
  {Matsui} et~al.}}]{Maeno:2001}
\bibinfo{author}{\bibfnamefont{T.}~\bibnamefont{{Maeno}}},
  \bibinfo{author}{\bibfnamefont{S.}~\bibnamefont{{Orito}}},
  \bibinfo{author}{\bibfnamefont{H.}~\bibnamefont{{Matsunaga}}},
  \bibinfo{author}{\bibfnamefont{K.}~\bibnamefont{{Abe}}},
  \bibinfo{author}{\bibfnamefont{K.}~\bibnamefont{{Anraku}}},
  \bibinfo{author}{\bibfnamefont{Y.}~\bibnamefont{{Asaoka}}},
  \bibinfo{author}{\bibfnamefont{M.}~\bibnamefont{{Fujikawa}}},
  \bibinfo{author}{\bibfnamefont{M.}~\bibnamefont{{Imori}}},
  \bibinfo{author}{\bibfnamefont{Y.}~\bibnamefont{{Makida}}},
  \bibinfo{author}{\bibfnamefont{N.}~\bibnamefont{{Matsui}}},
  \bibnamefont{et~al.}, \bibinfo{journal}{Astroparticle Physics}
  \textbf{\bibinfo{volume}{16}}, \bibinfo{pages}{121} (\bibinfo{year}{2001}),
  \eprint{arXiv:astro-ph/0010381}.

\bibitem[{\citenamefont{{Asaoka} et~al.}(2002)\citenamefont{{Asaoka},
  {Shikaze}, {Abe}, {Anraku}, {Fujikawa}, {Fuke}, {Haino}, {Imori}, {Izumi},
  {Maeno} et~al.}}]{Asaoka:2002}
\bibinfo{author}{\bibfnamefont{Y.}~\bibnamefont{{Asaoka}}},
  \bibinfo{author}{\bibfnamefont{Y.}~\bibnamefont{{Shikaze}}},
  \bibinfo{author}{\bibfnamefont{K.}~\bibnamefont{{Abe}}},
  \bibinfo{author}{\bibfnamefont{K.}~\bibnamefont{{Anraku}}},
  \bibinfo{author}{\bibfnamefont{M.}~\bibnamefont{{Fujikawa}}},
  \bibinfo{author}{\bibfnamefont{H.}~\bibnamefont{{Fuke}}},
  \bibinfo{author}{\bibfnamefont{S.}~\bibnamefont{{Haino}}},
  \bibinfo{author}{\bibfnamefont{M.}~\bibnamefont{{Imori}}},
  \bibinfo{author}{\bibfnamefont{K.}~\bibnamefont{{Izumi}}},
  \bibinfo{author}{\bibfnamefont{T.}~\bibnamefont{{Maeno}}},
  \bibnamefont{et~al.}, \bibinfo{journal}{Physical Review Letters}
  \textbf{\bibinfo{volume}{88}}, \bibinfo{eid}{051101} (\bibinfo{year}{2002}),
  \eprint{arXiv:astro-ph/0109007}.

\bibitem[{\citenamefont{{Adriani} et~al.}(2010)\citenamefont{{Adriani},
  {Barbarino}, {Bazilevskaya}, {Bellotti}, {Boezio}, {Bogomolov}, {Bonechi},
  {Bongi}, {Bonvicini}, {Borisov} et~al.}}]{Adriani:2010}
\bibinfo{author}{\bibfnamefont{O.}~\bibnamefont{{Adriani}}},
  \bibinfo{author}{\bibfnamefont{G.~C.} \bibnamefont{{Barbarino}}},
  \bibinfo{author}{\bibfnamefont{G.~A.} \bibnamefont{{Bazilevskaya}}},
  \bibinfo{author}{\bibfnamefont{R.}~\bibnamefont{{Bellotti}}},
  \bibinfo{author}{\bibfnamefont{M.}~\bibnamefont{{Boezio}}},
  \bibinfo{author}{\bibfnamefont{E.~A.} \bibnamefont{{Bogomolov}}},
  \bibinfo{author}{\bibfnamefont{L.}~\bibnamefont{{Bonechi}}},
  \bibinfo{author}{\bibfnamefont{M.}~\bibnamefont{{Bongi}}},
  \bibinfo{author}{\bibfnamefont{V.}~\bibnamefont{{Bonvicini}}},
  \bibinfo{author}{\bibfnamefont{S.}~\bibnamefont{{Borisov}}},
  \bibnamefont{et~al.}, \bibinfo{journal}{Physical Review Letters}
  \textbf{\bibinfo{volume}{105}}, \bibinfo{eid}{121101} (\bibinfo{year}{2010}),
  \eprint{1007.0821}.

\bibitem[{\citenamefont{{Bergstr{\"o}m}
  et~al.}(1999)\citenamefont{{Bergstr{\"o}m}, {Edsj{\"o}}, and
  {Ullio}}}]{Bergstrom:1999}
\bibinfo{author}{\bibfnamefont{L.}~\bibnamefont{{Bergstr{\"o}m}}},
  \bibinfo{author}{\bibfnamefont{J.}~\bibnamefont{{Edsj{\"o}}}},
  \bibnamefont{and} \bibinfo{author}{\bibfnamefont{P.}~\bibnamefont{{Ullio}}},
  \bibinfo{journal}{\apj} \textbf{\bibinfo{volume}{526}}, \bibinfo{pages}{215}
  (\bibinfo{year}{1999}), \eprint{arXiv:astro-ph/9902012}.

\bibitem[{\citenamefont{{Donato} et~al.}(2009)\citenamefont{{Donato}, {Maurin},
  {Brun}, {Delahaye}, and {Salati}}}]{Donato:2009}
\bibinfo{author}{\bibfnamefont{F.}~\bibnamefont{{Donato}}},
  \bibinfo{author}{\bibfnamefont{D.}~\bibnamefont{{Maurin}}},
  \bibinfo{author}{\bibfnamefont{P.}~\bibnamefont{{Brun}}},
  \bibinfo{author}{\bibfnamefont{T.}~\bibnamefont{{Delahaye}}},
  \bibnamefont{and} \bibinfo{author}{\bibfnamefont{P.}~\bibnamefont{{Salati}}},
  \bibinfo{journal}{Physical Review Letters} \textbf{\bibinfo{volume}{102}},
  \bibinfo{eid}{071301} (\bibinfo{year}{2009}), \eprint{0810.5292}.

\bibitem[{\citenamefont{{Cholis}}(2011)}]{Cholis:2011}
\bibinfo{author}{\bibfnamefont{I.}~\bibnamefont{{Cholis}}},
  \bibinfo{journal}{\jcap} \textbf{\bibinfo{volume}{9}}, \bibinfo{eid}{007}
  (\bibinfo{year}{2011}), \eprint{1007.1169}.

\bibitem[{\citenamefont{{Garny} et~al.}(2011)\citenamefont{{Garny}, {Ibarra},
  and {Vogl}}}]{Garny:2011}
\bibinfo{author}{\bibfnamefont{M.}~\bibnamefont{{Garny}}},
  \bibinfo{author}{\bibfnamefont{A.}~\bibnamefont{{Ibarra}}}, \bibnamefont{and}
  \bibinfo{author}{\bibfnamefont{S.}~\bibnamefont{{Vogl}}},
  \bibinfo{journal}{\jcap} \textbf{\bibinfo{volume}{7}}, \bibinfo{eid}{028}
  (\bibinfo{year}{2011}), \eprint{1105.5367}.

\bibitem[{\citenamefont{Evoli et~al.}(2012{\natexlab{a}})\citenamefont{Evoli,
  Cholis, Grasso, Maccione, and Ullio}}]{Evoli:2011id}
\bibinfo{author}{\bibfnamefont{C.}~\bibnamefont{Evoli}},
  \bibinfo{author}{\bibfnamefont{I.}~\bibnamefont{Cholis}},
  \bibinfo{author}{\bibfnamefont{D.}~\bibnamefont{Grasso}},
  \bibinfo{author}{\bibfnamefont{L.}~\bibnamefont{Maccione}}, \bibnamefont{and}
  \bibinfo{author}{\bibfnamefont{P.}~\bibnamefont{Ullio}},
  \bibinfo{journal}{Phys.Rev.} \textbf{\bibinfo{volume}{D85}},
  \bibinfo{pages}{123511} (\bibinfo{year}{2012}{\natexlab{a}}),
  \eprint{1108.0664}.

\bibitem[{\citenamefont{Ciafaloni et~al.}(2011)\citenamefont{Ciafaloni,
  Comelli, Riotto, Sala, Strumia et~al.}}]{Ciafaloni:2010ti}
\bibinfo{author}{\bibfnamefont{P.}~\bibnamefont{Ciafaloni}},
  \bibinfo{author}{\bibfnamefont{D.}~\bibnamefont{Comelli}},
  \bibinfo{author}{\bibfnamefont{A.}~\bibnamefont{Riotto}},
  \bibinfo{author}{\bibfnamefont{F.}~\bibnamefont{Sala}},
  \bibinfo{author}{\bibfnamefont{A.}~\bibnamefont{Strumia}},
  \bibnamefont{et~al.}, \bibinfo{journal}{JCAP}
  \textbf{\bibinfo{volume}{1103}}, \bibinfo{pages}{019} (\bibinfo{year}{2011}),
  \eprint{1009.0224}.

\bibitem[{\citenamefont{Ackermann
  et~al.}(2012{\natexlab{b}})}]{FermiLAT:2011ab}
\bibinfo{author}{\bibfnamefont{M.}~\bibnamefont{Ackermann}}
  \bibnamefont{et~al.} (\bibinfo{collaboration}{Fermi LAT Collaboration}),
  \bibinfo{journal}{Phys.Rev.Lett.} \textbf{\bibinfo{volume}{108}},
  \bibinfo{pages}{011103} (\bibinfo{year}{2012}{\natexlab{b}}),
  \eprint{1109.0521}.

\bibitem[{\citenamefont{Yuan et~al.}(2013)\citenamefont{Yuan, Bi, Chen, Guo,
  Lin et~al.}}]{Yuan:2013eja}
\bibinfo{author}{\bibfnamefont{Q.}~\bibnamefont{Yuan}},
  \bibinfo{author}{\bibfnamefont{X.-J.} \bibnamefont{Bi}},
  \bibinfo{author}{\bibfnamefont{G.-M.} \bibnamefont{Chen}},
  \bibinfo{author}{\bibfnamefont{Y.-Q.} \bibnamefont{Guo}},
  \bibinfo{author}{\bibfnamefont{S.-J.} \bibnamefont{Lin}},
  \bibnamefont{et~al.} (\bibinfo{year}{2013}), \eprint{1304.1482}.

\bibitem[{\citenamefont{Linden and Profumo}(2013)}]{Linden:2013mqa}
\bibinfo{author}{\bibfnamefont{T.}~\bibnamefont{Linden}} \bibnamefont{and}
  \bibinfo{author}{\bibfnamefont{S.}~\bibnamefont{Profumo}},
  \bibinfo{journal}{Astrophys.J.} \textbf{\bibinfo{volume}{772}},
  \bibinfo{pages}{18} (\bibinfo{year}{2013}), \eprint{1304.1791}.

\bibitem[{\citenamefont{Yin et~al.}(2013)\citenamefont{Yin, Yu, Yuan, and
  Bi}}]{Yin:2013vaa}
\bibinfo{author}{\bibfnamefont{P.-F.} \bibnamefont{Yin}},
  \bibinfo{author}{\bibfnamefont{Z.-H.} \bibnamefont{Yu}},
  \bibinfo{author}{\bibfnamefont{Q.}~\bibnamefont{Yuan}}, \bibnamefont{and}
  \bibinfo{author}{\bibfnamefont{X.-J.} \bibnamefont{Bi}}
  (\bibinfo{year}{2013}), \eprint{1304.4128}.

\bibitem[{\citenamefont{Gaggero et~al.}(2013)\citenamefont{Gaggero, Maccione,
  Di~Bernardo, Evoli, and Grasso}}]{Gaggero:2013rya}
\bibinfo{author}{\bibfnamefont{D.}~\bibnamefont{Gaggero}},
  \bibinfo{author}{\bibfnamefont{L.}~\bibnamefont{Maccione}},
  \bibinfo{author}{\bibfnamefont{G.}~\bibnamefont{Di~Bernardo}},
  \bibinfo{author}{\bibfnamefont{C.}~\bibnamefont{Evoli}}, \bibnamefont{and}
  \bibinfo{author}{\bibfnamefont{D.}~\bibnamefont{Grasso}}
  (\bibinfo{year}{2013}), \eprint{1304.6718}.

\bibitem[{\citenamefont{Malyshev et~al.}(2009)\citenamefont{Malyshev, Cholis,
  and Gelfand}}]{Malyshev:2009tw}
\bibinfo{author}{\bibfnamefont{D.}~\bibnamefont{Malyshev}},
  \bibinfo{author}{\bibfnamefont{I.}~\bibnamefont{Cholis}}, \bibnamefont{and}
  \bibinfo{author}{\bibfnamefont{J.}~\bibnamefont{Gelfand}},
  \bibinfo{journal}{Phys.Rev.} \textbf{\bibinfo{volume}{D80}},
  \bibinfo{pages}{063005} (\bibinfo{year}{2009}), \eprint{0903.1310}.

\bibitem[{\citenamefont{{Navarro} et~al.}(2004)\citenamefont{{Navarro},
  {Hayashi}, {Power}, {Jenkins}, {Frenk}, {White}, {Springel}, {Stadel}, and
  {Quinn}}}]{Navarro:2004}
\bibinfo{author}{\bibfnamefont{J.~F.} \bibnamefont{{Navarro}}},
  \bibinfo{author}{\bibfnamefont{E.}~\bibnamefont{{Hayashi}}},
  \bibinfo{author}{\bibfnamefont{C.}~\bibnamefont{{Power}}},
  \bibinfo{author}{\bibfnamefont{A.~R.} \bibnamefont{{Jenkins}}},
  \bibinfo{author}{\bibfnamefont{C.~S.} \bibnamefont{{Frenk}}},
  \bibinfo{author}{\bibfnamefont{S.~D.~M.} \bibnamefont{{White}}},
  \bibinfo{author}{\bibfnamefont{V.}~\bibnamefont{{Springel}}},
  \bibinfo{author}{\bibfnamefont{J.}~\bibnamefont{{Stadel}}}, \bibnamefont{and}
  \bibinfo{author}{\bibfnamefont{T.~R.} \bibnamefont{{Quinn}}},
  \bibinfo{journal}{\mnras} \textbf{\bibinfo{volume}{349}},
  \bibinfo{pages}{1039} (\bibinfo{year}{2004}), \eprint{astro-ph/0311231}.

\bibitem[{\citenamefont{Graham et~al.}(2006{\natexlab{b}})\citenamefont{Graham,
  Merritt, Moore, Diemand, and Terzic}}]{Graham:2006ae}
\bibinfo{author}{\bibfnamefont{A.~W.} \bibnamefont{Graham}},
  \bibinfo{author}{\bibfnamefont{D.}~\bibnamefont{Merritt}},
  \bibinfo{author}{\bibfnamefont{B.}~\bibnamefont{Moore}},
  \bibinfo{author}{\bibfnamefont{J.}~\bibnamefont{Diemand}}, \bibnamefont{and}
  \bibinfo{author}{\bibfnamefont{B.}~\bibnamefont{Terzic}},
  \bibinfo{journal}{Astron.J.} \textbf{\bibinfo{volume}{132}},
  \bibinfo{pages}{2701} (\bibinfo{year}{2006}{\natexlab{b}}),
  \eprint{astro-ph/0608613}.

\bibitem[{\citenamefont{El-Zant et~al.}(2004)\citenamefont{El-Zant, Hoffman,
  Primack, Combes, and Shlosman}}]{ElZant:2003rp}
\bibinfo{author}{\bibfnamefont{A.~A.} \bibnamefont{El-Zant}},
  \bibinfo{author}{\bibfnamefont{Y.}~\bibnamefont{Hoffman}},
  \bibinfo{author}{\bibfnamefont{J.}~\bibnamefont{Primack}},
  \bibinfo{author}{\bibfnamefont{F.}~\bibnamefont{Combes}}, \bibnamefont{and}
  \bibinfo{author}{\bibfnamefont{I.}~\bibnamefont{Shlosman}},
  \bibinfo{journal}{Astrophys.J.} \textbf{\bibinfo{volume}{607}},
  \bibinfo{pages}{L75} (\bibinfo{year}{2004}), \eprint{astro-ph/0309412}.

\bibitem[{\citenamefont{Burkert}(1996)}]{Burkert:1995yz}
\bibinfo{author}{\bibfnamefont{A.}~\bibnamefont{Burkert}},
  \bibinfo{journal}{IAU Symp.} \textbf{\bibinfo{volume}{171}},
  \bibinfo{pages}{175} (\bibinfo{year}{1996}), \eprint{astro-ph/9504041}.

\bibitem[{\citenamefont{Gentile et~al.}(2007)\citenamefont{Gentile, Salucci,
  Klein, and Granato}}]{Gentile:2006hv}
\bibinfo{author}{\bibfnamefont{G.}~\bibnamefont{Gentile}},
  \bibinfo{author}{\bibfnamefont{P.}~\bibnamefont{Salucci}},
  \bibinfo{author}{\bibfnamefont{U.}~\bibnamefont{Klein}}, \bibnamefont{and}
  \bibinfo{author}{\bibfnamefont{G.~L.} \bibnamefont{Granato}},
  \bibinfo{journal}{Mon.Not.Roy.Astron.Soc.} \textbf{\bibinfo{volume}{375}},
  \bibinfo{pages}{199} (\bibinfo{year}{2007}), \eprint{astro-ph/0611355}.

\bibitem[{\citenamefont{{Gentile} et~al.}(2004)\citenamefont{{Gentile},
  {Salucci}, {Klein}, {Vergani}, and {Kalberla}}}]{Gentile:2004}
\bibinfo{author}{\bibfnamefont{G.}~\bibnamefont{{Gentile}}},
  \bibinfo{author}{\bibfnamefont{P.}~\bibnamefont{{Salucci}}},
  \bibinfo{author}{\bibfnamefont{U.}~\bibnamefont{{Klein}}},
  \bibinfo{author}{\bibfnamefont{D.}~\bibnamefont{{Vergani}}},
  \bibnamefont{and}
  \bibinfo{author}{\bibfnamefont{P.}~\bibnamefont{{Kalberla}}},
  \bibinfo{journal}{\mnras} \textbf{\bibinfo{volume}{351}},
  \bibinfo{pages}{903} (\bibinfo{year}{2004}), \eprint{arXiv:astro-ph/0403154}.

\bibitem[{\citenamefont{Finkbeiner}(2004{\natexlab{a}})}]{Finkbeiner:2003im}
\bibinfo{author}{\bibfnamefont{D.~P.} \bibnamefont{Finkbeiner}},
  \bibinfo{journal}{Astrophys.J.} \textbf{\bibinfo{volume}{614}},
  \bibinfo{pages}{186} (\bibinfo{year}{2004}{\natexlab{a}}),
  \eprint{astro-ph/0311547}.

\bibitem[{\citenamefont{Finkbeiner}(2004{\natexlab{b}})}]{Finkbeiner:2004us}
\bibinfo{author}{\bibfnamefont{D.~P.} \bibnamefont{Finkbeiner}}
  (\bibinfo{year}{2004}{\natexlab{b}}), \eprint{astro-ph/0409027}.

\bibitem[{\citenamefont{Hooper et~al.}(2007)\citenamefont{Hooper, Finkbeiner,
  and Dobler}}]{Hooper:2007kb}
\bibinfo{author}{\bibfnamefont{D.}~\bibnamefont{Hooper}},
  \bibinfo{author}{\bibfnamefont{D.~P.} \bibnamefont{Finkbeiner}},
  \bibnamefont{and} \bibinfo{author}{\bibfnamefont{G.}~\bibnamefont{Dobler}},
  \bibinfo{journal}{Phys. Rev.} \textbf{\bibinfo{volume}{D76}},
  \bibinfo{pages}{083012} (\bibinfo{year}{2007}), \eprint{arXiv:0705.3655
  [astro-ph]}.

\bibitem[{\citenamefont{Dobler and Finkbeiner}(2008)}]{Dobler:2007wv}
\bibinfo{author}{\bibfnamefont{G.}~\bibnamefont{Dobler}} \bibnamefont{and}
  \bibinfo{author}{\bibfnamefont{D.~P.} \bibnamefont{Finkbeiner}},
  \bibinfo{journal}{Astrophys.J.} \textbf{\bibinfo{volume}{680}},
  \bibinfo{pages}{1222} (\bibinfo{year}{2008}), \eprint{0712.1038}.

\bibitem[{\citenamefont{Linden et~al.}(2011)\citenamefont{Linden, Hooper, and
  Yusef-Zadeh}}]{Linden:2011au}
\bibinfo{author}{\bibfnamefont{T.}~\bibnamefont{Linden}},
  \bibinfo{author}{\bibfnamefont{D.}~\bibnamefont{Hooper}}, \bibnamefont{and}
  \bibinfo{author}{\bibfnamefont{F.}~\bibnamefont{Yusef-Zadeh}},
  \bibinfo{journal}{Astrophys.J.} \textbf{\bibinfo{volume}{741}},
  \bibinfo{pages}{95} (\bibinfo{year}{2011}), \eprint{1106.5493}.

\bibitem[{\citenamefont{Hooper and Linden}(2011)}]{Hooper:2011ti}
\bibinfo{author}{\bibfnamefont{D.}~\bibnamefont{Hooper}} \bibnamefont{and}
  \bibinfo{author}{\bibfnamefont{T.}~\bibnamefont{Linden}},
  \bibinfo{journal}{Phys.Rev.} \textbf{\bibinfo{volume}{D84}},
  \bibinfo{pages}{123005} (\bibinfo{year}{2011}), \bibinfo{note}{13 pages, 11
  figures}, \eprint{1110.0006}.

\bibitem[{\citenamefont{Abe et~al.}(2012)}]{Planck:2012fb}
\bibinfo{author}{\bibfnamefont{P.}~\bibnamefont{Abe}} \bibnamefont{et~al.}
  (\bibinfo{collaboration}{Planck Collaboration}) (\bibinfo{year}{2012}),
  \eprint{1208.5483}.

\bibitem[{\citenamefont{Linden et~al.}(2010)\citenamefont{Linden, Profumo, and
  Anderson}}]{Linden:2010eu}
\bibinfo{author}{\bibfnamefont{T.}~\bibnamefont{Linden}},
  \bibinfo{author}{\bibfnamefont{S.}~\bibnamefont{Profumo}}, \bibnamefont{and}
  \bibinfo{author}{\bibfnamefont{B.}~\bibnamefont{Anderson}},
  \bibinfo{journal}{Phys.Rev.} \textbf{\bibinfo{volume}{D82}},
  \bibinfo{pages}{063529} (\bibinfo{year}{2010}), \eprint{1004.3998}.

\bibitem[{\citenamefont{Fornengo et~al.}(2012)\citenamefont{Fornengo, Lineros,
  Regis, and Taoso}}]{Fornengo:2011iq}
\bibinfo{author}{\bibfnamefont{N.}~\bibnamefont{Fornengo}},
  \bibinfo{author}{\bibfnamefont{R.~A.} \bibnamefont{Lineros}},
  \bibinfo{author}{\bibfnamefont{M.}~\bibnamefont{Regis}}, \bibnamefont{and}
  \bibinfo{author}{\bibfnamefont{M.}~\bibnamefont{Taoso}},
  \bibinfo{journal}{JCAP} \textbf{\bibinfo{volume}{1201}}, \bibinfo{pages}{005}
  (\bibinfo{year}{2012}), \eprint{1110.4337}.

\bibitem[{\citenamefont{Mambrini et~al.}(2012)\citenamefont{Mambrini, Tytgat,
  Zaharijas, and Zaldivar}}]{Mambrini:2012ue}
\bibinfo{author}{\bibfnamefont{Y.}~\bibnamefont{Mambrini}},
  \bibinfo{author}{\bibfnamefont{M.~H.} \bibnamefont{Tytgat}},
  \bibinfo{author}{\bibfnamefont{G.}~\bibnamefont{Zaharijas}},
  \bibnamefont{and} \bibinfo{author}{\bibfnamefont{B.}~\bibnamefont{Zaldivar}},
  \bibinfo{journal}{JCAP} \textbf{\bibinfo{volume}{1211}}, \bibinfo{pages}{038}
  (\bibinfo{year}{2012}), \eprint{1206.2352}.

\bibitem[{\citenamefont{Galli et~al.}(2009)\citenamefont{Galli, Iocco, Bertone,
  and Melchiorri}}]{Galli:2009zc}
\bibinfo{author}{\bibfnamefont{S.}~\bibnamefont{Galli}},
  \bibinfo{author}{\bibfnamefont{F.}~\bibnamefont{Iocco}},
  \bibinfo{author}{\bibfnamefont{G.}~\bibnamefont{Bertone}}, \bibnamefont{and}
  \bibinfo{author}{\bibfnamefont{A.}~\bibnamefont{Melchiorri}},
  \bibinfo{journal}{Phys.Rev.} \textbf{\bibinfo{volume}{D80}},
  \bibinfo{pages}{023505} (\bibinfo{year}{2009}), \eprint{0905.0003}.

\bibitem[{\citenamefont{Slatyer et~al.}(2009)\citenamefont{Slatyer,
  Padmanabhan, and Finkbeiner}}]{Slatyer:2009yq}
\bibinfo{author}{\bibfnamefont{T.~R.} \bibnamefont{Slatyer}},
  \bibinfo{author}{\bibfnamefont{N.}~\bibnamefont{Padmanabhan}},
  \bibnamefont{and} \bibinfo{author}{\bibfnamefont{D.~P.}
  \bibnamefont{Finkbeiner}}, \bibinfo{journal}{Phys.Rev.}
  \textbf{\bibinfo{volume}{D80}}, \bibinfo{pages}{043526}
  (\bibinfo{year}{2009}), \eprint{0906.1197}.

\bibitem[{\citenamefont{Galli et~al.}(2011)\citenamefont{Galli, Iocco, Bertone,
  and Melchiorri}}]{Galli:2011rz}
\bibinfo{author}{\bibfnamefont{S.}~\bibnamefont{Galli}},
  \bibinfo{author}{\bibfnamefont{F.}~\bibnamefont{Iocco}},
  \bibinfo{author}{\bibfnamefont{G.}~\bibnamefont{Bertone}}, \bibnamefont{and}
  \bibinfo{author}{\bibfnamefont{A.}~\bibnamefont{Melchiorri}},
  \bibinfo{journal}{Phys.Rev.} \textbf{\bibinfo{volume}{D84}},
  \bibinfo{pages}{027302} (\bibinfo{year}{2011}), \eprint{1106.1528}.

\bibitem[{\citenamefont{Evoli et~al.}(2012{\natexlab{b}})\citenamefont{Evoli,
  Pandolfi, and Ferrara}}]{Evoli:2012qh}
\bibinfo{author}{\bibfnamefont{C.}~\bibnamefont{Evoli}},
  \bibinfo{author}{\bibfnamefont{S.}~\bibnamefont{Pandolfi}}, \bibnamefont{and}
  \bibinfo{author}{\bibfnamefont{A.}~\bibnamefont{Ferrara}}
  (\bibinfo{year}{2012}{\natexlab{b}}), \eprint{1210.6845}.

\bibitem[{\citenamefont{Lopez-Honorez et~al.}(2013)\citenamefont{Lopez-Honorez,
  Mena, Palomares-Ruiz, and Vincent}}]{Lopez-Honorez:2013cua}
\bibinfo{author}{\bibfnamefont{L.}~\bibnamefont{Lopez-Honorez}},
  \bibinfo{author}{\bibfnamefont{O.}~\bibnamefont{Mena}},
  \bibinfo{author}{\bibfnamefont{S.}~\bibnamefont{Palomares-Ruiz}},
  \bibnamefont{and} \bibinfo{author}{\bibfnamefont{A.~C.}
  \bibnamefont{Vincent}}, \bibinfo{journal}{JCAP}
  \textbf{\bibinfo{volume}{1307}}, \bibinfo{pages}{046} (\bibinfo{year}{2013}),
  \eprint{1303.5094}.

\bibitem[{\citenamefont{Galli et~al.}(2013)\citenamefont{Galli, Slatyer,
  Valdes, and Iocco}}]{Galli:2013dna}
\bibinfo{author}{\bibfnamefont{S.}~\bibnamefont{Galli}},
  \bibinfo{author}{\bibfnamefont{T.~R.} \bibnamefont{Slatyer}},
  \bibinfo{author}{\bibfnamefont{M.}~\bibnamefont{Valdes}}, \bibnamefont{and}
  \bibinfo{author}{\bibfnamefont{F.}~\bibnamefont{Iocco}}
  (\bibinfo{year}{2013}), \eprint{1306.0563}.

\bibitem[{\citenamefont{Evans et~al.}(2004)\citenamefont{Evans, Ferrer, and
  Sarkar}}]{Evans:2003sc}
\bibinfo{author}{\bibfnamefont{N.}~\bibnamefont{Evans}},
  \bibinfo{author}{\bibfnamefont{F.}~\bibnamefont{Ferrer}}, \bibnamefont{and}
  \bibinfo{author}{\bibfnamefont{S.}~\bibnamefont{Sarkar}},
  \bibinfo{journal}{Phys.Rev.} \textbf{\bibinfo{volume}{D69}},
  \bibinfo{pages}{123501} (\bibinfo{year}{2004}), \eprint{astro-ph/0311145}.

\bibitem[{\citenamefont{Colafrancesco et~al.}(2007)\citenamefont{Colafrancesco,
  Profumo, and Ullio}}]{Colafrancesco:2006he}
\bibinfo{author}{\bibfnamefont{S.}~\bibnamefont{Colafrancesco}},
  \bibinfo{author}{\bibfnamefont{S.}~\bibnamefont{Profumo}}, \bibnamefont{and}
  \bibinfo{author}{\bibfnamefont{P.}~\bibnamefont{Ullio}},
  \bibinfo{journal}{Phys.Rev.} \textbf{\bibinfo{volume}{D75}},
  \bibinfo{pages}{023513} (\bibinfo{year}{2007}), \eprint{astro-ph/0607073}.

\bibitem[{\citenamefont{Strigari et~al.}(2007)\citenamefont{Strigari,
  Koushiappas, Bullock, and Kaplinghat}}]{Strigari:2006rd}
\bibinfo{author}{\bibfnamefont{L.~E.} \bibnamefont{Strigari}},
  \bibinfo{author}{\bibfnamefont{S.~M.} \bibnamefont{Koushiappas}},
  \bibinfo{author}{\bibfnamefont{J.~S.} \bibnamefont{Bullock}},
  \bibnamefont{and}
  \bibinfo{author}{\bibfnamefont{M.}~\bibnamefont{Kaplinghat}},
  \bibinfo{journal}{Phys.Rev.} \textbf{\bibinfo{volume}{D75}},
  \bibinfo{pages}{083526} (\bibinfo{year}{2007}), \eprint{astro-ph/0611925}.

\bibitem[{\citenamefont{Bovy}(2009)}]{Bovy:2009zs}
\bibinfo{author}{\bibfnamefont{J.}~\bibnamefont{Bovy}},
  \bibinfo{journal}{Phys.Rev.} \textbf{\bibinfo{volume}{D79}},
  \bibinfo{pages}{083539} (\bibinfo{year}{2009}), \eprint{0903.0413}.

\bibitem[{\citenamefont{Scott et~al.}(2010)\citenamefont{Scott, Conrad, Edsjo,
  Bergstrom, Farnier et~al.}}]{Scott:2009jn}
\bibinfo{author}{\bibfnamefont{P.}~\bibnamefont{Scott}},
  \bibinfo{author}{\bibfnamefont{J.}~\bibnamefont{Conrad}},
  \bibinfo{author}{\bibfnamefont{J.}~\bibnamefont{Edsjo}},
  \bibinfo{author}{\bibfnamefont{L.}~\bibnamefont{Bergstrom}},
  \bibinfo{author}{\bibfnamefont{C.}~\bibnamefont{Farnier}},
  \bibnamefont{et~al.}, \bibinfo{journal}{JCAP}
  \textbf{\bibinfo{volume}{1001}}, \bibinfo{pages}{031} (\bibinfo{year}{2010}),
  \eprint{0909.3300}.

\bibitem[{\citenamefont{Cholis and Salucci}(2012)}]{Cholis:2012am}
\bibinfo{author}{\bibfnamefont{I.}~\bibnamefont{Cholis}} \bibnamefont{and}
  \bibinfo{author}{\bibfnamefont{P.}~\bibnamefont{Salucci}},
  \bibinfo{journal}{Phys.Rev.} \textbf{\bibinfo{volume}{D86}},
  \bibinfo{pages}{023528} (\bibinfo{year}{2012}), \eprint{1203.2954}.

\bibitem[{\citenamefont{Charbonnier et~al.}(2011)\citenamefont{Charbonnier,
  Combet, Daniel, Funk, Hinton et~al.}}]{Charbonnier:2011ft}
\bibinfo{author}{\bibfnamefont{A.}~\bibnamefont{Charbonnier}},
  \bibinfo{author}{\bibfnamefont{C.}~\bibnamefont{Combet}},
  \bibinfo{author}{\bibfnamefont{M.}~\bibnamefont{Daniel}},
  \bibinfo{author}{\bibfnamefont{S.}~\bibnamefont{Funk}},
  \bibinfo{author}{\bibfnamefont{J.}~\bibnamefont{Hinton}},
  \bibnamefont{et~al.}, \bibinfo{journal}{Mon.Not.Roy.Astron.Soc.}
  \textbf{\bibinfo{volume}{418}}, \bibinfo{pages}{1526} (\bibinfo{year}{2011}),
  \eprint{1104.0412}.

\bibitem[{\citenamefont{Geringer-Sameth and
  Koushiappas}(2011)}]{GeringerSameth:2011iw}
\bibinfo{author}{\bibfnamefont{A.}~\bibnamefont{Geringer-Sameth}}
  \bibnamefont{and} \bibinfo{author}{\bibfnamefont{S.~M.}
  \bibnamefont{Koushiappas}}, \bibinfo{journal}{Phys.Rev.Lett.}
  \textbf{\bibinfo{volume}{107}}, \bibinfo{pages}{241303}
  (\bibinfo{year}{2011}), \eprint{1108.2914}.

\bibitem[{\citenamefont{Aliu et~al.}(2012)}]{Aliu:2012ga}
\bibinfo{author}{\bibfnamefont{E.}~\bibnamefont{Aliu}} \bibnamefont{et~al.}
  (\bibinfo{collaboration}{VERITAS Collaboration}),
  \bibinfo{journal}{Phys.Rev.} \textbf{\bibinfo{volume}{D85}},
  \bibinfo{pages}{062001} (\bibinfo{year}{2012}), \eprint{1202.2144}.

\bibitem[{\citenamefont{Aleksic et~al.}(2011)}]{Aleksic:2011jx}
\bibinfo{author}{\bibfnamefont{J.}~\bibnamefont{Aleksic}} \bibnamefont{et~al.}
  (\bibinfo{collaboration}{MAGIC Collaboration}), \bibinfo{journal}{JCAP}
  \textbf{\bibinfo{volume}{1106}}, \bibinfo{pages}{035} (\bibinfo{year}{2011}),
  \eprint{1103.0477}.

\bibitem[{\citenamefont{Pinzke et~al.}(2011)\citenamefont{Pinzke, Pfrommer, and
  Bergstrom}}]{Pinzke:2011ek}
\bibinfo{author}{\bibfnamefont{A.}~\bibnamefont{Pinzke}},
  \bibinfo{author}{\bibfnamefont{C.}~\bibnamefont{Pfrommer}}, \bibnamefont{and}
  \bibinfo{author}{\bibfnamefont{L.}~\bibnamefont{Bergstrom}},
  \bibinfo{journal}{Phys.Rev.} \textbf{\bibinfo{volume}{D84}},
  \bibinfo{pages}{123509} (\bibinfo{year}{2011}), \eprint{1105.3240}.

\bibitem[{\citenamefont{Ando and Nagai}(2012)}]{Ando:2012vu}
\bibinfo{author}{\bibfnamefont{S.}~\bibnamefont{Ando}} \bibnamefont{and}
  \bibinfo{author}{\bibfnamefont{D.}~\bibnamefont{Nagai}},
  \bibinfo{journal}{JCAP} \textbf{\bibinfo{volume}{1207}}, \bibinfo{pages}{017}
  (\bibinfo{year}{2012}), \eprint{1201.0753}.

\bibitem[{\citenamefont{Han et~al.}(2012)\citenamefont{Han, Frenk, Eke, Gao,
  and White}}]{Han:2012au}
\bibinfo{author}{\bibfnamefont{J.}~\bibnamefont{Han}},
  \bibinfo{author}{\bibfnamefont{C.~S.} \bibnamefont{Frenk}},
  \bibinfo{author}{\bibfnamefont{V.~R.} \bibnamefont{Eke}},
  \bibinfo{author}{\bibfnamefont{L.}~\bibnamefont{Gao}}, \bibnamefont{and}
  \bibinfo{author}{\bibfnamefont{S.~D.} \bibnamefont{White}}
  (\bibinfo{year}{2012}), \eprint{1201.1003}.

\bibitem[{\citenamefont{Hektor et~al.}(2013)\citenamefont{Hektor, Raidal, and
  Tempel}}]{Hektor:2012kc}
\bibinfo{author}{\bibfnamefont{A.}~\bibnamefont{Hektor}},
  \bibinfo{author}{\bibfnamefont{M.}~\bibnamefont{Raidal}}, \bibnamefont{and}
  \bibinfo{author}{\bibfnamefont{E.}~\bibnamefont{Tempel}},
  \bibinfo{journal}{Astrophys.J.} \textbf{\bibinfo{volume}{762}},
  \bibinfo{pages}{L22} (\bibinfo{year}{2013}), \eprint{1207.4466}.

\bibitem[{\citenamefont{Aharonian}(2009)}]{Aharonian:2009bc}
\bibinfo{author}{\bibfnamefont{.~F.} \bibnamefont{Aharonian}}
  (\bibinfo{collaboration}{HESS Collaboration}) (\bibinfo{year}{2009}),
  \eprint{0907.0727}.

\bibitem[{\citenamefont{Aleksic et~al.}(2010)}]{Aleksic:2009ir}
\bibinfo{author}{\bibfnamefont{J.}~\bibnamefont{Aleksic}} \bibnamefont{et~al.}
  (\bibinfo{collaboration}{MAGIC Collaboration}),
  \bibinfo{journal}{Astrophys.J.} \textbf{\bibinfo{volume}{710}},
  \bibinfo{pages}{634} (\bibinfo{year}{2010}), \eprint{0909.3267}.

\bibitem[{\citenamefont{Dugger et~al.}(2010)\citenamefont{Dugger, Jeltema, and
  Profumo}}]{Dugger:2010ys}
\bibinfo{author}{\bibfnamefont{L.}~\bibnamefont{Dugger}},
  \bibinfo{author}{\bibfnamefont{T.~E.} \bibnamefont{Jeltema}},
  \bibnamefont{and} \bibinfo{author}{\bibfnamefont{S.}~\bibnamefont{Profumo}},
  \bibinfo{journal}{JCAP} \textbf{\bibinfo{volume}{1012}}, \bibinfo{pages}{015}
  (\bibinfo{year}{2010}), \eprint{1009.5988}.

\bibitem[{\citenamefont{Zimmer et~al.}(2011)\citenamefont{Zimmer, Conrad, and
  Pinzke}}]{Zimmer:2011vy}
\bibinfo{author}{\bibfnamefont{S.}~\bibnamefont{Zimmer}},
  \bibinfo{author}{\bibfnamefont{J.}~\bibnamefont{Conrad}}, \bibnamefont{and}
  \bibinfo{author}{\bibfnamefont{A.}~\bibnamefont{Pinzke}}
  (\bibinfo{collaboration}{Fermi-LAT Collaboration}) (\bibinfo{year}{2011}),
  \eprint{1110.6863}.

\bibitem[{\citenamefont{Huang et~al.}(2012)\citenamefont{Huang, Vertongen, and
  Weniger}}]{Huang:2011xr}
\bibinfo{author}{\bibfnamefont{X.}~\bibnamefont{Huang}},
  \bibinfo{author}{\bibfnamefont{G.}~\bibnamefont{Vertongen}},
  \bibnamefont{and} \bibinfo{author}{\bibfnamefont{C.}~\bibnamefont{Weniger}},
  \bibinfo{journal}{JCAP} \textbf{\bibinfo{volume}{1201}}, \bibinfo{pages}{042}
  (\bibinfo{year}{2012}), \eprint{1110.1529}.

\bibitem[{\citenamefont{Vertongen and Weniger}(2011)}]{Vertongen:2011mu}
\bibinfo{author}{\bibfnamefont{G.}~\bibnamefont{Vertongen}} \bibnamefont{and}
  \bibinfo{author}{\bibfnamefont{C.}~\bibnamefont{Weniger}},
  \bibinfo{journal}{JCAP} \textbf{\bibinfo{volume}{1105}}, \bibinfo{pages}{027}
  (\bibinfo{year}{2011}), \eprint{1101.2610}.

\bibitem[{\citenamefont{Bringmann et~al.}(2012)\citenamefont{Bringmann, Huang,
  Ibarra, Vogl, and Weniger}}]{Bringmann:2012vr}
\bibinfo{author}{\bibfnamefont{T.}~\bibnamefont{Bringmann}},
  \bibinfo{author}{\bibfnamefont{X.}~\bibnamefont{Huang}},
  \bibinfo{author}{\bibfnamefont{A.}~\bibnamefont{Ibarra}},
  \bibinfo{author}{\bibfnamefont{S.}~\bibnamefont{Vogl}}, \bibnamefont{and}
  \bibinfo{author}{\bibfnamefont{C.}~\bibnamefont{Weniger}},
  \bibinfo{journal}{JCAP} \textbf{\bibinfo{volume}{1207}}, \bibinfo{pages}{054}
  (\bibinfo{year}{2012}), \eprint{1203.1312}.

\bibitem[{\citenamefont{Weniger}(2012)}]{Weniger:2012tx}
\bibinfo{author}{\bibfnamefont{C.}~\bibnamefont{Weniger}},
  \bibinfo{journal}{JCAP} \textbf{\bibinfo{volume}{1208}}, \bibinfo{pages}{007}
  (\bibinfo{year}{2012}), \eprint{1204.2797}.

\bibitem[{\citenamefont{Su and Finkbeiner}(2012{\natexlab{b}})}]{Su:2012ft}
\bibinfo{author}{\bibfnamefont{M.}~\bibnamefont{Su}} \bibnamefont{and}
  \bibinfo{author}{\bibfnamefont{D.~P.} \bibnamefont{Finkbeiner}}
  (\bibinfo{year}{2012}{\natexlab{b}}), \eprint{1206.1616}.

\bibitem[{\citenamefont{Rajaraman et~al.}(2012)\citenamefont{Rajaraman, Tait,
  and Whiteson}}]{Rajaraman:2012db}
\bibinfo{author}{\bibfnamefont{A.}~\bibnamefont{Rajaraman}},
  \bibinfo{author}{\bibfnamefont{T.~M.} \bibnamefont{Tait}}, \bibnamefont{and}
  \bibinfo{author}{\bibfnamefont{D.}~\bibnamefont{Whiteson}},
  \bibinfo{journal}{JCAP} \textbf{\bibinfo{volume}{1209}}, \bibinfo{pages}{003}
  (\bibinfo{year}{2012}), \eprint{1205.4723}.

\bibitem[{\citenamefont{Tempel et~al.}(2012)\citenamefont{Tempel, Hektor, and
  Raidal}}]{Tempel:2012ey}
\bibinfo{author}{\bibfnamefont{E.}~\bibnamefont{Tempel}},
  \bibinfo{author}{\bibfnamefont{A.}~\bibnamefont{Hektor}}, \bibnamefont{and}
  \bibinfo{author}{\bibfnamefont{M.}~\bibnamefont{Raidal}},
  \bibinfo{journal}{JCAP} \textbf{\bibinfo{volume}{1209}}, \bibinfo{pages}{032}
  (\bibinfo{year}{2012}), \eprint{1205.1045}.

\bibitem[{Fermi-LAT Collaboration()}]{Fermi-LAT:2013uma}
Fermi-LAT Collaboration (\bibinfo{year}{2013}), \eprint{1305.5597}.

\bibitem[{\citenamefont{Buchmuller and Garny}(2012)}]{Buchmuller:2012rc}
\bibinfo{author}{\bibfnamefont{W.}~\bibnamefont{Buchmuller}} \bibnamefont{and}
  \bibinfo{author}{\bibfnamefont{M.}~\bibnamefont{Garny}},
  \bibinfo{journal}{JCAP} \textbf{\bibinfo{volume}{1208}}, \bibinfo{pages}{035}
  (\bibinfo{year}{2012}), \eprint{1206.7056}.

\bibitem[{\citenamefont{Cohen et~al.}(2012)\citenamefont{Cohen, Lisanti,
  Slatyer, and Wacker}}]{Cohen:2012me}
\bibinfo{author}{\bibfnamefont{T.}~\bibnamefont{Cohen}},
  \bibinfo{author}{\bibfnamefont{M.}~\bibnamefont{Lisanti}},
  \bibinfo{author}{\bibfnamefont{T.~R.} \bibnamefont{Slatyer}},
  \bibnamefont{and} \bibinfo{author}{\bibfnamefont{J.~G.}
  \bibnamefont{Wacker}}, \bibinfo{journal}{JHEP}
  \textbf{\bibinfo{volume}{1210}}, \bibinfo{pages}{134} (\bibinfo{year}{2012}),
  \eprint{1207.0800}.

\bibitem[{\citenamefont{Buckley and Hooper}(2012)}]{Buckley:2012ws}
\bibinfo{author}{\bibfnamefont{M.~R.} \bibnamefont{Buckley}} \bibnamefont{and}
  \bibinfo{author}{\bibfnamefont{D.}~\bibnamefont{Hooper}},
  \bibinfo{journal}{Phys.Rev.} \textbf{\bibinfo{volume}{D86}},
  \bibinfo{pages}{043524} (\bibinfo{year}{2012}), \eprint{1205.6811}.

\bibitem[{\citenamefont{Fan and Reece}(2012)}]{Fan:2012gr}
\bibinfo{author}{\bibfnamefont{J.}~\bibnamefont{Fan}} \bibnamefont{and}
  \bibinfo{author}{\bibfnamefont{M.}~\bibnamefont{Reece}}
  (\bibinfo{year}{2012}), \eprint{1209.1097}.

\bibitem[{\citenamefont{Asano et~al.}(2013)\citenamefont{Asano, Bringmann,
  Sigl, and Vollmann}}]{Asano:2012zv}
\bibinfo{author}{\bibfnamefont{M.}~\bibnamefont{Asano}},
  \bibinfo{author}{\bibfnamefont{T.}~\bibnamefont{Bringmann}},
  \bibinfo{author}{\bibfnamefont{G.}~\bibnamefont{Sigl}}, \bibnamefont{and}
  \bibinfo{author}{\bibfnamefont{M.}~\bibnamefont{Vollmann}},
  \bibinfo{journal}{Phys.Rev.} \textbf{\bibinfo{volume}{D87}},
  \bibinfo{pages}{103509} (\bibinfo{year}{2013}), \eprint{1211.6739}.

\bibitem[{\citenamefont{Weiner and Yavin}(2013)}]{Weiner:2012gm}
\bibinfo{author}{\bibfnamefont{N.}~\bibnamefont{Weiner}} \bibnamefont{and}
  \bibinfo{author}{\bibfnamefont{I.}~\bibnamefont{Yavin}},
  \bibinfo{journal}{Phys.Rev.} \textbf{\bibinfo{volume}{D87}},
  \bibinfo{pages}{023523} (\bibinfo{year}{2013}), \eprint{1209.1093}.

\bibitem[{\citenamefont{Abdo et~al.}(2010{\natexlab{b}})}]{Abdo:2010dk}
\bibinfo{author}{\bibfnamefont{A.}~\bibnamefont{Abdo}} \bibnamefont{et~al.}
  (\bibinfo{collaboration}{Fermi-LAT Collaboration}), \bibinfo{journal}{JCAP}
  \textbf{\bibinfo{volume}{1004}}, \bibinfo{pages}{014}
  (\bibinfo{year}{2010}{\natexlab{b}}), \eprint{1002.4415}.

\bibitem[{\citenamefont{Hutsi et~al.}(2010)\citenamefont{Hutsi, Hektor, and
  Raidal}}]{Hutsi:2010ai}
\bibinfo{author}{\bibfnamefont{G.}~\bibnamefont{Hutsi}},
  \bibinfo{author}{\bibfnamefont{A.}~\bibnamefont{Hektor}}, \bibnamefont{and}
  \bibinfo{author}{\bibfnamefont{M.}~\bibnamefont{Raidal}},
  \bibinfo{journal}{JCAP} \textbf{\bibinfo{volume}{1007}}, \bibinfo{pages}{008}
  (\bibinfo{year}{2010}), \eprint{1004.2036}.

\bibitem[{\citenamefont{Calore et~al.}(2012)\citenamefont{Calore, De~Romeri,
  and Donato}}]{Calore:2011bt}
\bibinfo{author}{\bibfnamefont{F.}~\bibnamefont{Calore}},
  \bibinfo{author}{\bibfnamefont{V.}~\bibnamefont{De~Romeri}},
  \bibnamefont{and} \bibinfo{author}{\bibfnamefont{F.}~\bibnamefont{Donato}},
  \bibinfo{journal}{Phys.Rev.} \textbf{\bibinfo{volume}{D85}},
  \bibinfo{pages}{023004} (\bibinfo{year}{2012}), \eprint{1105.4230}.

\bibitem[{\citenamefont{Singal et~al.}(2012)\citenamefont{Singal, Petrosian,
  and Ajello}}]{Singal:2011yi}
\bibinfo{author}{\bibfnamefont{J.}~\bibnamefont{Singal}},
  \bibinfo{author}{\bibfnamefont{V.}~\bibnamefont{Petrosian}},
  \bibnamefont{and} \bibinfo{author}{\bibfnamefont{M.}~\bibnamefont{Ajello}},
  \bibinfo{journal}{Astrophys.J.} \textbf{\bibinfo{volume}{753}},
  \bibinfo{pages}{45} (\bibinfo{year}{2012}), \eprint{1106.3111}.

\bibitem[{\citenamefont{Hensley et~al.}(2010)\citenamefont{Hensley,
  Siegal-Gaskins, and Pavlidou}}]{Hensley:2009gh}
\bibinfo{author}{\bibfnamefont{B.~S.} \bibnamefont{Hensley}},
  \bibinfo{author}{\bibfnamefont{J.~M.} \bibnamefont{Siegal-Gaskins}},
  \bibnamefont{and} \bibinfo{author}{\bibfnamefont{V.}~\bibnamefont{Pavlidou}},
  \bibinfo{journal}{Astrophys.J.} \textbf{\bibinfo{volume}{723}},
  \bibinfo{pages}{277} (\bibinfo{year}{2010}), \eprint{0912.1854}.

\bibitem[{\citenamefont{Cuoco et~al.}(2011)\citenamefont{Cuoco, Sellerholm,
  Conrad, and Hannestad}}]{Cuoco:2010jb}
\bibinfo{author}{\bibfnamefont{A.}~\bibnamefont{Cuoco}},
  \bibinfo{author}{\bibfnamefont{A.}~\bibnamefont{Sellerholm}},
  \bibinfo{author}{\bibfnamefont{J.}~\bibnamefont{Conrad}}, \bibnamefont{and}
  \bibinfo{author}{\bibfnamefont{S.}~\bibnamefont{Hannestad}},
  \bibinfo{journal}{Mon.Not.Roy.Astron.Soc.} \textbf{\bibinfo{volume}{414}},
  \bibinfo{pages}{2040} (\bibinfo{year}{2011}), \eprint{1005.0843}.

\bibitem[{\citenamefont{Cuoco et~al.}(2012{\natexlab{a}})}]{Cuoco:2011ng}
\bibinfo{author}{\bibfnamefont{A.}~\bibnamefont{Cuoco}} \bibnamefont{et~al.}
  (\bibinfo{collaboration}{Fermi-LAT Collaboration}),
  \bibinfo{journal}{Nucl.Instrum.Meth.} \textbf{\bibinfo{volume}{A692}},
  \bibinfo{pages}{127} (\bibinfo{year}{2012}{\natexlab{a}}),
  \eprint{1110.1047}.

\bibitem[{\citenamefont{Ackermann
  et~al.}(2012{\natexlab{c}})}]{Ackermann:2012uf}
\bibinfo{author}{\bibfnamefont{M.}~\bibnamefont{Ackermann}}
  \bibnamefont{et~al.} (\bibinfo{collaboration}{Fermi LAT Collaboration}),
  \bibinfo{journal}{Phys.Rev.} \textbf{\bibinfo{volume}{D85}},
  \bibinfo{pages}{083007} (\bibinfo{year}{2012}{\natexlab{c}}),
  \eprint{1202.2856}.

\bibitem[{\citenamefont{Fornasa et~al.}(2013)\citenamefont{Fornasa, Zavala,
  Sanchez-Conde, Siegal-Gaskins, Delahaye et~al.}}]{Fornasa:2012gu}
\bibinfo{author}{\bibfnamefont{M.}~\bibnamefont{Fornasa}},
  \bibinfo{author}{\bibfnamefont{J.}~\bibnamefont{Zavala}},
  \bibinfo{author}{\bibfnamefont{M.~A.} \bibnamefont{Sanchez-Conde}},
  \bibinfo{author}{\bibfnamefont{J.~M.} \bibnamefont{Siegal-Gaskins}},
  \bibinfo{author}{\bibfnamefont{T.}~\bibnamefont{Delahaye}},
  \bibnamefont{et~al.}, \bibinfo{journal}{MNRAS, 429,}
  \textbf{\bibinfo{volume}{1529}} (\bibinfo{year}{2013}), \eprint{1207.0502}.

\bibitem[{\citenamefont{Diemand
  et~al.}(2007{\natexlab{a}})\citenamefont{Diemand, Kuhlen, and
  Madau}}]{Diemand:2006ik}
\bibinfo{author}{\bibfnamefont{J.}~\bibnamefont{Diemand}},
  \bibinfo{author}{\bibfnamefont{M.}~\bibnamefont{Kuhlen}}, \bibnamefont{and}
  \bibinfo{author}{\bibfnamefont{P.}~\bibnamefont{Madau}},
  \bibinfo{journal}{Astrophys.J.} \textbf{\bibinfo{volume}{657}},
  \bibinfo{pages}{262} (\bibinfo{year}{2007}{\natexlab{a}}),
  \eprint{astro-ph/0611370}.

\bibitem[{\citenamefont{Diemand
  et~al.}(2007{\natexlab{b}})\citenamefont{Diemand, Kuhlen, and
  Madau}}]{Diemand:2007qr}
\bibinfo{author}{\bibfnamefont{J.}~\bibnamefont{Diemand}},
  \bibinfo{author}{\bibfnamefont{M.}~\bibnamefont{Kuhlen}}, \bibnamefont{and}
  \bibinfo{author}{\bibfnamefont{P.}~\bibnamefont{Madau}},
  \bibinfo{journal}{Astrophys.J.} \textbf{\bibinfo{volume}{667}},
  \bibinfo{pages}{859} (\bibinfo{year}{2007}{\natexlab{b}}),
  \eprint{astro-ph/0703337}.

\bibitem[{\citenamefont{Boylan-Kolchin
  et~al.}(2009)\citenamefont{Boylan-Kolchin, Springel, White, Jenkins, and
  Lemson}}]{BoylanKolchin:2009nc}
\bibinfo{author}{\bibfnamefont{M.}~\bibnamefont{Boylan-Kolchin}},
  \bibinfo{author}{\bibfnamefont{V.}~\bibnamefont{Springel}},
  \bibinfo{author}{\bibfnamefont{S.~D.} \bibnamefont{White}},
  \bibinfo{author}{\bibfnamefont{A.}~\bibnamefont{Jenkins}}, \bibnamefont{and}
  \bibinfo{author}{\bibfnamefont{G.}~\bibnamefont{Lemson}},
  \bibinfo{journal}{Mon.Not.Roy.Astron.Soc.} \textbf{\bibinfo{volume}{398}},
  \bibinfo{pages}{1150} (\bibinfo{year}{2009}), \eprint{0903.3041}.

\bibitem[{\citenamefont{Aharonian et~al.}(2006)}]{Aharonian:2005gh}
\bibinfo{author}{\bibfnamefont{F.}~\bibnamefont{Aharonian}}
  \bibnamefont{et~al.} (\bibinfo{collaboration}{H.E.S.S. Collaboration}),
  \bibinfo{journal}{Nature} \textbf{\bibinfo{volume}{440}},
  \bibinfo{pages}{1018} (\bibinfo{year}{2006}), \eprint{astro-ph/0508073}.

\bibitem[{\citenamefont{Franceschini et~al.}(2008)\citenamefont{Franceschini,
  Rodighiero, and Vaccari}}]{Franceschini:2008tp}
\bibinfo{author}{\bibfnamefont{A.}~\bibnamefont{Franceschini}},
  \bibinfo{author}{\bibfnamefont{G.}~\bibnamefont{Rodighiero}},
  \bibnamefont{and} \bibinfo{author}{\bibfnamefont{M.}~\bibnamefont{Vaccari}},
  \bibinfo{journal}{Astron.Astrophys.} \textbf{\bibinfo{volume}{487}},
  \bibinfo{pages}{837} (\bibinfo{year}{2008}), \eprint{0805.1841}.

\bibitem[{\citenamefont{Gilmore et~al.}(2011)\citenamefont{Gilmore, Somerville,
  Primack, and Dominguez}}]{Gilmore:2011ks}
\bibinfo{author}{\bibfnamefont{R.}~\bibnamefont{Gilmore}},
  \bibinfo{author}{\bibfnamefont{R.}~\bibnamefont{Somerville}},
  \bibinfo{author}{\bibfnamefont{J.}~\bibnamefont{Primack}}, \bibnamefont{and}
  \bibinfo{author}{\bibfnamefont{A.}~\bibnamefont{Dominguez}}
  (\bibinfo{year}{2011}), \eprint{1104.0671}.

\bibitem[{\citenamefont{Cavadini et~al.}(2011)\citenamefont{Cavadini,
  Salvaterra, and Haardt}}]{Cavadini:2011ig}
\bibinfo{author}{\bibfnamefont{M.}~\bibnamefont{Cavadini}},
  \bibinfo{author}{\bibfnamefont{R.}~\bibnamefont{Salvaterra}},
  \bibnamefont{and} \bibinfo{author}{\bibfnamefont{F.}~\bibnamefont{Haardt}}
  (\bibinfo{year}{2011}), \eprint{1105.4613}.

\bibitem[{\citenamefont{Gomez-Vargas et~al.}(2013)}]{Gomez-Vargas:2013cna}
\bibinfo{author}{\bibfnamefont{G.}~\bibnamefont{Gomez-Vargas}}
  \bibnamefont{et~al.} (\bibinfo{collaboration}{Collaboration Fermi-LAT})
  (\bibinfo{year}{2013}), \eprint{1303.2154}.

\bibitem[{\citenamefont{Siegal-Gaskins
  et~al.}(2011)\citenamefont{Siegal-Gaskins, Reesman, Pavlidou, Profumo, and
  Walker}}]{SiegalGaskins:2010mp}
\bibinfo{author}{\bibfnamefont{J.~M.} \bibnamefont{Siegal-Gaskins}},
  \bibinfo{author}{\bibfnamefont{R.}~\bibnamefont{Reesman}},
  \bibinfo{author}{\bibfnamefont{V.}~\bibnamefont{Pavlidou}},
  \bibinfo{author}{\bibfnamefont{S.}~\bibnamefont{Profumo}}, \bibnamefont{and}
  \bibinfo{author}{\bibfnamefont{T.~P.} \bibnamefont{Walker}},
  \bibinfo{journal}{Mon.Not.Roy.Astron.Soc.} \textbf{\bibinfo{volume}{415}},
  \bibinfo{pages}{1074S} (\bibinfo{year}{2011}), \eprint{1011.5501}.

\bibitem[{\citenamefont{Cuoco et~al.}(2012{\natexlab{b}})\citenamefont{Cuoco,
  Komatsu, and Siegal-Gaskins}}]{Cuoco:2012yf}
\bibinfo{author}{\bibfnamefont{A.}~\bibnamefont{Cuoco}},
  \bibinfo{author}{\bibfnamefont{E.}~\bibnamefont{Komatsu}}, \bibnamefont{and}
  \bibinfo{author}{\bibfnamefont{J.}~\bibnamefont{Siegal-Gaskins}},
  \bibinfo{journal}{Phys.Rev.} \textbf{\bibinfo{volume}{D86}},
  \bibinfo{pages}{063004} (\bibinfo{year}{2012}{\natexlab{b}}),
  \eprint{1202.5309}.

\bibitem[{\citenamefont{Gorski et~al.}(2005)\citenamefont{Gorski, Hivon,
  Banday, Wandelt, Hansen et~al.}}]{Gorski:2004by}
\bibinfo{author}{\bibfnamefont{K.}~\bibnamefont{Gorski}},
  \bibinfo{author}{\bibfnamefont{E.}~\bibnamefont{Hivon}},
  \bibinfo{author}{\bibfnamefont{A.}~\bibnamefont{Banday}},
  \bibinfo{author}{\bibfnamefont{B.}~\bibnamefont{Wandelt}},
  \bibinfo{author}{\bibfnamefont{F.}~\bibnamefont{Hansen}},
  \bibnamefont{et~al.}, \bibinfo{journal}{Astrophys.J.}
  \textbf{\bibinfo{volume}{622}}, \bibinfo{pages}{759} (\bibinfo{year}{2005}),
  \eprint{astro-ph/0409513}.

\bibitem[{\citenamefont{{Binney} and {Tremaine}}(2008)}]{2008gady.book.....B}
\bibinfo{author}{\bibfnamefont{J.}~\bibnamefont{{Binney}}} \bibnamefont{and}
  \bibinfo{author}{\bibfnamefont{S.}~\bibnamefont{{Tremaine}}},
  \emph{\bibinfo{title}{{Galactic Dynamics: Second Edition}}}
  (\bibinfo{publisher}{Princeton University Press}, \bibinfo{year}{2008}).

\bibitem[{\citenamefont{Dwek et~al.}(1995)\citenamefont{Dwek, Arendt, Hauser,
  Kelsall, Lisse et~al.}}]{Dwek:1995xu}
\bibinfo{author}{\bibfnamefont{E.}~\bibnamefont{Dwek}},
  \bibinfo{author}{\bibfnamefont{R.}~\bibnamefont{Arendt}},
  \bibinfo{author}{\bibfnamefont{M.}~\bibnamefont{Hauser}},
  \bibinfo{author}{\bibfnamefont{T.}~\bibnamefont{Kelsall}},
  \bibinfo{author}{\bibfnamefont{C.}~\bibnamefont{Lisse}},
  \bibnamefont{et~al.}, \bibinfo{journal}{Astrophys.J.}
  \textbf{\bibinfo{volume}{445}}, \bibinfo{pages}{716} (\bibinfo{year}{1995}),
  \bibinfo{note}{fermilab Library Only}.

\bibitem[{\citenamefont{Babusiaux and Gilmore}(2005)}]{Babusiaux:2005zi}
\bibinfo{author}{\bibfnamefont{C.}~\bibnamefont{Babusiaux}} \bibnamefont{and}
  \bibinfo{author}{\bibfnamefont{G.}~\bibnamefont{Gilmore}},
  \bibinfo{journal}{Mon. Not. Roy. Astron. Soc.}
  \textbf{\bibinfo{volume}{358}}, \bibinfo{pages}{1309} (\bibinfo{year}{2005}),
  \eprint{astro-ph/0501383}.

\bibitem[{\citenamefont{Zoccali}(2009)}]{Zoccali:2009hu}
\bibinfo{author}{\bibfnamefont{M.}~\bibnamefont{Zoccali}}
  (\bibinfo{year}{2009}), \eprint{0910.5133}.

\bibitem[{\citenamefont{Malyshev et~al.}(2010)\citenamefont{Malyshev, Cholis,
  and Gelfand}}]{Malyshev:2010xc}
\bibinfo{author}{\bibfnamefont{D.}~\bibnamefont{Malyshev}},
  \bibinfo{author}{\bibfnamefont{I.}~\bibnamefont{Cholis}}, \bibnamefont{and}
  \bibinfo{author}{\bibfnamefont{J.~D.} \bibnamefont{Gelfand}},
  \bibinfo{journal}{Astrophys. J.} \textbf{\bibinfo{volume}{722}},
  \bibinfo{pages}{1939} (\bibinfo{year}{2010}), \eprint{1002.0587}.

\bibitem[{\citenamefont{Juric et~al.}(2008)}]{Juric:2005zr}
\bibinfo{author}{\bibfnamefont{M.}~\bibnamefont{Juric}} \bibnamefont{et~al.}
  (\bibinfo{collaboration}{SDSS Collaboration}),
  \bibinfo{journal}{Astrophys.J.} \textbf{\bibinfo{volume}{673}},
  \bibinfo{pages}{864} (\bibinfo{year}{2008}), \eprint{astro-ph/0510520}.

\bibitem[{\citenamefont{Strong et~al.}(2000)\citenamefont{Strong, Moskalenko,
  and Reimer}}]{Strong:1998fr}
\bibinfo{author}{\bibfnamefont{A.~W.} \bibnamefont{Strong}},
  \bibinfo{author}{\bibfnamefont{I.~V.} \bibnamefont{Moskalenko}},
  \bibnamefont{and} \bibinfo{author}{\bibfnamefont{O.}~\bibnamefont{Reimer}},
  \bibinfo{journal}{Astrophys.J.} \textbf{\bibinfo{volume}{537}},
  \bibinfo{pages}{763} (\bibinfo{year}{2000}), \eprint{astro-ph/9811296}.

\bibitem[{\citenamefont{Moskalenko et~al.}(2006)\citenamefont{Moskalenko,
  Porter, and Strong}}]{Moskalenko:2005ng}
\bibinfo{author}{\bibfnamefont{I.~V.} \bibnamefont{Moskalenko}},
  \bibinfo{author}{\bibfnamefont{T.~A.} \bibnamefont{Porter}},
  \bibnamefont{and} \bibinfo{author}{\bibfnamefont{A.~W.}
  \bibnamefont{Strong}}, \bibinfo{journal}{Astrophys.J.}
  \textbf{\bibinfo{volume}{640}}, \bibinfo{pages}{L155} (\bibinfo{year}{2006}),
  \eprint{astro-ph/0511149}.

\bibitem[{\citenamefont{Porter et~al.}(2006)\citenamefont{Porter, Moskalenko,
  and Strong}}]{Porter:2006tb}
\bibinfo{author}{\bibfnamefont{T.~A.} \bibnamefont{Porter}},
  \bibinfo{author}{\bibfnamefont{I.~V.} \bibnamefont{Moskalenko}},
  \bibnamefont{and} \bibinfo{author}{\bibfnamefont{A.~W.}
  \bibnamefont{Strong}}, \bibinfo{journal}{Astrophys.J.}
  \textbf{\bibinfo{volume}{648}}, \bibinfo{pages}{L29} (\bibinfo{year}{2006}),
  \eprint{astro-ph/0607344}.

\bibitem[{\citenamefont{Porter et~al.}(2008)\citenamefont{Porter, Moskalenko,
  Strong, Orlando, and Bouchet}}]{Porter:2008ve}
\bibinfo{author}{\bibfnamefont{T.~A.} \bibnamefont{Porter}},
  \bibinfo{author}{\bibfnamefont{I.~V.} \bibnamefont{Moskalenko}},
  \bibinfo{author}{\bibfnamefont{A.~W.} \bibnamefont{Strong}},
  \bibinfo{author}{\bibfnamefont{E.}~\bibnamefont{Orlando}}, \bibnamefont{and}
  \bibinfo{author}{\bibfnamefont{L.}~\bibnamefont{Bouchet}},
  \bibinfo{journal}{Astrophys. J.} \textbf{\bibinfo{volume}{682}},
  \bibinfo{pages}{400} (\bibinfo{year}{2008}), \eprint{0804.1774}.

\bibitem[{\citenamefont{Collaboration}(2010)}]{Collaboration:2010gqa}
\bibinfo{author}{\bibfnamefont{T.~F.-L.} \bibnamefont{Collaboration}},
  \bibinfo{journal}{Astrophys.J.} \textbf{\bibinfo{volume}{720}},
  \bibinfo{pages}{435} (\bibinfo{year}{2010}), \eprint{1003.0895}.

\bibitem[{\citenamefont{Faucher-Giguere and
  Loeb}(2010)}]{FaucherGiguere:2009df}
\bibinfo{author}{\bibfnamefont{C.-A.} \bibnamefont{Faucher-Giguere}}
  \bibnamefont{and} \bibinfo{author}{\bibfnamefont{A.}~\bibnamefont{Loeb}},
  \bibinfo{journal}{JCAP} \textbf{\bibinfo{volume}{1001}}, \bibinfo{pages}{005}
  (\bibinfo{year}{2010}), \eprint{0904.3102}.

\bibitem[{\citenamefont{Abdo et~al.}(2010{\natexlab{c}})}]{Abdo:2009ax}
\bibinfo{author}{\bibfnamefont{A.}~\bibnamefont{Abdo}} \bibnamefont{et~al.}
  (\bibinfo{collaboration}{Fermi LAT collaboration}),
  \bibinfo{journal}{Astrophys.J.Suppl.} \textbf{\bibinfo{volume}{187}},
  \bibinfo{pages}{460} (\bibinfo{year}{2010}{\natexlab{c}}),
  \bibinfo{note}{astrophysical Journal Supplement 187, 460-494 (2010 April). 85
  pages. Erratum corrected in this version, 2010 December (radio-gamma phase
  offset for PSR J1124-5916)}, \eprint{0910.1608}.

\bibitem[{\citenamefont{Makiya et~al.}(2011)\citenamefont{Makiya, Totani, and
  Kobayashi}}]{Makiya:2010zt}
\bibinfo{author}{\bibfnamefont{R.}~\bibnamefont{Makiya}},
  \bibinfo{author}{\bibfnamefont{T.}~\bibnamefont{Totani}}, \bibnamefont{and}
  \bibinfo{author}{\bibfnamefont{M.}~\bibnamefont{Kobayashi}},
  \bibinfo{journal}{Astrophys.J.} \textbf{\bibinfo{volume}{728}},
  \bibinfo{pages}{158} (\bibinfo{year}{2011}), \eprint{1005.1390}.

\bibitem[{\citenamefont{Fields et~al.}(2010)\citenamefont{Fields, Pavlidou, and
  Prodanovic}}]{Fields:2010bw}
\bibinfo{author}{\bibfnamefont{B.~D.} \bibnamefont{Fields}},
  \bibinfo{author}{\bibfnamefont{V.}~\bibnamefont{Pavlidou}}, \bibnamefont{and}
  \bibinfo{author}{\bibfnamefont{T.}~\bibnamefont{Prodanovic}},
  \bibinfo{journal}{Astrophys.J.} \textbf{\bibinfo{volume}{722}},
  \bibinfo{pages}{L199} (\bibinfo{year}{2010}), \eprint{1003.3647}.

\bibitem[{\citenamefont{Inoue}(2011)}]{Inoue:2011bm}
\bibinfo{author}{\bibfnamefont{Y.}~\bibnamefont{Inoue}},
  \bibinfo{journal}{Astrophys.J.} \textbf{\bibinfo{volume}{733}},
  \bibinfo{pages}{66} (\bibinfo{year}{2011}), \eprint{1103.3946}.

\bibitem[{\citenamefont{Berezinsky et~al.}(2011)\citenamefont{Berezinsky,
  Gazizov, Kachelriess, and Ostapchenko}}]{Berezinsky:2010xa}
\bibinfo{author}{\bibfnamefont{V.}~\bibnamefont{Berezinsky}},
  \bibinfo{author}{\bibfnamefont{A.}~\bibnamefont{Gazizov}},
  \bibinfo{author}{\bibfnamefont{M.}~\bibnamefont{Kachelriess}},
  \bibnamefont{and}
  \bibinfo{author}{\bibfnamefont{S.}~\bibnamefont{Ostapchenko}},
  \bibinfo{journal}{Phys.Lett.} \textbf{\bibinfo{volume}{B695}},
  \bibinfo{pages}{13} (\bibinfo{year}{2011}), \eprint{1003.1496}.

\bibitem[{\citenamefont{Le and Dermer}(2008)}]{Le:2008au}
\bibinfo{author}{\bibfnamefont{T.}~\bibnamefont{Le}} \bibnamefont{and}
  \bibinfo{author}{\bibfnamefont{C.~D.} \bibnamefont{Dermer}}
  (\bibinfo{year}{2008}), \eprint{0807.0355}.

\bibitem[{\citenamefont{Thompson et~al.}(2006)\citenamefont{Thompson, Quataert,
  and Waxman}}]{Thompson:2006qd}
\bibinfo{author}{\bibfnamefont{T.~A.} \bibnamefont{Thompson}},
  \bibinfo{author}{\bibfnamefont{E.}~\bibnamefont{Quataert}}, \bibnamefont{and}
  \bibinfo{author}{\bibfnamefont{E.}~\bibnamefont{Waxman}},
  \bibinfo{journal}{Astrophys.J.} \textbf{\bibinfo{volume}{654}},
  \bibinfo{pages}{219} (\bibinfo{year}{2006}), \eprint{astro-ph/0606665}.

\bibitem[{\citenamefont{Pfrommer et~al.}(2007)\citenamefont{Pfrommer, Ensslin,
  and Springel}}]{Pfrommer:2007sz}
\bibinfo{author}{\bibfnamefont{C.}~\bibnamefont{Pfrommer}},
  \bibinfo{author}{\bibfnamefont{T.~A.} \bibnamefont{Ensslin}},
  \bibnamefont{and} \bibinfo{author}{\bibfnamefont{V.}~\bibnamefont{Springel}}
  (\bibinfo{year}{2007}), \eprint{0707.1707}.

\bibitem[{\citenamefont{Blasi et~al.}(2007)\citenamefont{Blasi, Gabici, and
  Brunetti}}]{Blasi:2007pm}
\bibinfo{author}{\bibfnamefont{P.}~\bibnamefont{Blasi}},
  \bibinfo{author}{\bibfnamefont{S.}~\bibnamefont{Gabici}}, \bibnamefont{and}
  \bibinfo{author}{\bibfnamefont{G.}~\bibnamefont{Brunetti}},
  \bibinfo{journal}{Int.J.Mod.Phys.} \textbf{\bibinfo{volume}{A22}},
  \bibinfo{pages}{681} (\bibinfo{year}{2007}), \eprint{astro-ph/0701545}.

\bibitem[{\citenamefont{Keshet et~al.}(2003)\citenamefont{Keshet, Waxman, Loeb,
  Springel, and Hernquist}}]{Keshet:2002sw}
\bibinfo{author}{\bibfnamefont{U.}~\bibnamefont{Keshet}},
  \bibinfo{author}{\bibfnamefont{E.}~\bibnamefont{Waxman}},
  \bibinfo{author}{\bibfnamefont{A.}~\bibnamefont{Loeb}},
  \bibinfo{author}{\bibfnamefont{V.}~\bibnamefont{Springel}}, \bibnamefont{and}
  \bibinfo{author}{\bibfnamefont{L.}~\bibnamefont{Hernquist}},
  \bibinfo{journal}{Astrophys.J.} \textbf{\bibinfo{volume}{585}},
  \bibinfo{pages}{128} (\bibinfo{year}{2003}), \eprint{astro-ph/0202318}.

\bibitem[{\citenamefont{Gabici and Blasi}(2003)}]{Gabici:2002fg}
\bibinfo{author}{\bibfnamefont{S.}~\bibnamefont{Gabici}} \bibnamefont{and}
  \bibinfo{author}{\bibfnamefont{P.}~\bibnamefont{Blasi}},
  \bibinfo{journal}{Astropart.Phys.} \textbf{\bibinfo{volume}{19}},
  \bibinfo{pages}{679} (\bibinfo{year}{2003}), \eprint{astro-ph/0211573}.

\end{thebibliography}
\bibliographystyle{apsrev}

\end{document}